%% file: upgrade-p2-v5.tex
\documentclass[twocolumn,epjc3]{svjour3}  
\smartqed  
\RequirePackage{graphicx}
\usepackage[bookmarks=false]{hyperref}
\usepackage{color}

\input{gerda-abbreviations.tex}   
\newcommand{\enrCoax}         {{$^{\rm enr}$Coax}}
\newcommand{\enrBEGe}         {{$^{\rm enr}$BEGe}}
\journalname{Eur. Phys. J. C}
\begin{document}

\title{Upgrade for Phase~II of the {\mbox{\protect{\sc{Gerda}}}} Experiment}
 
\titlerunning{{\mbox{{\textsc GERDA}}}  upgrade}  

\author{
The \mbox{\protect{\sc{Gerda}}} collaboration\thanksref{corrauthor}
        \and  \\[4mm]
M.~Agostini\thanksref{TUM} \and
A.M.~Bakalyarov\thanksref{KU} \and
M.~Balata\thanksref{ALNGS} \and
I.~Barabanov\thanksref{INR} \and
L.~Baudis\thanksref{UZH} \and
C.~Bauer\thanksref{HD} \and
E.~Bellotti\thanksref{MIBF,MIBINFN} \and
S.~Belogurov\thanksref{ITEP,INR,alsoMEPHI} \and
S.T.~Belyaev\thanksref{KU,nowDEC} \and
G.~Benato\thanksref{UZH}
% ...
}
\authorrunning{the \textsc{Gerda} collaboration}
\thankstext{corrauthor}{INFN Laboratori Nazionali del Gran Sasso, Italy.\\
\emph{Correspondence},
                                email: gerda-eb@mpi-hd.mpg.de}
\thankstext{alsoMEPHI}{\emph{also at:} NRNU MEPhI, Moscow, Russia}
\thankstext{nowDEC}{deceased}
\thankstext{nowHNI}{\emph{also at:} The Henryk Niewodniczanski Institute of Nuclear Physics PAS, Krakow, Poland}
\thankstext{nowMos}{\emph{also at:} Moscow Institute for Physics and Technology, Moscow, Russia}
\thankstext{nowIKZ}{\emph{present address:} IKZ, Dresden, Germany}
\thankstext{nowCarlton}{\emph{present address:} Carleton University, Ottawa, Canada}
\thankstext{nowParis}{\emph{present address:} LAL, CNRS/IN2P3, Universit{\'e} Paris-Saclay, Orsay, France}
\thankstext{nowQLAQUILA}{\emph{also at:} Dipartimento di Scienze Fisiche e
    Chimiche, Universita{\`a} degli Studi di L'Aquila, L'Aquila, Italy}

\institute{%
INFN Laboratori Nazionali del Gran Sasso and Gran Sasso Science Institute,
Assergi, Italy\label{ALNGS} \and
INFN Laboratori Nazionali del Sud, Catania, Italy\label{LNS} \and
Institute of Physics, Jagiellonian University, Cracow, Poland\label{CR} \and
Institut f{\"u}r Kern- und Teilchenphysik, Technische Universit{\"a}t Dresden,
      Dresden, Germany\label{DD} \and
Joint Institute for Nuclear Research, Dubna, Russia\label{JINR} \and
European Commission, JRC-Geel, Geel, Belgium\label{GEEL} \and
Max-Planck-Institut f{\"u}r Kernphysik, Heidelberg, Germany\label{HD} \and
Dipartimento di Fisica, Universit{\`a} Milano Bicocca,
     Milano, Italy\label{MIBF} \and
INFN Milano Bicocca, Milano, Italy\label{MIBINFN} \and
Dipartimento di Fisica, Universit{\`a} degli Studi di Milano e INFN Milano,
    Milano, Italy\label{MILUINFN} \and
Institute for Nuclear Research of the Russian Academy of Sciences,
    Moscow, Russia\label{INR} \and
Institute for Theoretical and Experimental Physics, NRC ``Kurchatov Institute'',
    Moscow, Russia\label{ITEP} \and
National Research Centre ``Kurchatov Institute'', Moscow, Russia\label{KU} \and
Max-Planck-Institut f{\"ur} Physik, M{\"u}nchen, Germany\label{MPIP} \and
Physik Department and Excellence Cluster Universe,
    Technische  Universit{\"a}t M{\"u}nchen, M\"unchen, Germany\label{TUM} \and
Dipartimento di Fisica e Astronomia dell{`}Universit{\`a} di Padova,
    Padova, Italy\label{PDUNI} \and
INFN  Padova, Padova, Italy\label{PDINFN} \and
Physikalisches Institut, Eberhard Karls Universit{\"a}t T{\"u}bingen,
    T{\"u}bingen, Germany\label{TU} \and
Physik Institut der Universit{\"a}t Z{\"u}rich, Z{\"u}rich,
    Switzerland\label{UZH}
}

\date{File: upgrade-p2-v5.tex, compiled: \today}

\maketitle
\begin{abstract}
The \gerda\ collaboration is performing a sensitive search for neutrinoless
double beta decay of \gesix\ at the INFN Laboratori Nazionali del Gran Sasso,
Italy.  The upgrade of the \GERDA\ experiment from Phase~I to Phase~II has
been concluded in December 2015.  The first Phase~II data release shows that
the goal to suppress the background by one order of magnitude compared to
Phase~I has been achieved.  \GERDA\ is thus the first experiment that will
remain background-free up to its design exposure (100 kg$\cdot$yr). It will
reach thereby a half-life sensitivity of more than 10$^{26}$ yr within 3 years
of data collection.  This paper describes in detail the modifications and
improvements of the experimental setup for Phase~II and discusses the
performance of individual detector components.
\end{abstract}
\section{Introduction} \label{sec:intro}
Neutrinoless double beta (\onbb ) decay is a hypothetical lepton number
violating process, (A,Z)$\rightarrow$(A,Z+2)\,+\,2e$^-$, where inside a
nucleus two neutrons convert into two protons and two electrons. Its
observation would establish the neutrino to be its own anti-particle (Majorana
particle), provide access to the absolute mass scale of neutrinos, and support
extensions of the Standard Model of particle physics which try to explain the
dominance of baryonic matter over anti-matter in our
universe~\cite{moha}. Recent experiments have established the half-life of
\onbb\ decay to be larger than 10$^{25}$\,yr~\cite{pdg}, and hence its
detection requires the utmost suppression of any kind of background.

The GERmanium Detector Array (\gerda) collaboration searches at the INFN
Laboratori Nazionali del Gran Sasso (\lngs) for \onbb\ decay of \gesix, \gesix
$\rightarrow$ $^{76}$Se\,+\,2e$^-$. The \gerda\ experiment has been conceived
in two phases.  Here, a short overview of the experimental setup of Phase~I
which lasted from November 2011 until September 2013 is presented, while a
detailed description has been given elsewhere~\cite{g1-instr}.
\begin{figure}[htb!]
\begin{center}
\includegraphics[width=0.95\columnwidth]{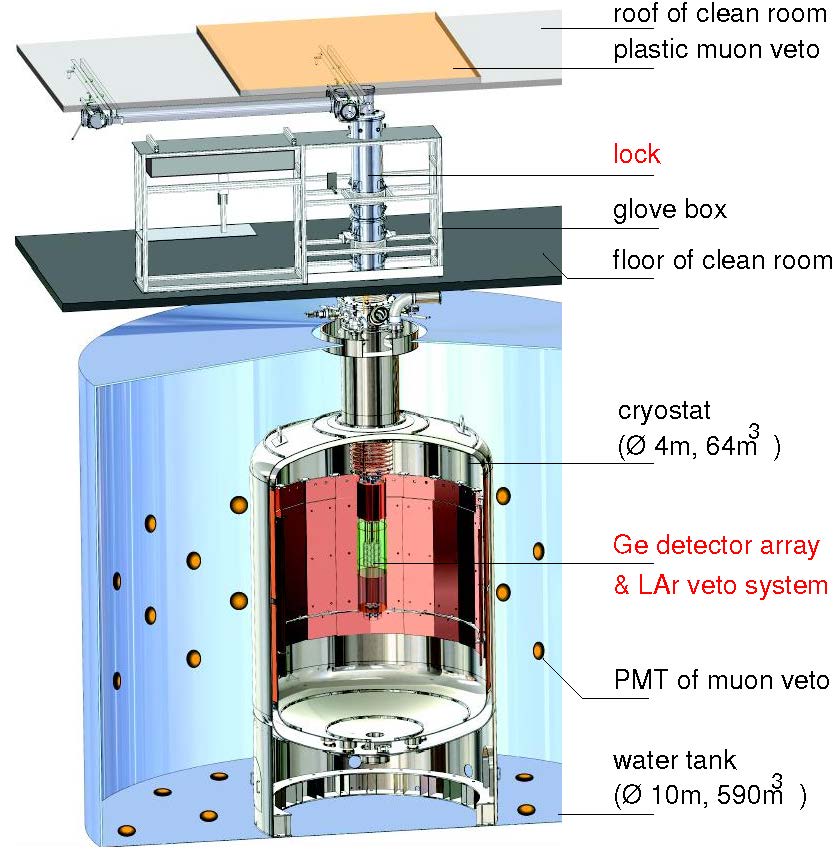}
   \caption{\label{fig:g-setup}
         \gerda\ setup. The new Phase~II components are labeled in red.
}
\end{center}
\end{figure}

The \gerda\ experiment is located underground below a rock overburden of about
3500\,m water equivalent that eliminates the hadronic component of cosmic ray
showers and reduces the muon flux to $\sim$1.25/(m$^{2}\cdot$h).  \gerda\ uses
high purity germanium (HPGe) detectors enriched in \gesix\ which are arranged
in strings within a cryostat filled with 64\,m$^3$ of liquid argon (quality
5.0), see Fig.~\ref{fig:g-setup}.  The liquid argon (LAr) acts both as cooling
and shielding medium. The cryostat itself is enclosed by a large tank
containing 590\,m$^3$ of ultra-pure water produced by the \borex\ water
plant~\cite{borexwater}; the 2-3\,m thick water layer serves both as
additional passive shield as well as the medium for a Cherenkov veto system
with 66 photomultiplier tubes (PMTs) against muons.  A clean room on top of
the cryostat and water tank houses a glove box and the lock for assembly and
deployment of the Ge detectors.

The Phase~1 detector array, 4 strings in total, consisted of 8 enriched
semi-coaxial Ge detectors with a total mass of 15.6\,kg and 3 semi-coaxial Ge
detectors from low-background natural material.  The one string of natural Ge
detectors was replaced in July 2012 by 5 Broad Energy Germanium (BEGe)
detectors with a total mass of 3.6\,kg; these diodes served as prototypes for
Phase~II.

The physics results of Phase~I~\cite{g1-prl} were based on an exposure of
21.6\,kg$\cdot$yr. The energy scale was determined by (bi)weekly calibrations
with \Th\ sources.  In the region of interest (ROI) around Q$_{\beta\beta}$\,=
2039\,keV, the interpolated exposure-averaged energy resolution of the
enriched semi-coaxial and BEGe detectors was determined to be
4.8(2)\,keV and 3.2(2)\,keV in terms of
full-width-at-half-maximum (FWHM), respectively.  A background index (BI) of
about 10$^{-2}$\,\ctsper \ was achieved, one order of magnitude lower than in
the best previous \onbb\ decay searches with \gesix.  No signal was found for
\onbb\ decay, and a new 90\,\% confidence level (CL) limit of
\thalfzero\,$>\,$2.1$\cdot10^{25}$\,yr was derived (median sensitivity
$2.4\cdot10^{25}$\,yr) that strongly disfavored a previous claim of
observation~\cite{Klap}.  Further Phase~I results include a much improved
half-life for \nnbb\ decay of \gesix\ and improved limits for Majoron
$\beta\beta$ decay modes~\cite{g1-majoron}, as well as \nnbb\ decays of
\gesix\ into excited states of $^{76}$Se~\cite{g1-nnbb_exSe}.

\begin{figure}[htb]
\begin{center}
\includegraphics[width=0.8\columnwidth]{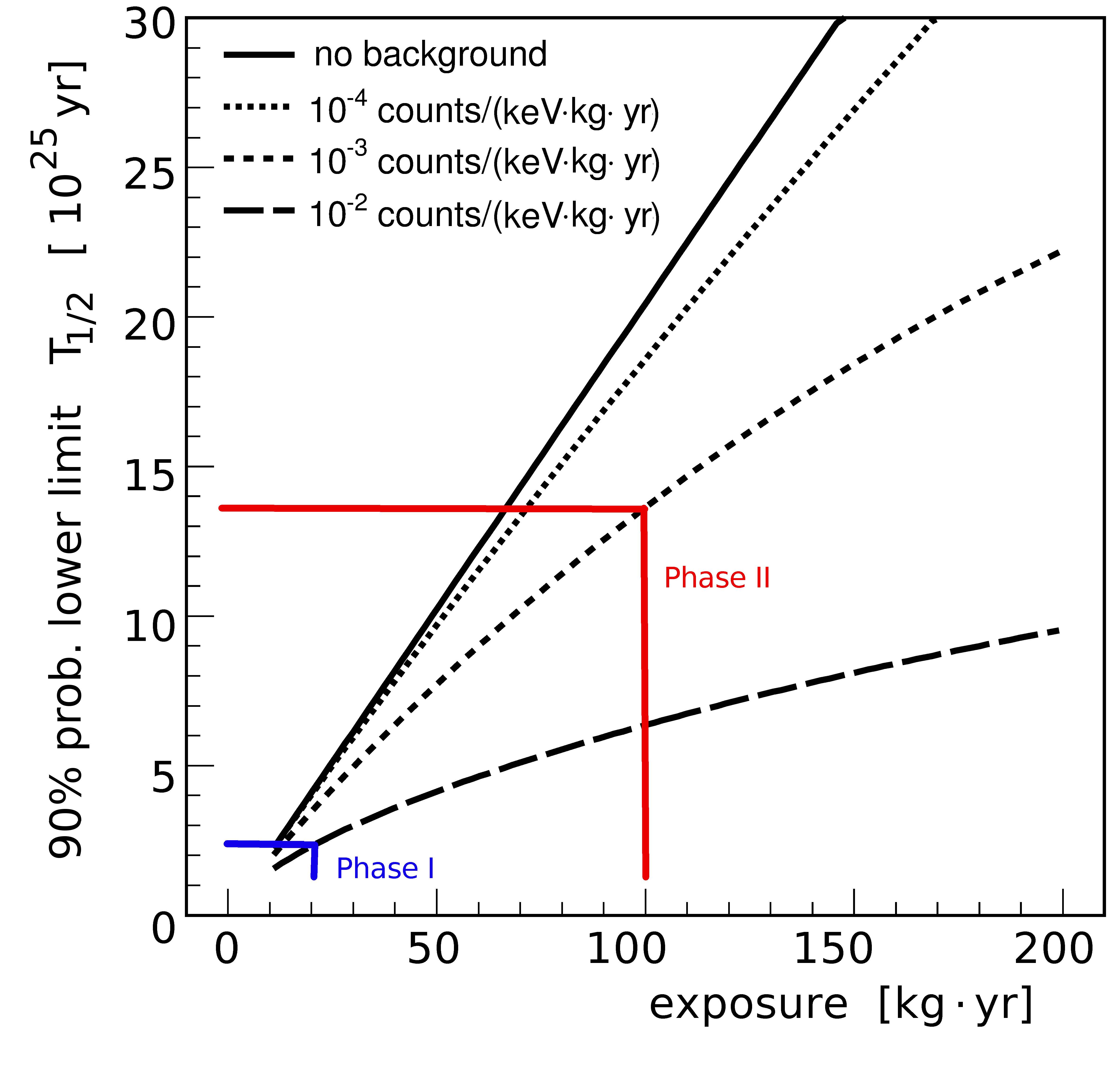}
\caption{\label{fig:g2-sensitivity} 
    Estimated sensitivity of the \gerda\ experiment as a function of exposure
    for various background indices. The scenarios for \gerda\ Phase~I and II
    are indicated~\cite{Al+Kev}.
}
\end{center}
\end{figure}

Phase~II of \gerda\ was designed to improve the sensitivity on the half-life
of \onbb\ decay by about one order of magnitude. At the end of Phase~I
\gerda\ had left the background-free regime%
\footnote{The background-free regime is effective if the background
  contribution is less than 1 in the energy region of interest
  (\qbb$\pm$0.5\,FWHM).}  
where sensitivity scales linearly with exposure ${\cal E}$\,=\,M$\cdot$t, the
product of detector mass M and measurement period t, and entered the
background i.e. statistical fluctuation limited scenario where it scales
approximately with the square root of exposure divided by the background index
BI and the energy resolution $\Delta$E, $\sqrt{({\cal E}/(BI\cdot \Delta E)}$.
An efficient upgrade requires thus to re-enter the background-free re\-gime,
that means to not only increase exposure (detector mass) but also to reduce
correspondingly the background (see Fig.~\ref{fig:g2-sensitivity}).
Improvements of the already excellent resolution are possible but limited to a
factor of about 2 for the given technology.
\gerda\ thus needs to achieve a BI of \pIIbi\ in Phase~II in order to reach
the desired sensitivity beyond 10$^{26}$\,yr at an exposure of about
100\,\kgyr\ (see Fig.~\ref{fig:g2-sensitivity}).

The analysis of the Phase~I data showed that most background events were due
to radioactive isotopes in materials close to the
detectors~\cite{g1-back}. The straightforward consequence was to further
reduce material close to the detectors and/or to replace it by material of
higher radiopurity.  The major BI reduction had to come, however, from a
largely improved discrimination of background events taking full advantage of
their different event topology. While \onbb\ events normally deposit energy in
a confined volume (a few mm$^3$) of the detector, the background events can
also deposit energy in the LAr around the detector, at the detector surface,
or scatter at several locations in the detector.  Events can thus be
identified as background by coincident scintillation light in the LAr, by
coincidences within the detector array and/or by the analysis of the signal
pulse shape.  \GERDA\ has taken full advantage of all these options in
Phase~II: the additional batch of 20\,kg of enriched Ge detectors consists of
diodes of the novel BEGe type exhibiting superior pulse shape discrimination
(PSD)~\cite{g1-psd} and energy resolution; a larger and more densely packed
detector array exhibits enhanced efficiency for detector-detector
(anti-)coincidences, and importantly, the LAr around the detector array has
been instrumented for the readout of scintillation light creating thus an
effective active LAr veto system.  The efficacy of this approach has indeed
been proven by the first results obtained with the upgraded
\gerda\ experiment. Started in December 2015, the Phase~II physics run reached
in June 2016 the exposure of 10.8\,\kgyr. These accumulated data have been
already sufficient to demonstrate that the projected background level of
\dctsper\ has been achieved and, to extract in combination with the Phase~I
data set a new lower limit for the \onbb\ decay half-life of \gesix\ of
$>5.3\cdot10^{25}$\,yr at 90\,\% CL~\cite{g2-nature}.

The following sections describe the modifications of the \gerda\ experimental
setup for Phase~II including the new detector components and their
performance.  Section~2 presents an overview of the properties of the coaxial
and BEGe detectors making up the Phase~II detector array; it provides also
details of the new mechanical mounts, cabling, electrical contacts and the
cold electronic front end.  A major part of this paper, Section~3, is devoted
to the LAr veto system.  Section~4 discusses the modifications of the
infrastructure, in particular the new lock needed for the largely increased
detector array and the LAr veto system.  Section~5 summarizes the screening
results for the newly introduced components.  The performance of the
individual subsystems and the background level achieved in Phase~II until
April 2017 are presented in Section~6. Conclusions are given in Section~7.
\section{Germanium detectors} \label{sec:detectors-mounting} 
 
\subsection{Characteristics of Phase~II detectors}

The \GERDA\ Phase~II detector array includes 7 strings, which carry 40
detectors in total. The detectors can be divided into three groups: the newly
produced BEGe detectors, the semi-coaxial ANG and RG detectors, and the
semi-coaxial GTF detectors~\cite{g1-instr}.  The detectors of the first two
groups are made of germanium enriched in $^{76}$Ge (\enrBEGe, \enrCoax), while
those of the third group are made of germanium with natural isotopic abundance
(\natcoax).  The main properties of the individual detectors groups are
discussed below, the properties of the individual detectors are listed in the
Appendix (see Table~\ref{table:detproperties}).

\subsubsection{The semi-coaxial detectors}
The 7 semi-coaxial $^{76}$Ge enriched detectors (ANG, RG), which originated
from the former Heidelberg-Moscow and IGEX experiments, represented the core
of \GERDA\ Phase~I~\cite{g1-instr}. In \GERDA\ Phase~II, they have again
been included.

The $^{76}$Ge enrichment fractions of the ANG and RG detectors are in the
range of 85.5\,\% to 88.3\,\%. All \enrCoax\ detectors have masses greater than
2\,kg, except ANG1. The total mass is 15.578(7)\,kg.  The applied bias
voltages coincide nearly with those $V_{rec}$ recommended by the
manufacturer~\cite{PhDmarik}.  The $^{76}$Ge content of the 3 natural GTF
detectors corresponds to the natural abundance of 7.8\,\%.

\subsubsection{The BEGe detectors}
In order to increase the exposure (via increase of total detector mass) and to
improve the background index (via an enhanced pulse shape performance), the
\GERDA\ collaboration opted for the production of 30 new detectors following
the BEGe design of the company Canberra~\cite{g2-begeprod}. After extensive
preparation, 30 detectors were delivered. Only one detector, GD02D, turned out
to be rather a p-n junction than of p-type material, and thus suitable only
for anti-coincidence studies in \GERDA.

The 30 \enrBEGe\ detectors have an enrichment fraction of 87.8\,\%.  The
diameters of the detectors range from 58.3(1)\,mm to 79.0(1)\,mm, and their
heights from 22.9(3) to 35.3(1)\,mm.  In 21 cases the detectors are cylindric,
in 9 cases they have a conical shape. The latter shape was tolerated in order
to maximize the number of crystal slices that can be obtained from one single
crystal ingot.  The total mass is 20.024(30)\,kg.  Herein, the $\pm1$\,g error
from weighing was assumed to be correlated for all detectors. Neglecting the
detector GD02D, the total detector mass is reduced to 19.362(29)\,kg.

The average active volume fraction $f_{av}$ and the total active mass $M_{av}$
of all 29 fully operational \textsc{Gerda} Phase~II BEGe detectors have been
determined combining the full charge collection depth (FCCD) results from
$^{241}${Am} and $^{133}${Ba} source measurements conducted in the HADES
underground laboratory~\cite{heroica}. Moreover, an increase of the FCCD of
0.2 to 0.3\,mm due to storage at room temperature over a period of nearly 3
years has been considered. All in all this led to:
\begin{equation}
f_{av} =  0.885^{+0.016}_{-0.015} (uncorr)\ \ \ ^{+0.006}_{-0.003} (corr) 
\end{equation}
\begin{equation}
M_{av} =  17.132^{+0.315}_{-0.294} (uncorr)\ \ \ ^{+0.123}_{-0.063} (corr)\,\,{\rm{kg}}.	
\end{equation}

Compared to the initial purified Ge powder used for crystal pulling, a crystal
mass yield of 50.1\,\% and an active mass yield of 48.2\,\% were achieved (see
Table~\ref{table:mass-yield}). Considering that approximately 25\,\% of the
kerf material could be recovered, the achieved total mass yield is high.
\begin{table}[ht]
\begin{center}
\caption{\label{tab:mass-yield}
    Crystal and active mass yield in \% from enriched germanium to the final 30 BEGe
    detectors for \gerda\ Phase~II. The mass transfer fractions are given
    relative to the original enriched GeO$_2$ material (3rd column) and to the
    purified  metallic Ge used for crystal growth (4th column).
}
\label{table:mass-yield}
\begin{tabular}{lcrr}
\hline
\multicolumn{1}{c}{~~~~~~~~germanium}   & mass    &\multicolumn{2}{c}{relative fraction}	\\
{operation}			&	(kg)
    &\multicolumn{1}{c}{ 
  [\%]}	&\multicolumn{1}{c}{ [\%]}\\
\hline 
Ge in GeO$_2$ after enrichment    		& 37.5 & 100.0 & --\textcolor{white}{....}\\
purified Ge for crystal growth    		& 35.5 &  94.1\ & 100.0\\
30 produced diodes		             	& 20.0 &  53.3\ &  56.3\\
29 operational diodes	             		& 19.4 &  51.7\ &  54.6\\
active mass of 29 diodes	 		& 17.1 &  45.6\ &  48.2\\
\hline
\end{tabular}
\end{center}
\end{table}
 
The bias voltages applied on the detectors in \GERDA\ coincide typically with
the values $V_{rec}$ recommended by the manufacturer (see Appendix
Table~\ref{table:detproperties}).  The collaboration, however, performed
detailed voltage scans, in which depletion voltages and new operational
voltages were determined which still guarantee uncompromised detector
response.  The new values are 600\,V lower, on average. Indeed,
\GERDA\ operates in a few cases the detectors at these lower values,
preventing thus unwanted high leakage currents or other instabilities.

The \enrBEGe\ detectors were characterized in vacuum cryostats within the HADES
underground laboratory~\cite{heroica}.  Note that these measurements were
performed with `passivated' detectors where the groove between the p$^+$ and
n$^+$ electrode is covered by an insulating silicon monoxide (`passivation')
layer~\cite{g2-begeprod}.  The energy resolution of all detectors turned out
to be excellent.  At the 1333\,keV $^{60}$Co $\gamma$-line the FWHM energy
resolution is 1.72(7)\,keV, with best and worst values of 1.59\,keV
(GD89A) and 1.87\,keV (GD79C).  While a small dependence on the detector mass
became visible, no dependence on the detector shape could be observed,
i.e. cylindric and conical ones have the same performance.

The pulse shape discrimination power was deduced from $^{228}$Th source
measurements.  For an event selection criterion keeping 90\,\% of signal-like
events (double-escape peak events from the \Tl\ line), the following survival
probabilities for background-like event populations were found: the
single-escape peak is reduced to (5-12)\,\%, the full-energy peaks at 2615 and
1620\,keV survive at (6-19)\,\% and (9-19)\,\%, respectively, and the
Compton-events in the ROI are in the range of (32-48)\,\%.

\subsubsection{Electrical contacts}
Aluminum bonding pads of 600\,nm thickness have been evaporated on the p$^+$
signal and n$^+$ high-voltage electrodes of all Phase~II detectors. The
evaporation process had been developed in collaboration with the company
Canberra and subsequently integrated together with the dedicated evaporation
system into the manufacturer's production chain~\cite{PhDtobias}.

\subsubsection{Activation by cosmic rays}
Great attention has been paid to minimize the activation of the newly produced
BEGe detectors by cosmic rays~\cite{g2-begeprod}.  As of January 1 2017, we
expect from cosmic activation in the 30 BEGes detectors (20.02 kg) about 24
$^{68}$Ge nuclei and 300 $^{60}$Co nuclei\footnote
{Activation rates at sea level vary from 1 to 13 and 1.6 to 6.7
  nuclei/(kg$\cdot$d) for $^{68}$Ge and $^{60}$Co nuclei, respectively. Here,
  we assume 5.8 and 3.3 nuclei/(kg$\cdot$d)~\cite{g2-begeprod}.};
for comparison, the saturation activity is $\sim$2300 $^{68}$Ge nuclei and
$\sim$9200 $^{60}$Co nuclei per kg germanium, respectively.  Simulations show
that during the year 2017 the decay of these nuclei will increase the background index
before LAr veto and PSD by 2.5$\cdot$\vctsper\ and 0.34$\cdot$\vctsper,
respectively.  For the semi-coaxial enriched detectors, the $^{68}$Ge
activation has vanished due to their long underground storage; the $^{60}$Co
contribution is about 1/3 of the expectation for the BEGe detectors.

\subsection{Detector Mount and Cabling}
The \gerda\ Phase~I background has been shown to originate predominantly from
sources close to the Ge detectors.  In addition, since the average mass ratio
of BEGe to semi-coaxial detectors is about 1:3, further optimization of the
detector mount for Phase~II required to reduce the amount of construction
materials and/or to improve their radiopurity.
\begin{figure}[htb]
\begin{center}
\includegraphics[width=0.5\columnwidth]{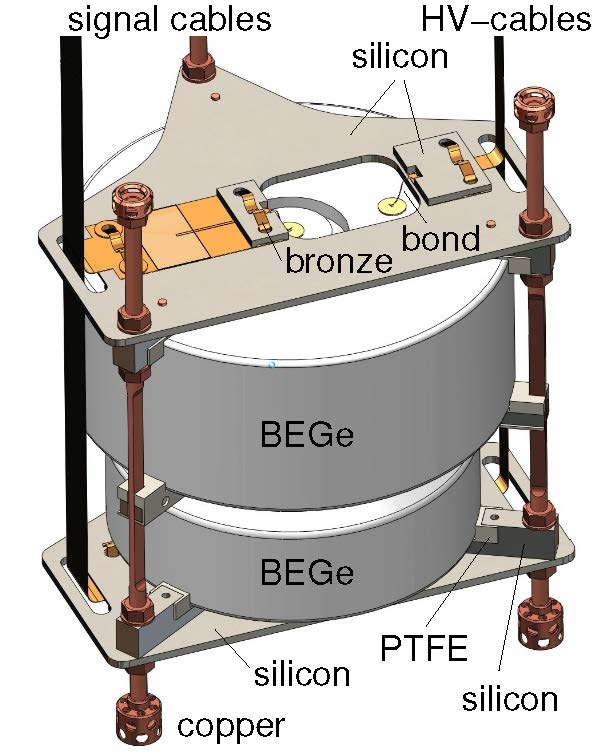}
\vskip0.7truecm
\includegraphics[scale=0.2]{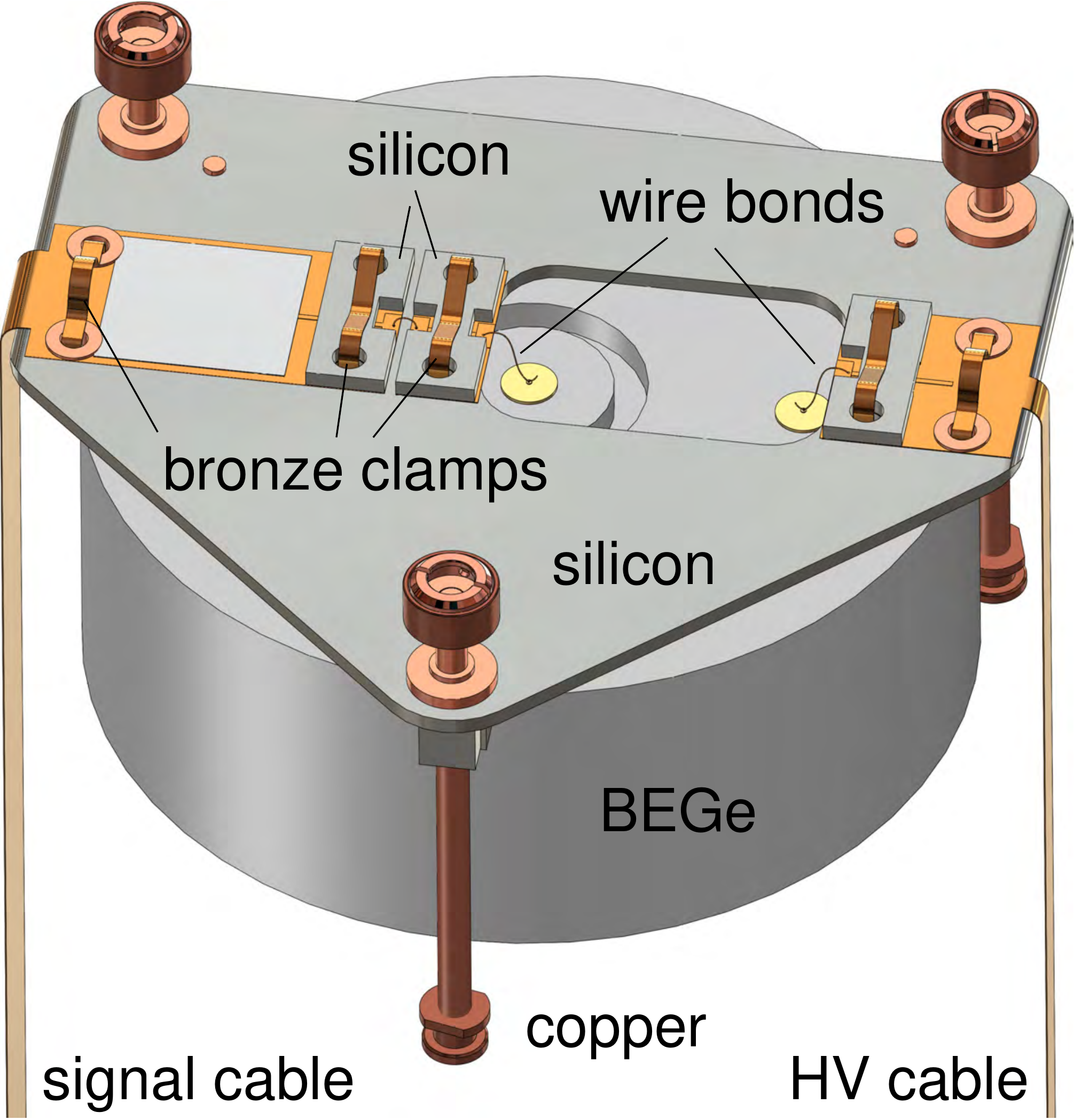}
\caption{\label{fig:bege-mount}
     Alternative mounts for pairs of and single BEGe detectors, and their
     connection to the signal and high voltage (HV) flexible cables. The
     single detector module is viewed from bottom.
}
\end{center}
\end{figure}
Fig.~\ref{fig:bege-mount} shows the original Phase~II BEGe detector module
which consists of 2 BEGe Ge diodes that are mounted back-to-back. The new
design replaces the Phase~I spring loaded contacts to the detector electrodes
by 25\,\mum\ diameter Al wire bonds.  This allowed the substitution of a
large part of the Phase~I copper material (and PTFE) by mono-crystalline
silicon which is less strong but intrinsically extremely radio-pure (see
Tables~\ref{tab:massperkg} and~\ref{tab:detholder}). The silicon plate serves
both to define the position of the vertical copper bars which take the weight
of the Ge detectors and to provide the substrate onto which signal and high
voltage cables are attached with bronze clamps. The top and bottom of the
copper bars carry bolts and nuts for the connection to another detector
module.

\begin{table}[htb]
\begin{center}
\caption{\label{tab:massperkg}
    Comparison of the masses $m$ of construction materials for the Phase~I and
    II detector holders, and of the masses $m'$ normalized to 1\,kg of
    detector mass assuming an average mass of 0.67\,kg for single BEGe and
    2.4\,kg for semi-coaxial detectors.
}
\begin{tabular}{lccccc}
\hline
Material  &  \multicolumn{2}{c}{Phase I (coaxial) }&\hphantom{q}&\multicolumn{2}{c}{Phase II (BEGe)} \\
\cline{2-3} \cline{5-6} \\ [-2ex]
	  &  $m$ (g) & $m'$ (g/kg)&   & $m$ (g) & $m'$ (g/kg)   \\
\hline    	  
Cu        &  84  & 35.0 &   &  13  & 19.4 \\   
Si        &   1  &  0.4 &   &  20  & 29.9 \\
PTFE      &   7  &  2.9 &   &   1  &  1.5 \\
CuSn6     &   -  &   -  &   & 0.5  & 0.75\\
\hline
total     &  92  & 38.3 &   & 34.5 & 51.55 \\
\hline
\end{tabular}
\end{center}
\end{table}
Another advantage of the new holder is that the detector mounting procedure
becomes easier and safer than in Phase~I, since all mounting steps except
bonding are being done without touching the diode. Thus the possibility to
scratch the very sensitive p$^+$ contact is minimized. Some detectors have
been mounted and dismounted in their Phase~II holders several times without
any deterioration of their performance. Bonding also solved the previous
problems of irreproducible HV contact quality.

During commissioning for Phase~II, so far unknown problems with detector
biasing and leakage currents showed up that were highly correlated with the
orientation of BEGe detectors such that detectors with the groove pointing
upwards, the `top' detectors in the mount, were much more affected than the
`bottom' detectors with the groove pointing downwards. In Phase~I all
detectors had been indeed mounted with the groove pointing downwards. The
problem could be attributed to microscopic particulates which had fallen into
the groove during the mounting procedure or during operation in the
LAr. Hence, the concept of mounting pairs of BEGe detectors was given up, and
single BEGe detectors were mounted individually like the semi-coaxial
detectors avoiding grooves pointing upwards (see Fig.~\ref{fig:bege-mount},
bottom). By the start of the Phase~II physics run, all problematic BEGe pair
assemblies and also the newly deployed detectors received the new holders
which virtually eliminated the previous problems. It is planned to replace the
mount of the 6 remaining BEGe pair assemblies in a forthcoming maintenance
break.
  \begin{figure}[!ht]
  \begin{center}
  \includegraphics[width=0.8\columnwidth]{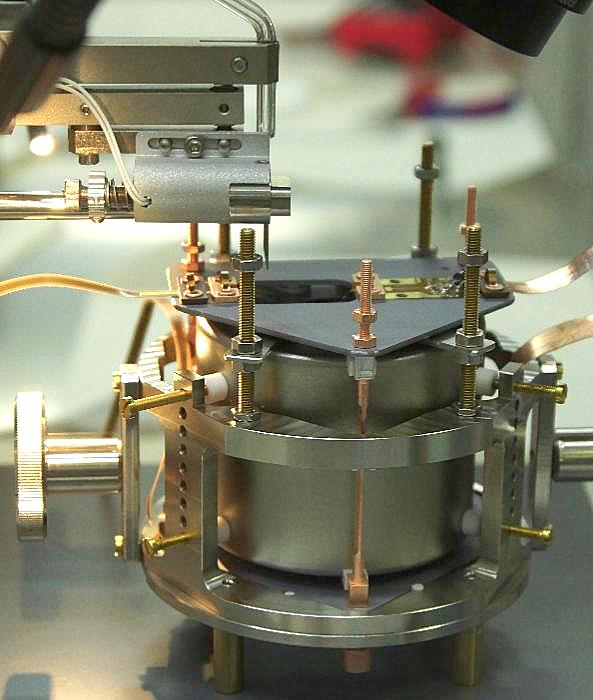}
  \caption{\label{fig:jigbonder}
          A detector module with two BEGe detectors within its jig. The top
          plate of the jig has been removed to provide full access for the
          deep-access bonder.
}
  \end{center}
  \end{figure}

The low-mass Phase~II detector mount with its silicon plate has no tolerance
against crack-producing stress, thus requires the detector assembly to be kept
in a jig until finally deployed in the detector string. The rather compact jig
is made from stainless steel and thoroughly cleaned in order to avoid
contamination of the detector module.  Fig.~\ref{fig:jigbonder} shows a jig
and detector module within a glove box, flushed with nitrogen, in front of a
deep-access bonder ready for connecting the flexible cables with the signal
and HV contact of the top BEGe detector.  After turning the jig for bonding
the second detector, jig and module are stored in their dedicated vacuum
container until deployment.

After mounting and bonding, each detector was tested for leakage current in
the \gerda\ Germanium Detector Laboratory (GDL) which is also located
underground at \LNGS.  All detectors that passed this test were integrated in
the \gerda\ setup. However, about half of the diodes showed again high leakage
currents already in the GDL test bench. They were dismounted from their
holders and sent to Canberra for reprocessing and, in some cases,
passivation. On return they were mounted again in the holders, tested in GDL
and added to the \gerda\ setup at the final stage of integration.

\subsection{Ge detector readout}
\begin{figure}[htb!]
\begin{center}
\includegraphics[width=0.98\columnwidth]{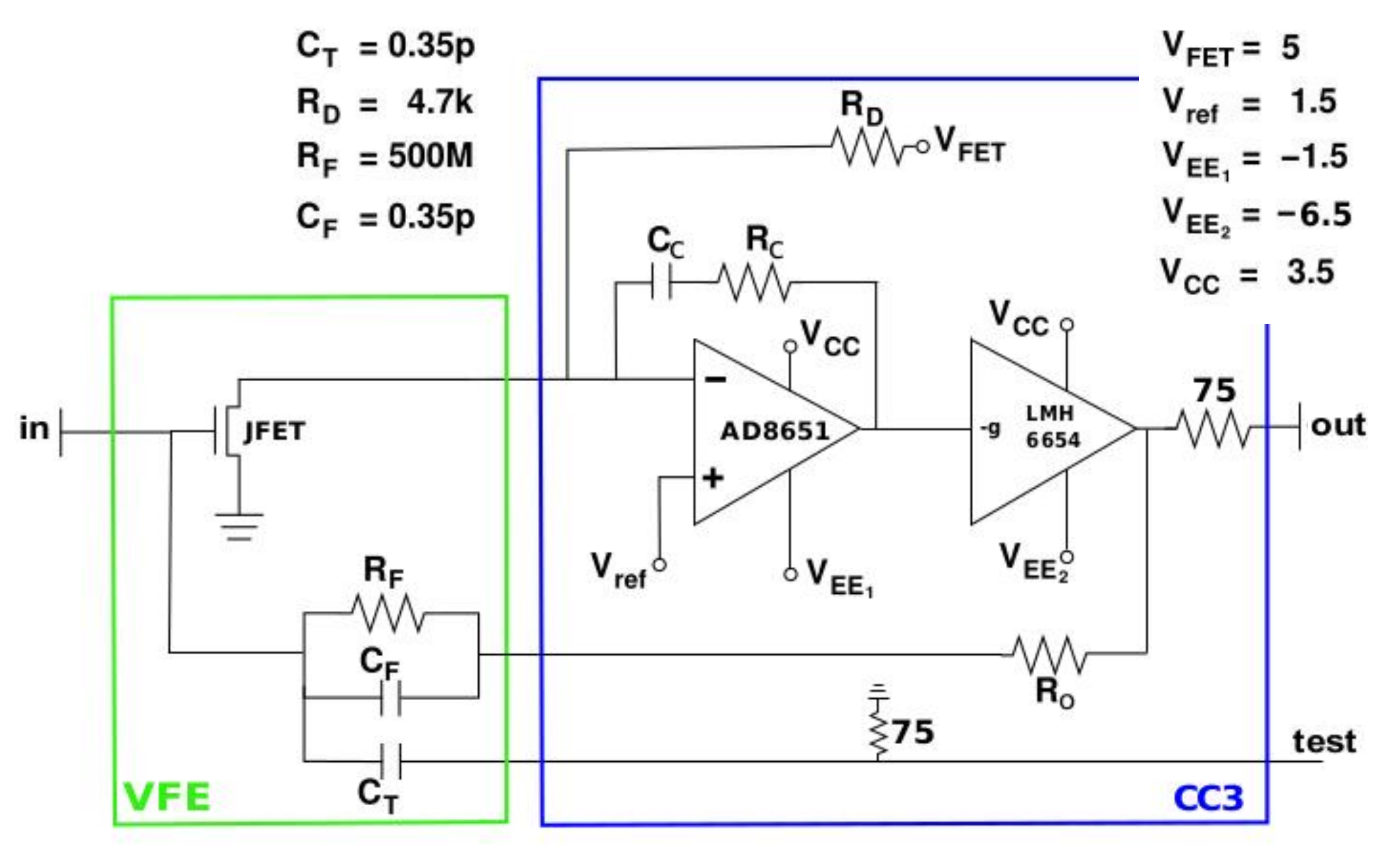}
\caption{\label{fig:cc3schem}
         Schematic of one channel of the CC3 Ge readout circuit. The green
         frame shows the very front end (JFET, feedback resistor and
         capacitor), the blue frame the following stages.
}
\end{center}
\end{figure}

The Ge detectors are read out with custom-produced preamplifiers called `CC3'.
Fig.~\ref{fig:cc3schem} shows the schematic of this cryogenic, low
radioactivity, 4-channel charge sensitive amplifier~\cite{cc2,cc3}.  It is a
continuous resistor reset amplifier, made of two amplifying stages based on
commercial CMOS operational amplifiers.
\begin{figure}[htb!]
\begin{center}
\includegraphics[scale=0.6]{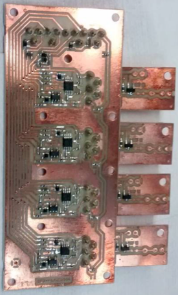}
\caption{\label{fig:cc3pict}
      Photo of the CC3 Ge 4-channel readout circuit.  Each circuit serves 4
      channels.  The very front end components (see Fig.~\ref{fig:cc3schem})
      are located on the 4 small PCBs which are connected with low activity
      pin-contacts to the main board; this allows us to replace individually
      for each channel the components of the very front end, and in particular
      the rather sensitive JFET.
}
\end{center}
\end{figure}
Miniaturized SMD components and Ta capacitors, mostly in 0402 size, are chosen
to minimize the radioactivity of the circuit (Fig.~\ref{fig:cc3pict}).  The
input stage is the BF862 JFET from NPX. The CC3 circuit has a sensitivity of
$\sim$150\,mV/MeV, a dynamic range of $\sim$15\,MeV, a rise-time
(10\,\%-90\,\%) of $<$100\,ns, $<$70\,mW/ch power consumption, and an
intrinsic noise of $\sim$0.8\,keV Ge-equi\-va\-lent.

The Ge readout electrode is connected to the JFET-PCB by a flexible flat cable
(FFC), made from Pyralux\textsuperscript{\textregistered} or
Cuflon\textsuperscript{\textregistered}. This allows for a detachable contact
between the detector and the main front end board, without any glueing or
soldering at the detector level, hence minimizing the activity close to the
detectors while maximizing the contact reliability.  Two different FFCs are
adopted for the signal and HV contact: the HV FFCs are made from 10 mils
Cuflon\textsuperscript{\textregistered}, or 3 mils
Pyralux\textsuperscript{\textregistered}, the signal FFCs from 3 mils
Cuflon\textsuperscript{\textregistered} or
Pyralux\textsuperscript{\textregistered}.  For all FFCs the Cu trace is 2\,mm
wide and 18\,\mum\ thick.
		      
\subsection{The seven detector strings}
\begin{figure}[htb!]
\begin{center}
\includegraphics[scale=0.5]{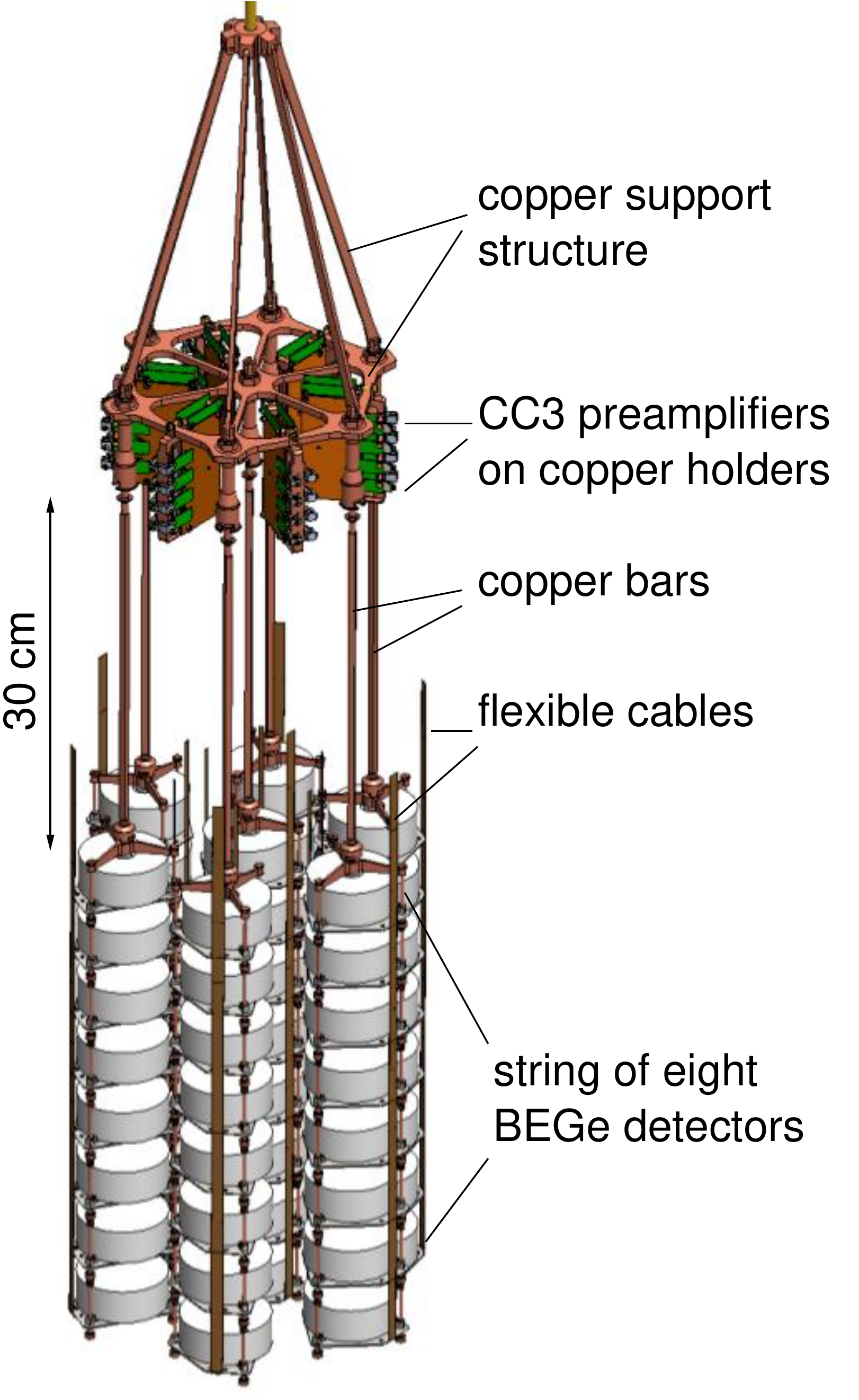}
\caption{\label{fig:stringscc3}
                Arrangement of the seven detector strings with preamplifiers.
}
\end{center}
\end{figure}

Fig.~\ref{fig:stringscc3} shows the Ge detector array together with the
electronic front end boards on top in about 30\,cm distance.  The height of
the array is 40\,cm, its diameter is about 30\,cm.
There are 7 strings with 40 detectors in total. Six strings consist either of
8 BEGe or 3 semi-coaxial (enriched or natural) detectors. One string is a
mixture of semi-coaxial and BEGe detectors.  A photo of the detector array is
shown in Fig.~\ref{fig:arrayfoto} (see Appendix).

Further figures in the Appendix show the array from the top including the
location of the calibration sources as well as the inner boundary of the LAr
veto system (Fig.~\ref{fig:arfromtop}) and the detailed arrangement of all
detectors (Fig.~\ref{fig:stringcontent}).  Detectors marked with blue are
passivated.  It should be noted that, contrary to Phase~I experience, no
leakage current increase has been found for neither passivated nor
non-passivated diodes after one year of operation (see
Section~\ref{sec:performance} and Fig.~\ref{fig:bslvst}).

\subsection{The mini-shrouds for mitigating \(^{42}\)K background}
The background due to $^{42}$Ar is prominent in \GERDA. $^{42}$Ar decays into
$^{42}$K, which is a $\beta$ emitter with an endpoint energy of 3.5\,MeV. A
copper cylinder, called `mini-shroud' (MS), placed around the detectors was
used for mitigation of $^{42}$K background in Phase~I~\cite{g1-instr}. The
MS screens the electric field of the detector and creates a mechanical barrier
which prevents the collection of $^{42}$K ions on the detector surface.  The
volume from which $^{42}$K collection takes place becomes much smaller and
thus the level of $^{42}$K background decreases. However, the copper MS cannot
be used in \GERDA\ Phase~II since the LAr scintillation veto is implemented
for the suppression of various backgrounds. Scintillation light generated
inside the copper MS would not be visible by the LAr instrumentation and the
efficiency of the LAr veto system would be considerably reduced. Another
reason for the development of a new MS is the higher demands on radiopurity
which would not have been met by the copper MS of Phase~I.

That is why for Phase~II a new MS made from ultra-pure nylon was
developed~\cite{Lubash}.  A photo of the detector array with each string
enclosed by its individual transparent MS is shown in Fig.~\ref{fig:arrayfoto}
of the Appendix.  Such a nylon MS does not screen the electric field of the
detector like a copper one, but serves just as a barrier that stops the drift
of $^{42}$K ions towards the detectors. The nylon films were provided by
Princeton University. They were fabricated for the \borexino\ internal
balloon~\cite{bor_vessel}.  The thickness of the films is 125\,\mum. Similar
to other plastics materials, nylon is almost opaque for the deep ultraviolet
radiation generated in LAr. Hence it is covered on both sides with a
wavelength shifter (WLS) based on tetra-phenyl-butadiene (TPB), that shifts
the 128\,nm scintillation light to wavelengths of about 450\,nm, suitable for
transport through the nylon and detection by the LAr veto system.

\begin{figure}[htb!]
\centering
\includegraphics[width=1\columnwidth]{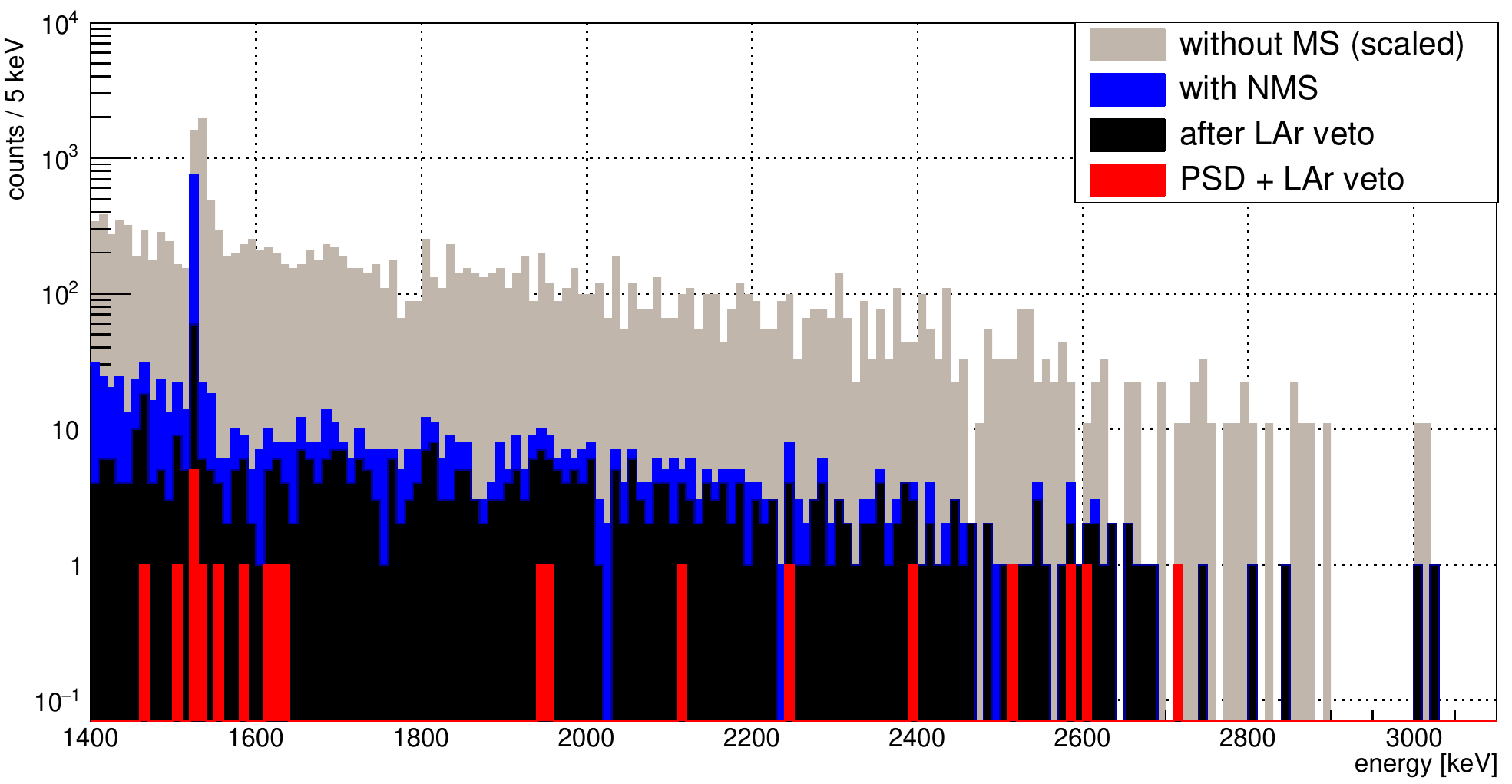}
\caption{\label{fig:42Kmitigation}
         Suppression of the events from $^{42}$Ar decays by the nylon
         mini-shroud (NMS), LAr veto and PSD: the
         grey area shows the scaled spectrum from a measurement with a bare
         BEGe detector without NMS, `blue' the measurement with NMS, `black'
         the events surviving the LAr veto cut, and `red' the remaining events
         after both the LAr veto and PSD cuts~\cite{Lubash}.
}
\end{figure}
The investigation and development of the background suppression methods were
done at the low-background test facility \LArGe~\cite{LArGe} in GDL. It was
demonstrated that the nylon MS is robust enough to be deployed into the
cryogenic liquid and that it does not deteriorate the LAr scintillation veto
performance. By reducing the collection of $^{42}$K atoms at the surface of
the Ge detector, and by combining the PSD technique with the scintillation
veto, it was possible to decrease the $^{42}$K background by more than a
factor of 10$^3$ (see Fig.~\ref{fig:42Kmitigation}).

\section{LAr veto system}	 

The liquid argon veto system (LAr veto) of \GERDA\ is a detector system
devised to detect argon scintillation light in the vicinity of the Ge detector
array.  It evolved from studies of scintillation light detection in LAr with
8" PMTs in the low-background facility \LArGe~\cite{LArGe} and silicon
photomultipliers (SiPMs) coupled to wave length shifting fibers for increasing
light detection efficiency~\cite{janis1}.  The goal is to reject those types
of background events in the Ge detectors that simultaneously deposit energy in
the surrounding LAr, and hence generate scintillation.  These background types
mainly include $\gamma$-ray background from Ra and Th decays in solid
materials inside and around the detectors. But also other types of background
can successfully be rejected, such as muons or decays from $^{42}$Ar/$^{42}$K.
The concept of a LAr anti-coincidence veto has been proven in \LArGe\ where
suppression factors of up to a few times $10^3$ were achieved depending on the
background type and source distance towards the Ge detectors~\cite{LArGe}.

\begin{figure}[h]
\begin{center}
\includegraphics[width=1.0\columnwidth]{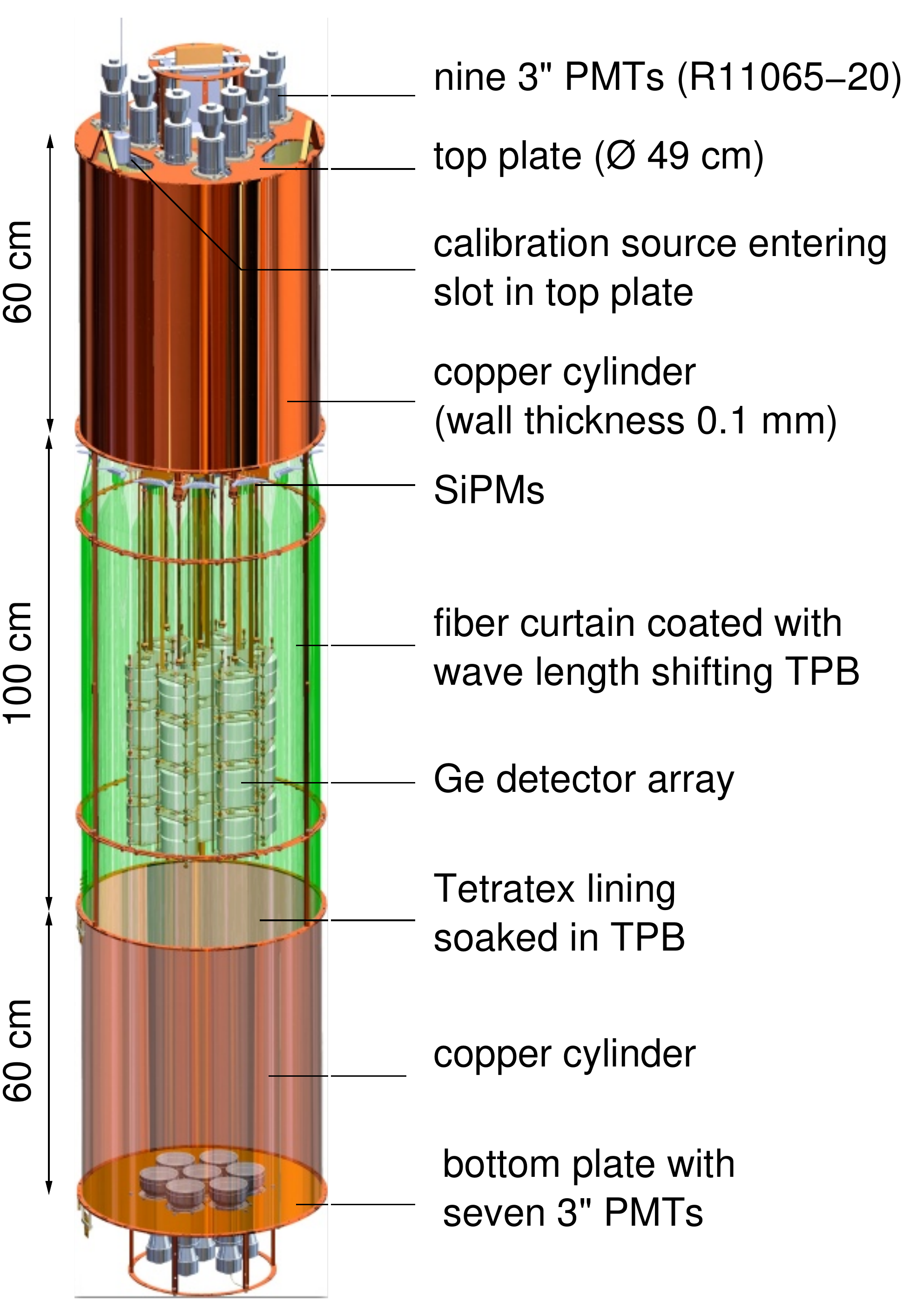}
\caption{\label{fig:LArSetup}
        The Ge detector array enclosed by the LAr veto system.
}
\end{center}
\end{figure}

The LAr light instrumentation in \GERDA\ is conceived as a hybrid system
comprising PMTs and WLS fibers with SiPM readout.  It is designed to be a
retractable unit that can be deployed together with the Ge detector array into
the cryostat through the lock system. The elaborated mounting scheme of the
veto system is described in the next section. The lock system constrains the
veto's geometry to a lengthy cylindrical shape with a diameter of
$\sim$0.50\,m and a total height of $\sim$2.6\,m.  A (CAD) technical drawing
depicting the complete LAr veto system is shown in Fig.~\ref{fig:LArSetup}.

\label{sec:lar-veto}

\subsection{PMT system}
\begin{figure}[h]
\begin{center}
\includegraphics[width=0.9\columnwidth]{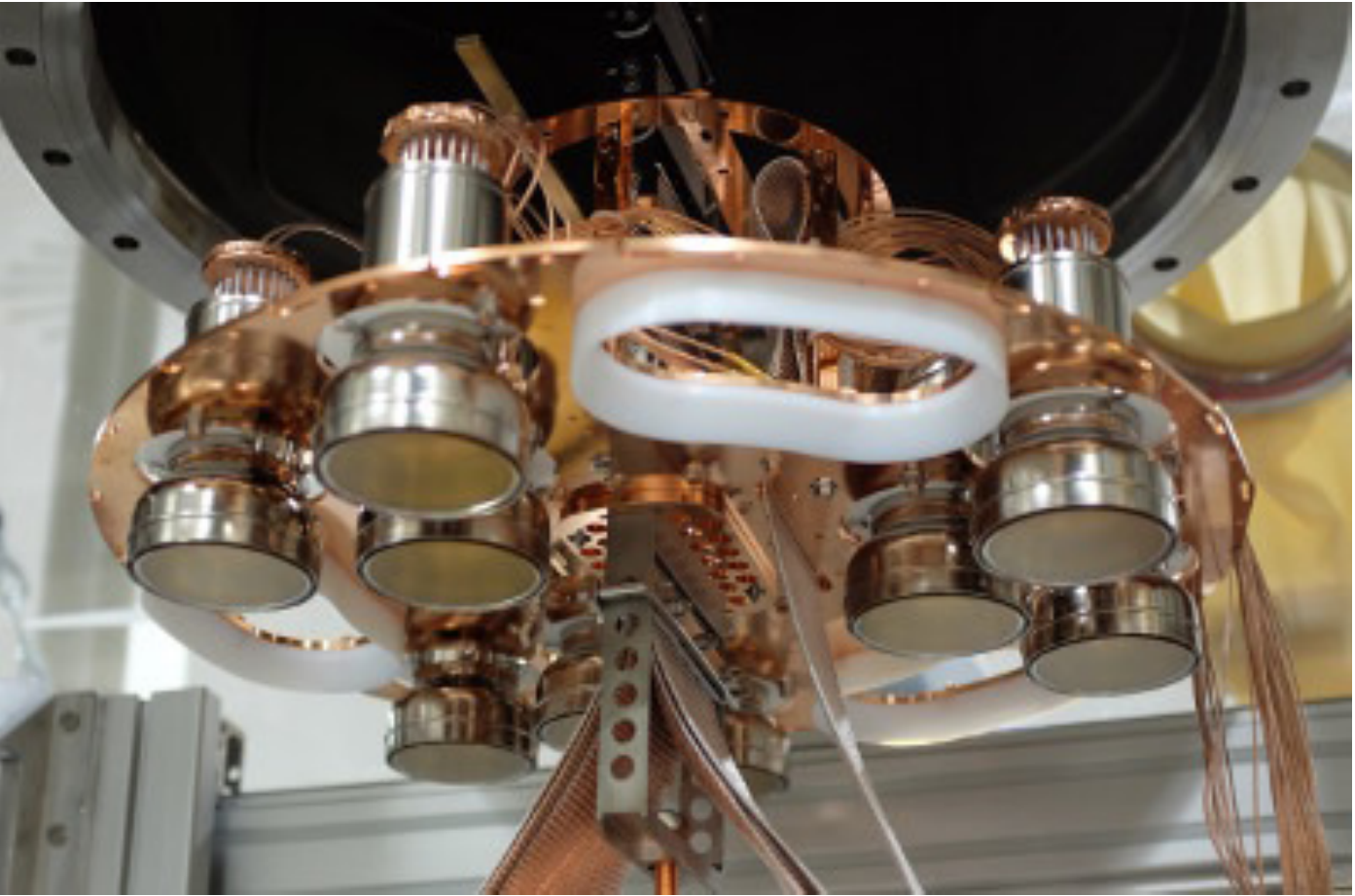}
\caption{\label{fig:PMTtop}
     Top PMT plate with 9 PMTs and three longitudinal slots with PTFE guides
     for the deployment of the calibration sources.
}
\end{center}
\end{figure}

\begin{figure}[h]
\begin{center}
\includegraphics[width=0.9\columnwidth]{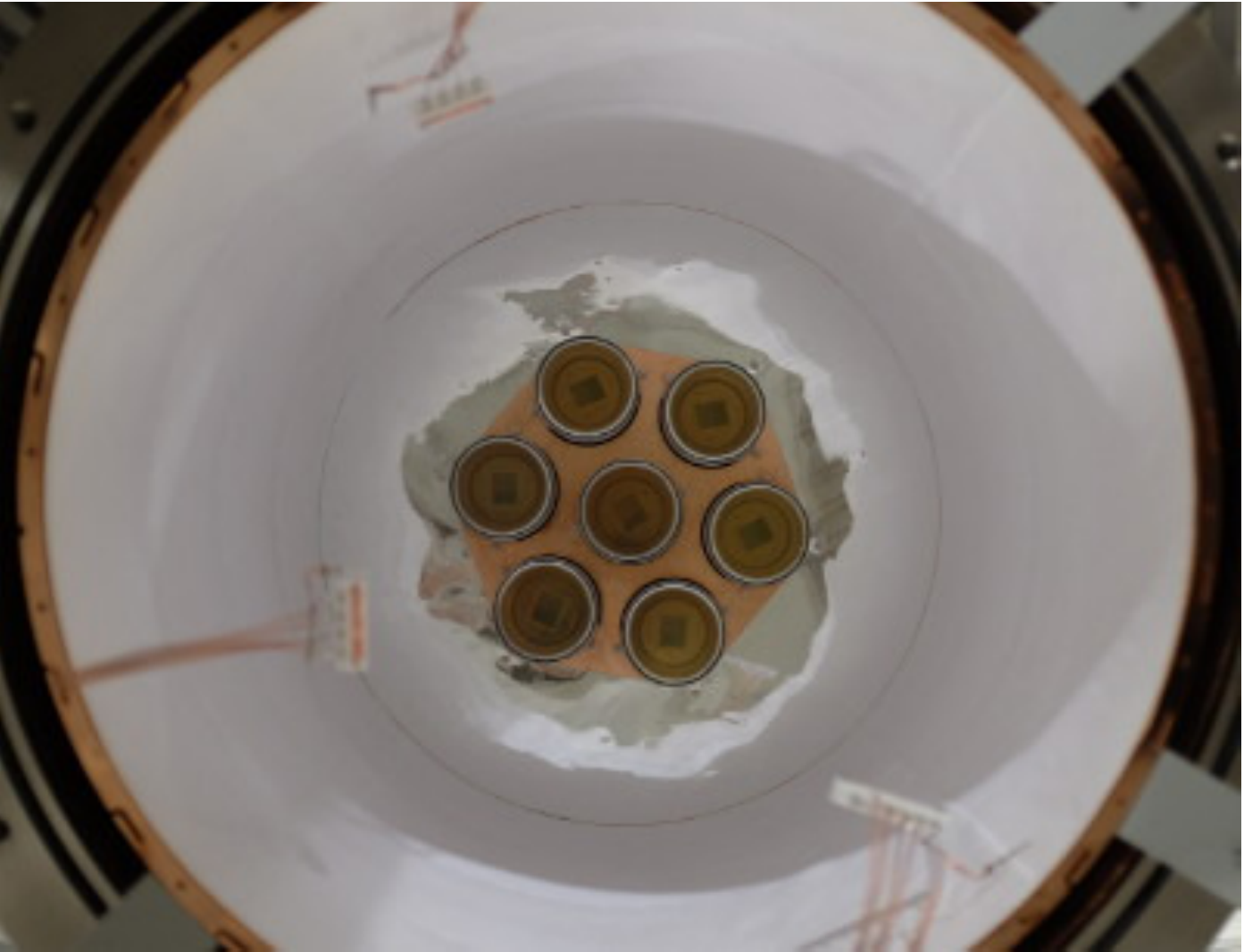}
\caption{\label{fig:PMTbottom}
     Bottom PMT plate with 7 PMTs  and copper cylinder. Plate and cylinder are
     covered by  wavelength shifting Tetratex\textsuperscript{\textregistered}
     foil soaked in TPB~\cite{tetra}.
}
\end{center}
\end{figure}

The PMT light readout system measures the scintillation light around the
detector array with nine PMTs from the top and seven PMTs from the bottom. The
3" PMTs are installed on copper plates at the two ends of the cylindrical LAr
volume facing inwards (see Figs.~\ref{fig:PMTtop}
and~\ref{fig:PMTbottom}). These end plates are separated from the central
fiber section by copper shrouds of 60\,cm height. The copper shrouds consist
of 100\,\mum\ thick copper foils which carry laser-welded flanges at both ends
for the connection to the PMT plates and the fiber section. They are lined
with Tetratex\textsuperscript{\textregistered} PTFE foil of 254\,\mum\
thickness from the inside.  The Tetratex\textsuperscript{\textregistered} foil
is impregnated with TPB, and thereby serves as a WLS of the scintillation and
diffuse reflector of the shifted light. The foil has been thoroughly tested
for mechanical and optical stability of the WLS~\cite{tetra}.
 
\begin{figure}[h]
\begin{center}
\includegraphics[width=0.9\columnwidth]{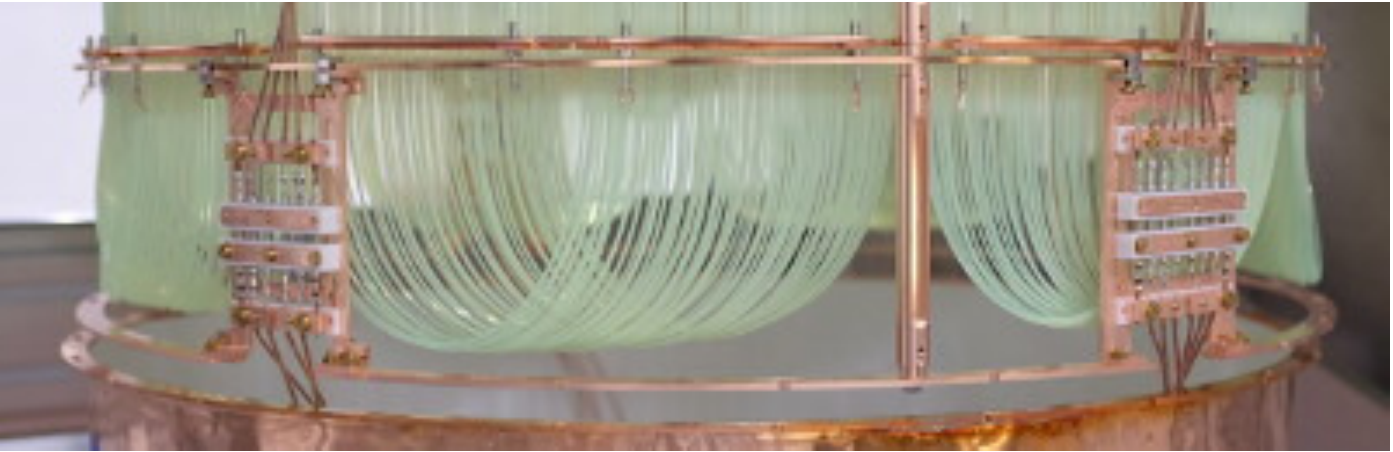}
\caption{\label{fig:PMTplug}
     PMT cable plugs custom made from copper and PTFE allow to separate the
     bottom LAr veto segment from the central fiber part.
}
\end{center}
\end{figure}

The cabling of the bottom PMTs runs along the outside of the LAr veto
system. The bottom shroud can be easily detached from the central fiber unit
within the glove box.  For this purpose custom made low-radioactivity cable
plugs for the bottom PMTs were placed on the connecting flange (see
Fig.~\ref{fig:PMTplug}).

The PMTs are of type Hamamatsu R11065-20 Mod. They have a bialkali
photocathode that reaches a quantum efficiency of about 40\,\% at the
wavelength of 420\,nm. Typically a peak-to-valley ratio of about four is
achieved for single photoelectrons.  To enable direct detection of
scintillation the photocathodes are coated with a 1-4\,\mum\ thick layer of
WLS, TPB (10\,\% by mass) embedded in polystyrene (90\,\%). In various test
series it was found that the first generation of R11065 PMTs was unstable
under cryogenic conditions. However, in cooperation with the manufacturer it
was possible to receive the current modified and improved version: each of the
PMTs has been continuously operated for at least six weeks in a LAr test stand
prior to the deployment in \GERDA, and has been operated stably in
\GERDA\ since the start of Phase~II more than one year ago (see Section 6).

\begin{figure}[h]
\begin{center}
\includegraphics[width=0.8\columnwidth]{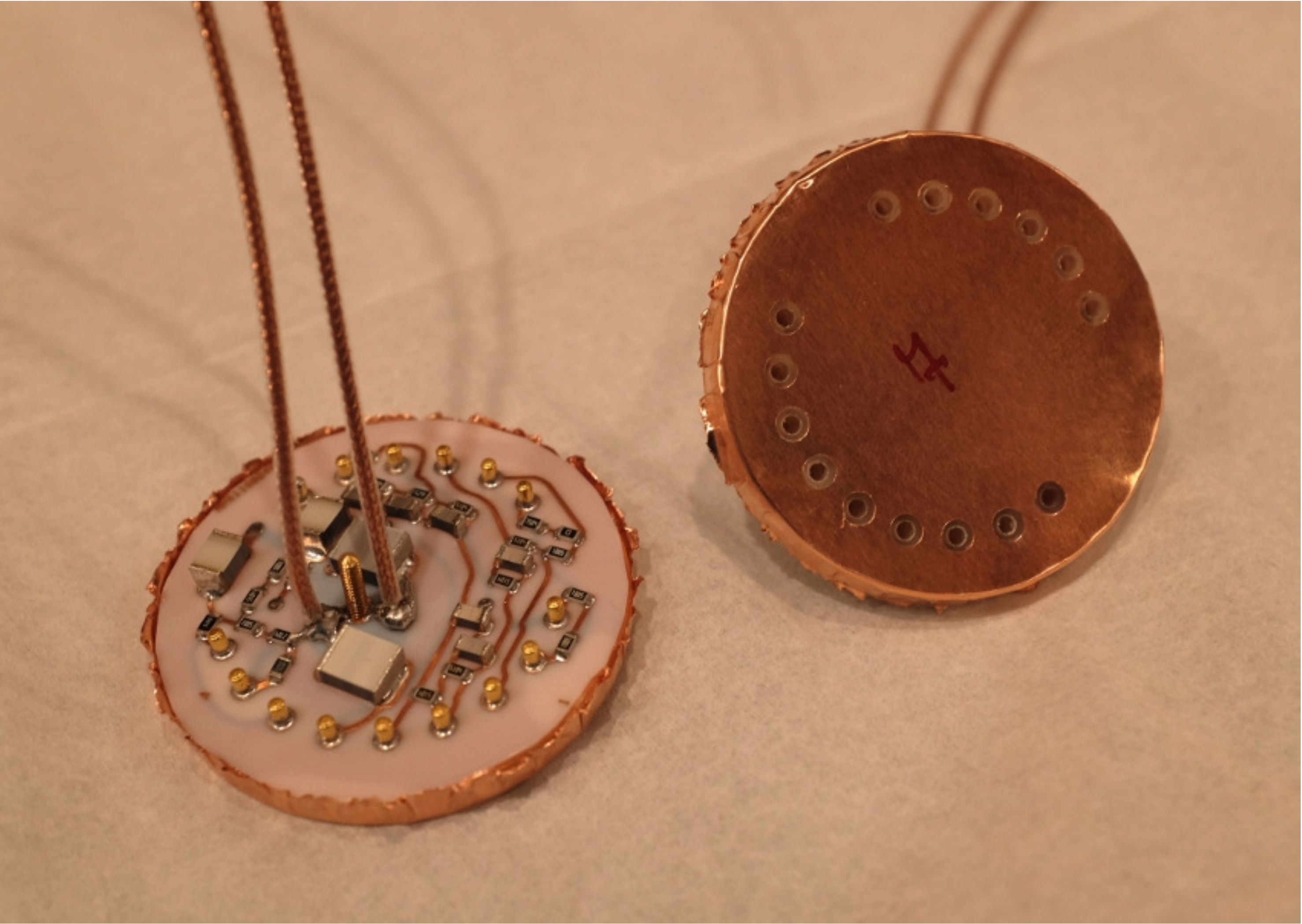}
\caption{\label{fig:PMTVD}
           Voltage divider potted into epoxy resin.
}
\end{center}
\end{figure}

The voltage dividers are designed for negative bias and high signal
quality. The electrical power consumption is kept low at about 20\,mW to
prevent the argon from boiling. For further protection against discharges due
to argon gas bubbles the PCB is potted into epoxy resin and a copper pot (see
Fig.~\ref{fig:PMTVD}).
Custom made SAMI RG178 coaxial cables are used for both signal and bias
voltage inside the \GERDA\ lock and cryostat.

The PMT signals are amplified outside of the lock by custom made shapers and
split into two branches; one output leads to a FADC for digitization, and a
second branch is used to independently monitor the count rates: the signals
are fed into a custom made scaler device that applies a threshold of
$\sim$20\,\% photoelectron amplitude to count hits and store the rate in a
database. The \GERDA\ Slow Control has access to this database and invokes an
immediate automatic safety ramp down of a PMT's bias voltage in case its rate
exceeds a limit of 20\,kHz. The bias power supply is the same CAEN SY1527
system as used for the Cherenkov veto PMTs~\cite{g1-instr}. High voltage
filters are mounted at the feedthroughs to the \GERDA\ lock to reduce
electronic pickup noise on the bias power.

The gain of the PMTs is calibrated to (2-3)$\cdot$10$^6$ with bias voltages in
the range from -1300\,V to -1550\,V. At these settings the peak-to-valley
ratio reaches its best values of 3.5 to 4, which is relevant to identify hits
above baseline noise at a low threshold. The gain is constantly monitored and
found to be stable throughout the operation in \GERDA\ (see
Section~\ref{sec:performance}).

\subsection{The fiber-SiPM system } \label{sec:sipm}

\begin{figure}[h]
\begin{center}
\includegraphics[width=0.90\columnwidth]{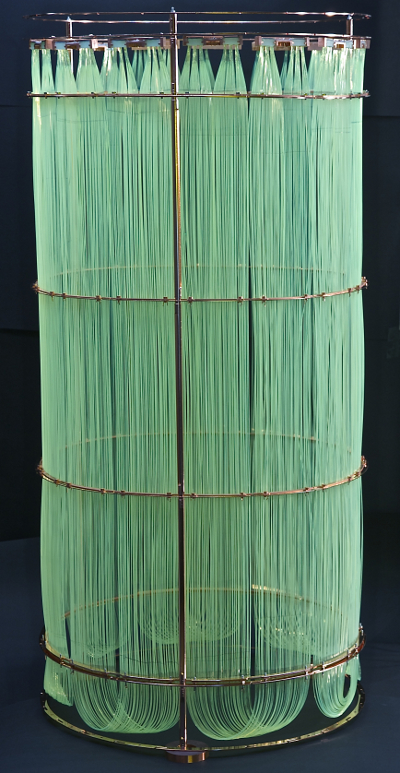}
\caption{\label{fig:shroud}
          The fiber curtain: height $\sim$1\,m, diameter $\sim$0.5\,m, 405
          fibers read out on both ends by 90 SiPMs.
}
\end{center}
\end{figure}
\begin{figure}[h]
\begin{center}
\includegraphics[width=0.9\columnwidth]{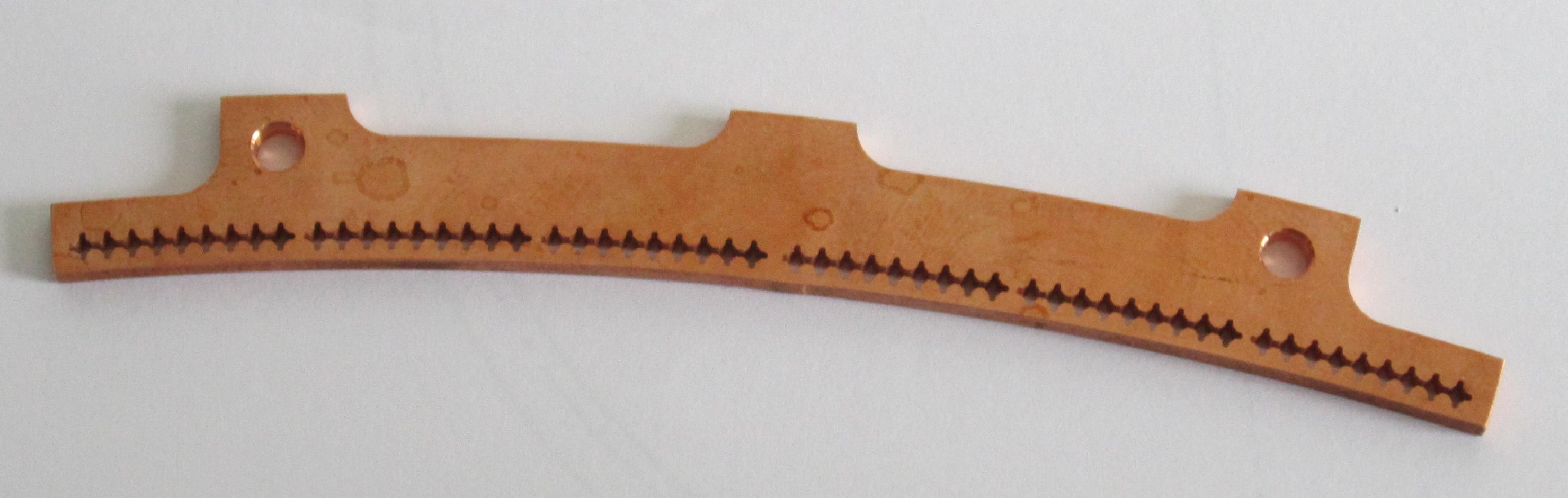}
\caption{\label{fig:fiberHolder}
    Fiber holder produced with wire erosion. Note the pattern of diagonally
    aligned square holes.  One holder covers an angular range of
    23.5$^\circ$.
}
\end{center}
\end{figure}
The middle section of the LAr veto setup (Fig.~\ref{fig:shroud}) consists of a
curtain ($\sim$50\,\% coverage) of WLS fibers which are read out with
SiPMs. 
Note that its field of view is limited by the copper radon shroud\footnote{
The radon shroud separates a central volume of about 3\,m height and 0.75\,m
diameter from the remaining volume of the cryostat in order to prevent that
radon, emanating e.g. from the walls of the vessel, may be transported by
convection close to the detector array~\cite{g1-instr}.
}
at diameter 0.75\,m.

\subsubsection{Fiber curtain}

The design goal of the fiber detector was to achieve the largest possible
coverage with light detectors while using the minimum amount of material hence
minimizing the radioactivity in the vicinity of the Ge detectors.
 
The BCF-91A multiclad fibers from Saint-Gobain with a cross section of
1$\times$1\,mm$^{2}$ were chosen.  The square cross section is needed for the
highest possible trapping efficiency and the absorption spectrum of the
BCF-91A fiber matches well the emission spectrum of the TPB.  The surface of
the fibers is coated with TPB by vacuum deposition.

The fibers are supported by a lightweight copper frame that also carries the
weight of the bottom PMT section of the setup.

The arrangement of the square fibers is such that their diagonal is tangential
to the circular flanges maximizing their surface turned towards the enclosed
volume.  The fibers are held in place by copper holders as the one seen in
Fig.~\ref{fig:fiberHolder}.  These holders bundle 54 fibers in one unit which
are connected to six 3$\times$3\,mm$^2$ SiPMs.

At the bottom part of the fiber shroud the fibers are bent around and fed
through the neighboring copper holder (see Fig.~\ref{fig:shroud}) such that
two pairs of copper holders form a double module.  Every single fiber is about
1.8 m long and both of its ends are instrumented with SiPMs at the top of the
cylinder.

The total amount of fibers is about 730\,m. This corresponds to a mass of
about 765\,g.  The total surface is about 2.9\,m$^2$ half of which is facing
inwards the enclosed volume.

\subsubsection{SiPMs}

For possible use in \gerda\ the radioactivity of commercial SiPMs is a big
concern because of the substrate that is either ceramic or ordinary glass
fiber PCB material. On the other hand the purity of the silicon wafers the
chips are made of is expected to be very high.  To have the radiopurity issue
under control it was decided to pack the SiPMs ourselves.  Therefore the
3$\times$3\,mm$^2$ SiPMs were purchased in die from Ketek GmbH.

\begin{figure}
\begin{center}
\includegraphics[width=0.9\columnwidth]{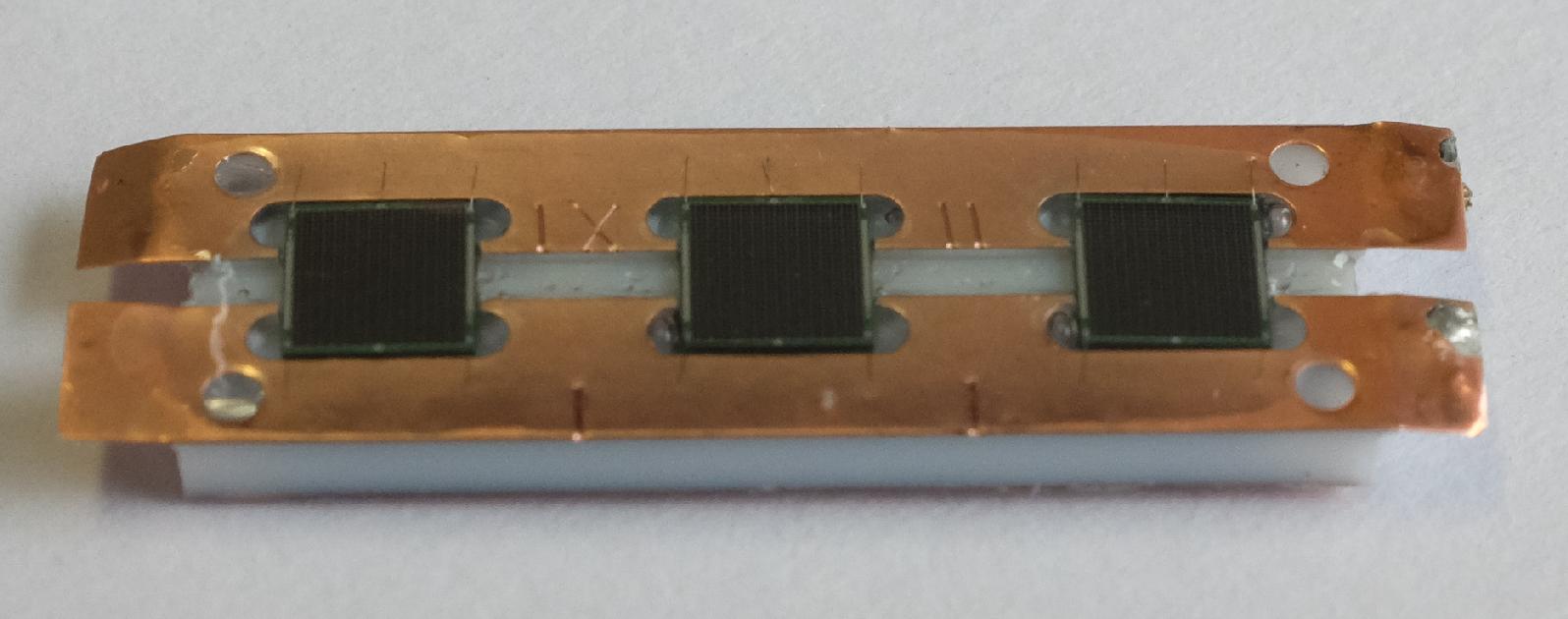}
\caption{\label{fig:cuflonHolder}
    Custom packaging of three 3$\times$3\,mm$^2$ SiPMs on a
    Cuflon\textsuperscript{\textregistered} holder.
}
\end{center}
\end{figure}
The packaging consists of a Cuflon\textsuperscript{\textregistered} PCB with
square holes machined into it for the SiPM chips.  Such a holder with SiPMs
already implanted is shown in Fig.~\ref{fig:cuflonHolder}.  Each holder has
place for three SiPMs. The top copper layer of the PCB material is divided
into two strips during the milling to form the two contacts of the SiPM array.
The SiPMs are placed in the holder and bonded to the copper stripes. Then the
holder is covered with a thin layer of transparent epoxy glue (Polytec EP601).

Each array was tested first at room temperature, then in liquid nitrogen
(LN). The arrays that passed the first test were assembled in double arrays of
six SiPMs and tested again in LN.  Only fully functional SiPM arrays with low
dark rate ($<1$\,Hz/mm$^2$ at about 2.7\,V overvoltage in LN) were accepted
for deployment in \gerda.
 
Six SiPMs are connected in parallel to one {50\,$\Omega$} cable in the cable
chain.  There is no active or passive electronic component in the LAr.  The
total cable length from the SiPMs to the amplifier input is about 20\,m.

The strongly temperature dependent quenching resistors of the SiPMs cause very
long pixel recharge times in the range of microseconds when the SiPMs are
submersed in LAr.  In addition, the capacity of the large array and the cable
reduces the peak amplitude of the signal significantly. The slow and small
amplitude signals suggest the use of charge sensitive amplifiers.

\begin{figure*}[htb!]
\begin{center}
\includegraphics[width=120mm]{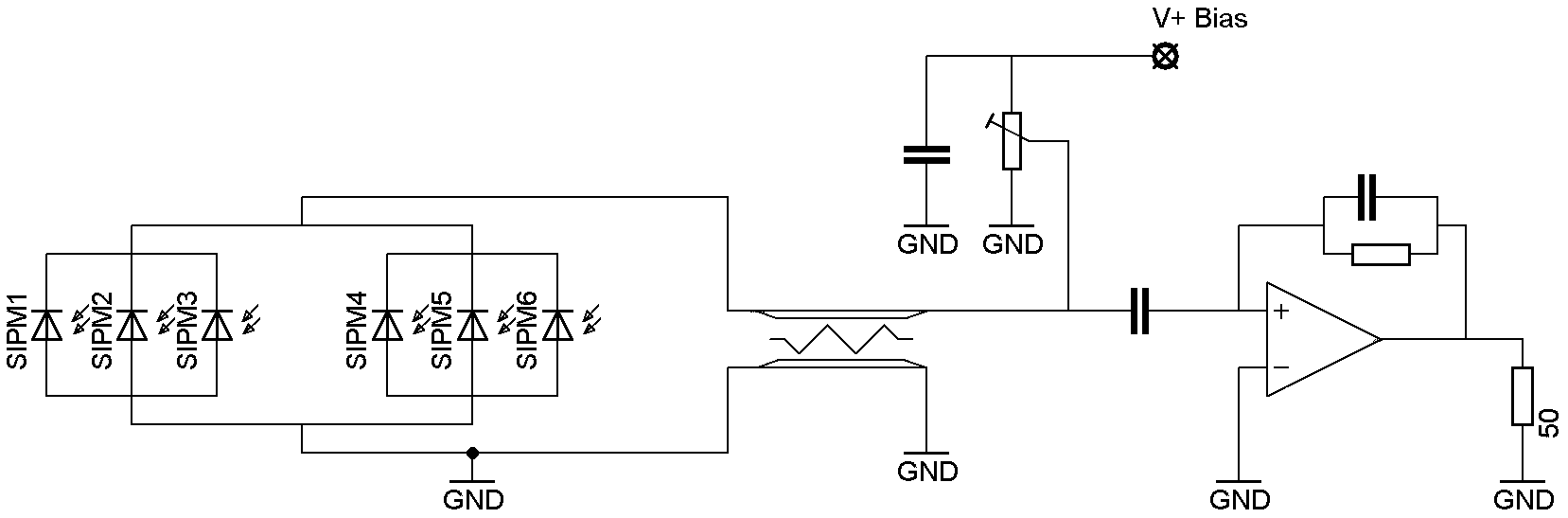}
\caption{\label{fig:sipm_readout}
        Circuit diagram of the SiPM readout. Only one channel is shown with
        six SiPMs in parallel which corresponds to an array of 54~mm$^2$.  The
        cable separating the SiPMs and the amplifier is about 20~m long. The
        charge sensitive amplifier is a Cremat-112.
}
\end{center}
\end{figure*}
Fig.~\ref{fig:sipm_readout} shows the readout circuit schematics. Each group
of three SiPMs corresponds to a unit shown in Fig.~\ref{fig:cuflonHolder}.
The transmission line in Fig.~\ref{fig:sipm_readout} stands for the 20~m cable
mentioned above.  The potentiometer regulates the bias voltage of the SiPMs
which is connected to the core of the coaxial cable.  The signal is decoupled
with a 100\,nF capacitor and connected to the charge sensitive amplifier (CR112
from Cremat).  The bias circuit and the charge amplifier are mounted in a
custom-made NIM module.
    
\subsection{Data acquisition and analysis}

\begin{figure}[htb]
\begin{center}
\includegraphics[width=0.9\columnwidth]{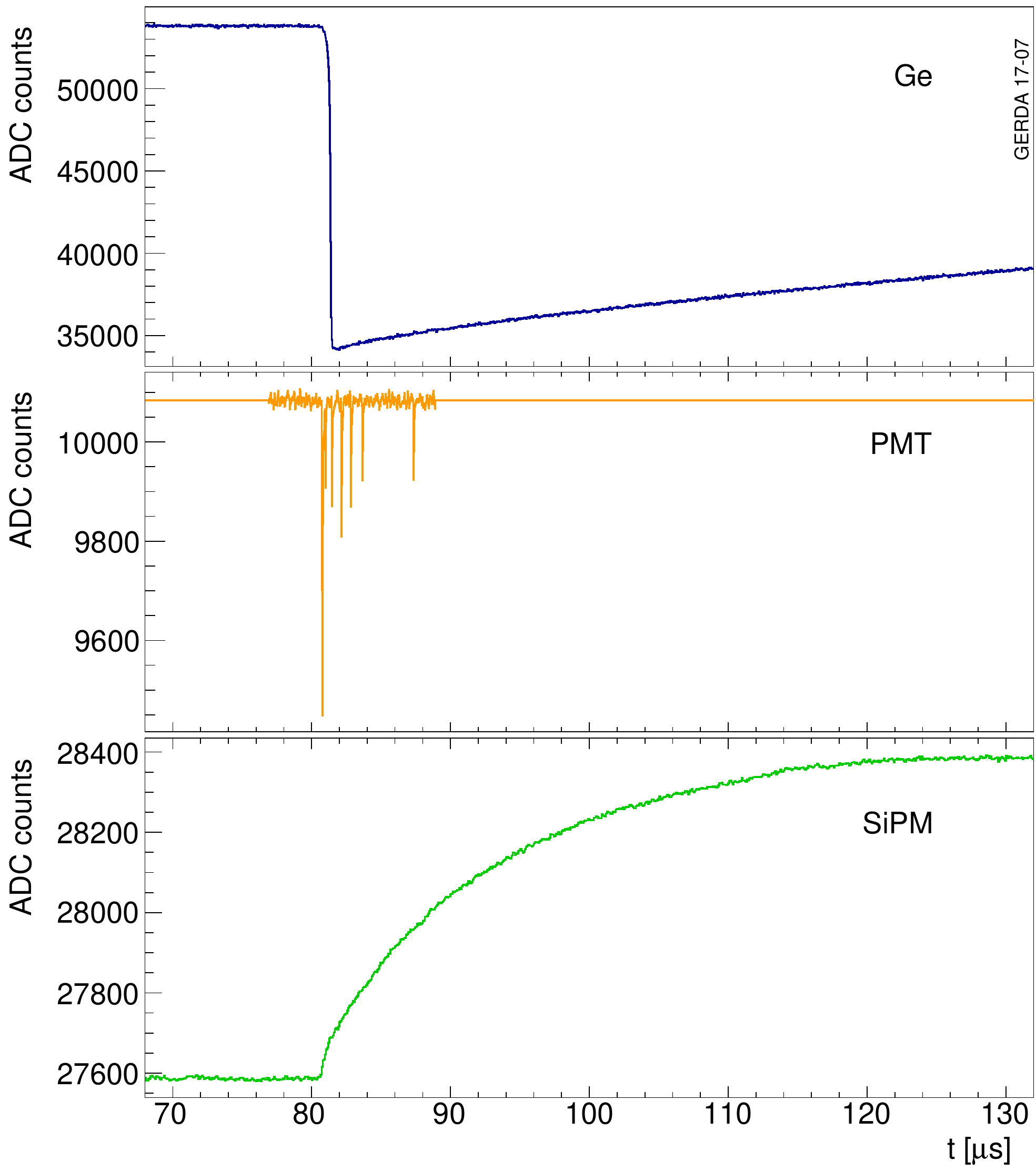}
\caption{\label{fig:GePMTSiPMT}
         Sample traces of an event with signals in a Ge detector and the LAr
         veto system.  On top the trace of the Ge detector that triggered the
         event; a PMT trace in the middle - the fast component of the
         scintillation light is followed by several smaller pulses from the
         triplet component; at bottom one of the SiPMs traces with the
         expected slower signal (see Subsection~\ref{sec:sipm}).
}
\end{center}
\end{figure}

The same FADC system (SIS 3301 Struck) used for the Ge
detectors~\cite{g1-instr} records the pulse shapes of the 16 PMT and 15 SiPM
channels and saves them for off-line analysis.  The PMT traces are digitized
with 100\,MS/s and for each channel a trace of 12\,\mus\ length is saved to
disk.  The resolution of the SiPM traces is reduced to 80~ns to save disk
space but traces of 120\,\mus\ length are recorded.  All LAr channels are read
out together with the Ge channels if at least one Ge detector has an energy
deposition above 100\,keV.  Fig.~\ref{fig:GePMTSiPMT} shows the traces of a
representative background event that has been triggered by a Ge detector and
that has produced in addition signals in the PMTs and SiPMs of the LAr veto
system.
PMT hits are reconstructed in the offline analysis following the procedure
described in~\cite{PhDanne}.  First, the baseline is determined using an
iterative method from~\cite{PhDmark}. Then a leading edge trigger with a
threshold of 3 baseline standard deviations is applied to identify up to
fifteen hits per trace. Amplitude and charge of each hit are computed. No
quality cuts or signal filtering is applied. The trigger positions are
converted into time differences relative to the first trigger found in the Ge
detector traces. Trigger positions and amplitudes are subsequently used
together with hits from the SiPM to test the LAr veto condition.

SiPM hits in the recorded traces are identified with a trigger finding
algorithm based on the trapezoidal filter.  The moving window deconvolution is
applied twice on the recorded traces.  In the first step the decay time of the
amplifier is deconvoluted (50~\mus) and in the second step the RC constant of
the SiPM given by the quenching resistor times pixel capacity is removed. To
determine the trigger time a trigger finding algorithm~\cite{PhDanne} is
employed on the resulting waveform, and the amplitude of the pulses is read
after a fixed delay following the trigger.

The algorithms were implemented in the \gelatio\ framework~\cite{gelatio} which
is used to process \gerda\ data.  Each event is characterized by the
calibrated energy deposited in the Ge diode, a data quality flag, the
classification as signal or background event from the PSD analysis, and veto
flags from the muon veto and LAr veto systems.

\section{Upgrade of infrastructure}
 \label{sec:armlock-cablechain}
The break after the end of the Phase~I run was used for both maintenance and
upgrade work.  After more than 3 years of operation the water tank was
emptied. Selected welds and surfaces of both cryostat and water tank were
inspected. No corrosion problems were observed, and the system safety of the
pressure equipment was certified by a notified body. With the Phase~I lock
dismounted, a $^{228}$Th calibration source of 20\,kBq was recovered that
dropped by accident during Phase~I to the bottom of the cryostat. While it did
not affect the Phase~I background index, its presence would have not been
tolerable in Phase~II. For this recovery the bottom of the radon shroud had to
be cut out. Both actions were performed with remotely controlled tools in the
LAr-filled and hermetically closed cryostat.

\subsection{Clean room upgrade}
In order to improve the temperature stability within the clean room, the
ventilation system was upgraded. An additional pump was installed that
regulates the cooling water supply for the \GERDA\ clean room ventilation
system according to the needs. Flux sensors have been installed to strategic
cold water tubes, allowing for real time monitoring of the cooling water
throughput.  These measures have led to an increase of temperature stability
inside the clean room. While in Phase~I the stability was about
$\pm$0.7$^{\circ}$\,C it could be now stabilized to
$\pm$0.2$^{\circ}$\,C. Additional temperature sensors have been installed to
the body of the lock system. These show that temperature fluctuations of the
experimental volume itself are smaller than $\pm$0.1$^{\circ}$C. This
improvement is relevant for the long term stability of the ohmic resistance of
the readout cables.

\subsection{Muon veto system}
The muon veto system~\cite{mv_tec} was slightly upgraded.  For the replacement
of the lock the plastic muon veto system had to be removed from the roof of
the clean room.  After reinstallation, a broken amplifier of the plastic veto
was replaced.

During the inspection of the cryostat the water tank was empty and thus a
refurbishment of the PMTs of the muon veto was possible.  Being accessible
without scaffolding, two of four broken PMTs could be replaced by spare ones.
At the beginning of Phase~II, in total still 3 out of 66 Cherenkov PMTs were
not working, 4 more failed during data collection.  The plastic panels on the
roof of the clean room (see Fig.~\ref{fig:g-setup}) are working
satisfactorily.
	    
\subsection{Lock system}
\begin{figure}[h!]
\begin{center}
\includegraphics[width=0.95\columnwidth]{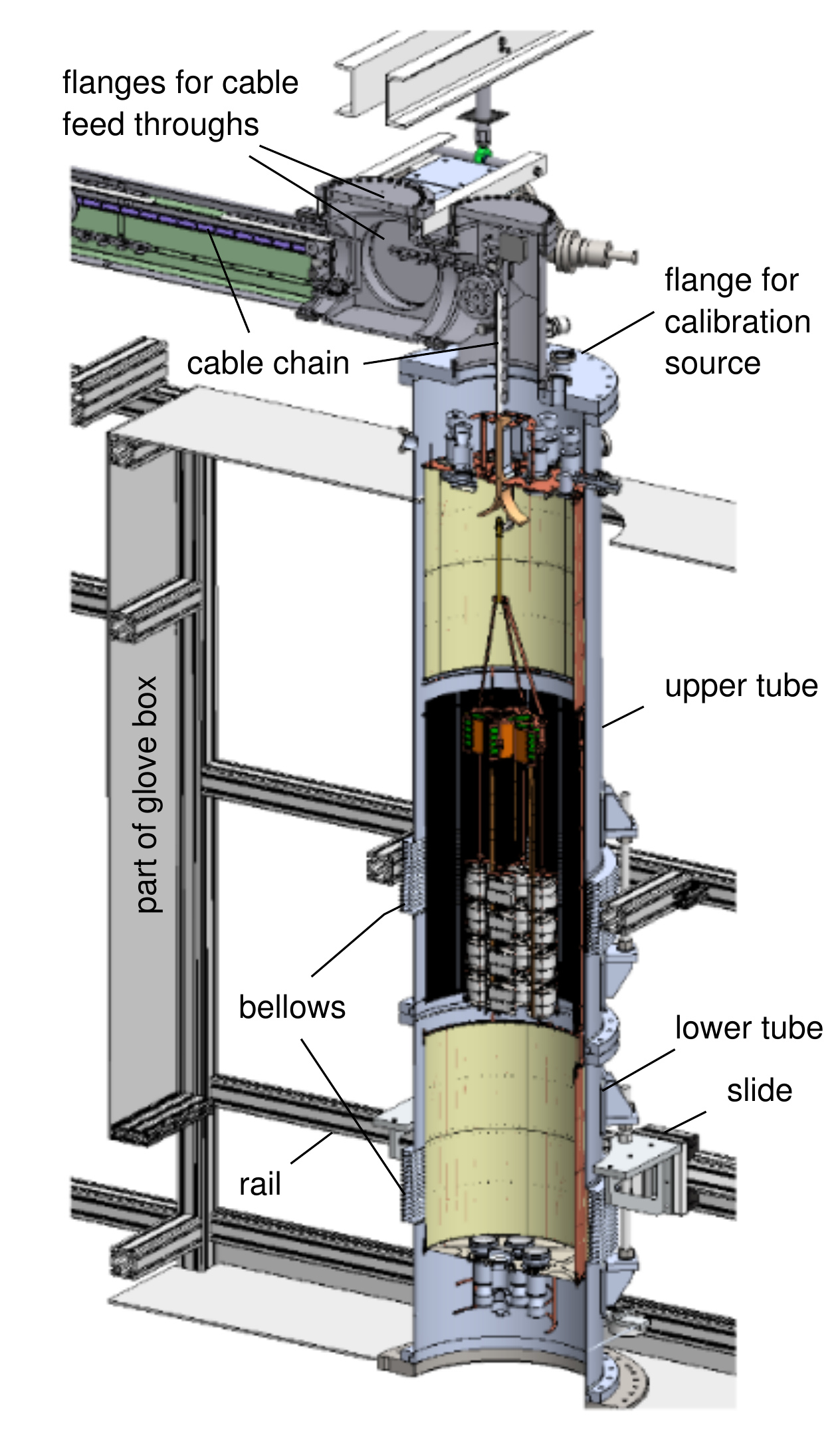}
\caption{\label{fig:g2-lock}
     Vertical part of the Phase~II lock with the Ge detector array and the LAr
     veto system in retracted position. Both upper and lower tube are enclosed
     by a glove box.  The lower tube can be laterally moved such that the
     resulting gap allows the installation and service of both the Ge detector
     array and the LAr veto system.
}
\end{center}
\end{figure}

Keeping the same functional principle, the Phase~II lock replaces the Phase~I
twin lock system by a single-arm lock with enlarged diameter (550\,mm) and
height (2682\,mm). This allows us to deploy both the 7 string detector array
and the complete assembly of the LAr veto instrumentation (see
Fig.~\ref{fig:g2-lock}).
 Both detector systems are suspended from a chain
which allows us to lower them by $\sim$6\,m down to the center of the
cryostat.  Like in Phase~I, the suspension chain also serves for guiding the
cables.

\subsubsection{Functionality of lock}
Two stainless steel tubes constitute the lock volume, an upper one of 1.75\,m
height, and a lower one of 0.93\,m height. They mount vertically directly on
the DN630 shutter which separates the cryostat's volume from
atmosphere. Bellows and set-screws allow us to vary the length of both tubes
by $\pm$5\,cm. The lock is opened by sliding the lower tube horizontally to
the side after its two flanges have been detached. The resulting gap allows to
introduce the three segments of the LAr veto system successively into the
lock. The top PMT plate represents the interface between cable chain and the
upper copper shroud; it is permanently installed in the lock resting just by
its weight on a keyed index plate which is attached at the end of the cable
chain. The position of the top PMT plate can be locked by three bolts at the
upper end of the top tube; thus the cable chain (and the Ge detector array)
can be moved independently downward. In standard operation, when only the Ge
detector array has to be serviced, the top and middle segments of the LAr veto
system are thus retained in the upper tube of the lock while its bottom
segment is moved together with the lower tube to the side.

All handling is done from the outside of the glove box via various
appropriately positioned glove ports.
 
\subsubsection{Cables}
The coaxial cables deployed inside the lock are custom produced to minimize
both the total radioactivity and the outgassing of nitrogen and radon in the
lock and in the LAr. In fact, nitrogen impurities are powerful LAr
scintillation light quenchers and Rn progenies can cause an increase of
background.  240 pieces of 12\,m long coaxial cables have been deployed to
connect the lock feedthroughs for signal, power supplies and HV to the front
end circuits, Ge detectors, and the SiPMs and PMTs of the LAr veto system.
Each Ge front end circuit (see Fig.~\ref{fig:cc3pict}) requires nine coaxial
cables: four for the circuit power supply, one for the pulser, and four for
the amplifier outputs. The cabling serves 11 front end circuits, 4 channels
each plus spares.  Three different cable types are deployed: RG179 for the Ge
high voltage, RG178 50 Ohm for the SiPMs and PMTs bias and readout, and 75 Ohm
coaxial cables for the signal outputs of the Ge charge sensitive
preamplifiers.  Table~\ref{tab:cables} shows the list of cables and their
specifications.
\begin{table}[htb!]
\begin{center}
\caption{\label{tab:cables}
        List and characteristics of the coaxial cables in the 5 cable bands
        deployed in the \GERDA\ Phase~II cable chain.
}
\begin{tabular}{lccc}
\hline
& RG178 & RG179 & 75$\Omega$ \\
\hline
purpose              &  LAr veto & Ge HV & Ge signal \\
band:no of cables    &  V:48  & I:35  & III:61  \\
                     &   & II:35 & IV:61 \\
\hline
AWG                   &  30   &  30   &   33 \\
conductor             &  Cu & Cu & Cu \\
dielectric            & PFA   & PFA   &  PFA \\
\O [mm]               & 1.8   & 2.55  &  1.4 \\
impedance [$\Omega$]  &  50   &  75   &   75 \\
attenuation [db/100m] & 95    & 68    &   95 \\
capacity [pF]         & 95    & 64    &   70 \\
weight [g/m]          &  7    & 14    &    4 \\
resistivity [$\Omega$/m]&0.37 & 0.37  &  1.5 \\
\hline
\end{tabular}
\end{center}
\end{table}
The constituent materials have been chosen to obey cryogenics specifications
and minimize space occupancy, radioactivity and outgassing rates. For the HV
cables, the dielectric strength of the insulator is relevant. The
perfluoroalkoxy alkane (PFA) polymer material has been chosen because of its
high dielectric strength.  All the cables have been custom produced by
SAMI\footnote{SAMI Conduttori Elettrici Speciali, Via Venezia snc, 20060
Liscate (MI)}: both central and shielding conductor braids are in bare
copper, and the dielectric and jacket material PFA has not been colored to
preserve radiopurity, Rn emanation and minimize cable outgassing.

The cable outgassing has been measured both in vacuum tests and by
chromatography on a RG179 sample of 100 m: thanks to high quality materials a
total outgassing of $10^{-6}$ mbar$\cdot\ell$/s was achieved after 24 hours of
pumping.  The measurements were performed at 40$^\circ$C, 100$^\circ$C, and
150$^\circ$C. Only at 100$^\circ$C and 150$^\circ$C, a few peaks of organic
and fluor-organic compounds, octane and butane decafluoro-butane, showed up at
few minutes arrival times.  No peaks corresponding to known NIST Library
nitrogen or nitrogen compounds were observed.

For deployment in the cable chain, the 240 cables have been woven in five
cable bands of 95\,mm width (see Table~\ref{tab:cables}).  Weaving was done
with weaving machines at a commercial company\footnote{PD Cable Systems,
http://www.pdcablesystems.de/index.html}.  Special care was taken to
properly clean the cables and the PTFE thread prior to weaving.  All parts of
the weaving machine the cables got in touch with during the weaving process
were disassembled and properly cleaned using ultrapure isopropanol and water.
After weaving, the cable bands were once more cleaned, dried and packed under
clean room conditions.

\subsubsection{Cable chain}
The cable chain and its supporting mechanics are designed to deploy a total
mass of up to 60\,kg. Made from stainless steel of 1.5\,mm thickness, its cross section (102\,mm width $\times$ 20\,mm
height, with a usable height of 12\,mm) is enlarged by a factor of 7.5
relative to Phase~I in order to accommodate the cable bands for the array and
LAr veto system.  A 1\,m long piece of the cable chain has been used for
stress tests applying a force up to 4.95\,kN, corresponding to a test load on
the pulley of $\sim$200\,kg. There was no sign of critical deformation.

When retracted, the cable chain is held inside a horizontal 3.9\,m long DN250
tube by a pulley that runs along the length of the tube guided by a linear
bearing.  The cable chain is deflected around this pulley by 180$^{\circ}$ and
above the cryostat neck by 90$^{\circ}$ vertically downwards.  The pulley is
connected to a metal band at the far side from the cryostat neck that can be
rolled up on a winch. By unrolling the metal band the pulley moves towards the
cryostat neck and the array suspended to the cable chain can be lowered into
the cryostat (see Fig.~10 in Ref.~\cite{g1-instr}).  At the far end from the
cryostat a cross is connected to the horizontal tube with two CF
flanges. These contain the feedthrough for the motor axle moving the
suspension system as well as the vacuum and gas ports. On the close end to the
cryostat two crosses with three CF250 flanges each are connected that carry
all the signal and HV feedthroughs including the HV filters.

For movement of the array a stepper motor EC60 with gear GP81 and encoder
provided by MAXON\footnote{maxon motor ag, http://www.maxonmotor.ch} is used.
A magnetofluid rotational feedthrough with 20\,mm axle provided by
VacSol\footnote{VacSol GmbH, http://www.vacsol.com} is used for transmission
of the motor movement onto the winch that is moving the cable chain, hence,
the Ge detector array and LAr veto system.

\subsubsection{PLC system}
A new PLC system based on a Simatic S7-300 
for the control of the array movements via the stepper motor, operation of the
DN630 shutter between cryostat and lock system, pumps and LED control has been
built.  The system has been designed to maximize safety during operation via
interlocks. Two redundant induction sensors on each side of the pulley system
serve as end switches allowing a calibration of the system in the up and down
position. The position of the array is obtained independently by the stepper
position of the motor and a measuring tape with holes. In case a significant
deviation of position measurements is determined, the PLC stops movement of
the array.

\subsubsection{Fabrication and commissioning}
All parts of the internals of the lock system were specially selected and
screened for radioactivity using low background HPGe detectors and/or for
radon emanation prior to mounting (see Section~\ref{sec:screening}). The tubes
have been designed and tested according to the European pressure vessel code.

The horizontal part of the Phase~II lock system has been fully assembled under
clean room ISO~5 conditions. All parts were cleaned in an ultrasonic bath
inside a wet bench using ultra-pure water and isopropanol prior to mounting.
After final mounting dry tests under clean room conditions were performed to
verify mechanical precision and reproducibility of the positioning of test
loads. Torsion of the cable chain leads to a reproducible rotation of the load
of $\sim$(10$\pm2)^{\circ}$ when fully lowered.

Once mounted at \lngs\ on top of the \GERDA\ cryostat the system was tested
for vacuum leaks. None could be found at the level of
10$^{-8}$\,mbar$\cdot\ell$/s.

Before opening the DN630 shutter between lock and LAr cryostat, the system has
been (and is) conditioned over 12 hours by several pumping cycles in order to
ensure a low enough outgassing rate from cables and other parts of the
internal mechanical components.

\subsection{Calibration system}
The well-proven Phase~I calibration system~\cite{calib1} has been integrated
into the Phase~II lock system.  New sources were produced for Phase~II and
characterized~\cite{calib2}.  The three individual calibration units are
mounted on the top flange of the lock. Their geometrical arrangement on a
mounting circle of 380\,mm diameter and an angular distance of 120$^\circ$ is
such (Appendix Fig.~\ref{fig:arfromtop}) that each source, when lowered, just
fits into the space between the cylinder of the LAr veto system and two
neighboring outer strings of the detector array; thereby the sources enter
the inner volume of the LAr veto system by three slots in the top PMT plate
(see Fig.~\ref{fig:PMTtop}). A modification of the source holders, the
insulation of the Ta absorber from the stainless steel band~\cite{calib1},
prevents a previously observed HV instability in several Ge detectors.

Individual gate valves allow the units to be decoupled from the cryostat such
that the respective calibration sources can be exchanged. Besides the standard
$^{228}$Th source of low neutron emission \cite{calib2}, available sources
include $^{226}$Ra and $^{56}$Co.

\section{Material screening results} 
\label{sec:screening}
Compared to Phase~I the amount of material deployed close to the detectors has
increased in Phase~II, and hence the constraints for the radiopurity of the
respective materials have become even more restrictive.  State-of-the-art
screening techniques have been used again to verify the desired
radiopurity. The available screening facilities have been discussed earlier in
some detail~\cite{g1-instr}. They include Ge $\gamma$ ray spectrometers in the
underground laboratories at MPI-K Heidelberg, HADES (IRMM) in Belgium,
and \lngs, Italy. Screening with Inductively Coupled Plasma Mass
spectrometers (ICPMS) has been performed at \lngs\ and at INR RAS,
Moscow. Radon emanation measurements were done with both the ultra-low
background proportional counters developed originally at MPI-K Heidelberg for
the Gallex experiment, and with the (less sensitive) permanently installed
$^{222}$Rn electrostatic counter at \gerda.

In the following Subsections the uncertainties are given as $\pm$1 standard
deviations and the limits are quoted at 68\,\% CL.

\subsection{Radon emanation inside the lock}
$^{222}$Rn emanated inside the lock may be dissolved in the LAr and contribute
to the background. Hence the selection of low-emanating construction materials
is crucial.  All non-metal materials mounted inside the lock have been
qualified by Rn emanation measurements prior to assembly. Additionally, all
components located close to the detector array have been screened using HPGe
detectors.  Like in Phase~I all flanges of the lock are sealed with metal
gasket against atmosphere when possible. Else Kalrez O-rings were chosen whose
low emanation rate had been established in Phase~I~\cite{g1-instr}.

The measured Rn emanation of the woven coaxial cables bands is given in
Table~\ref{tab:wovencables}.  The total emanation rate expected from all
cables is lower than 4.0\,mBq.  This is consistent with the total $^{226}$Ra
activity from the woven cable bands obtained from HPGe screening (see
Table~\ref{tab:cableact}) of $\sim$10\,mBq, considering the fact that only a
fraction of the $^{222}$Rn resulting from the decays of $^{226}$Ra does
actually emanate from the cables.
\begin{table}[htb]
\begin{center}
\caption{\label{tab:wovencables}
              Results from Radon emanation measurements of the woven cable bands
              (see Table~\ref{tab:cables}).
}
\begin{tabular}{lc}
\hline
Band                      &  activity [$\upmu$Bq\,/\,m] \\
\hline
48$\times$ RG178          & $<$266 \\
35$\times$ RG179          & $<$19  \\
61$\times$ 75\,$\Omega$   & 12\,$\pm$\,2\\
\hline
\end{tabular}
\end{center}
\end{table}

Radon emanation measurements of the fully equipped lock system showed rates
below the tolerable level. A saturation activity of (12.5$\pm$5)\,mBq was
measured that has to be compared to (55$\pm$3.5)\,mBq from the cryostat alone.

\subsection{Nylon mini-shrouds}
The radioactive purity of the coated nylon films was checked by ICPMS
measurements at \lngs~\cite{diVacri}.  Results of the measurements are shown
in Table~\ref{tab:minishrimpur}.
\begin{table}[htb]
\begin{center}
\caption{\label{tab:minishrimpur}
       Radioactive impurities of the components of one nylon mini-shroud (MS)
       from ICPMS measurements.  Uncertainties are estimated to be about
       30\,\%.
}
\begin{tabular}{lcccc}
\hline
Component           &  U & Th       & K      &    mass     \\
                    & [ppt] & [ppt] & [ppb]  &     [g]    \\
\hline
TPB                 & 10    & 9     &  65    &           \\
polystyrene         &$<$5   & 10    & 100    &          \\
glue                &$<$10  & $<10$ & 900    &      \\
nylon               & $<$10 & $<$15 & -      & 27.6 \\
nylon coated        & 11    &  18   & $<$25  &        \\
nylon glued         & 38    & 39    & 1200   &      \\
\hline
MS finished         & 6.1\,$\upmu$Bq & 2.6\,$\upmu$Bq & 242\,$\upmu$Bq & 28.1 \\
\hline
\end{tabular}
\end{center}
\end{table}
Differences in radiopurity of similar samples indicated that surface
contaminations play a big role; so it is important to prepare and keep foils
in a clean condition.  Coating of the nylon foils was performed in a clean
room by brushing: this allows to deposit a small amount of WLS with a good
enough coverage.  The deposited WLS was determined by weighing the nylon
before and after coating.  A typical mass of the coating was about 0.3
mg\,/\,cm$^2$ for the nylon film coated from both sides.
		    
To assess if the radiopurity level is acceptable, a detailed simulation of the
nylon MS in the \gerda Phase~II setup was performed~\cite{PhDbjoern}. The
expected contribution of all mini-shrouds to the background index is about
5$\cdot10^{-4}$\ctsper\ before LAr veto and PSD cuts, and two orders of
magnitude lower after these cuts.

\begin{table*}[htb!]
\begin{center}
\caption{\label{tab:cableact}
        The length-specific activities of the custom coaxial cables deployed
        in \gerda\ Phase~II compared to the standard cables used in
        Phase~I. All screening has been done with Ge $\gamma$
        spectrometers. Phase~II cables were woven in bands (see
        Table~\ref{tab:cables}).
}
\begin{tabular}{lccccccc}
\hline\noalign{\smallskip}
Cable   & $^{226}$Ra & $^{228}$Th & $^{40}$K & $^{60}$Co & $^{137}$Cs & $^{108m}$Ag & $^{110m}$Ag  \\
\cline{2-8} \\[-2ex]
&\multicolumn{7}{c}{($\upmu$Bq/m)} \\
\hline\noalign{\smallskip}
Phase I: \\
RG178      & $11\pm3$ & $13\pm5$ & 680 &  $< 0.4$ & - & $13\pm2$ & $8\pm1$ \\
TR11.18kV  & $134$ & $96$ & 3700 & - & n.a.  & n.a. & n.a. \\
HABIA50    & $<5$ &  $3\pm2$ & 1200 & $<1$ & - & $2\pm1$ & $4\pm2$\\
\hline\noalign{\smallskip}
Phase II: \\
75 Ohm      & $5\pm1$     & $5\pm1$   & $190\pm30$   &  $< 1$ & $3\pm1$ & - & - \\
RG178       & $4\pm1$     & $<5$      & $81\pm16$    &  $< 2$ & -       & - & - \\
RG179       & $< 5$       & $6\pm2$   & $74\pm15$    &  $<2$  & -       & - & -  \\
\hline
\end{tabular}
\end{center}
\end{table*}
\begin{table*}[htb!]
\begin{center}
\caption{\label{tab:feact}
    The activity of the \GERDA\ front-end cables and PCBs in Phase~I and II.
    The Phase~II front-end PCBs have been screened fully populated while the
    Phase~I samples had no contact pins yet.  The values of the Phase~I readout
    / HV `cable', a high-purity copper wire
    within a PTFE tube, refer to its insulation.
} 
\begin{tabular}{lcccccc}
\hline\noalign{\smallskip}
         & Mass & $^{226}$Ra & $^{228}$Th & $^{40}$K & $^{60}$Co & $^{137}$Cs   \\
\cline{3-7} \\[-2ex]	 
         & (g)  & \multicolumn{5}{c}{($\upmu$Bq\,/\,pc)} \\
\noalign{\smallskip}\hline\noalign{\smallskip}
Phase I: \\
readout / HV `cable' (1m) & 4.28  &  $4.7\pm0.9$  &  $<3.3$    &  $34\pm9$    & - & $< 1.5$   \\
3-ch front end PCB w/o pins &5.2   & $290\pm100$   & $140\pm60$ & $1900\pm700$ & $<56$ & $<42$  \\
\noalign{\smallskip}\hline\noalign{\smallskip} 
Phase II: & \\
Pyralux\textsuperscript{\textregistered} 3 mil (80\,cm)      & 0.4 &$3\pm1$ & $4\pm1$ & $34\pm15$ &  $<0.55$ & $<1.9$  \\
Cuflon\textsuperscript{\textregistered} 3 mil  (51\,cm)      & 0.5 &$25\pm5$ & $<11$ & $120\pm60$ &  $<9.2$ & $<5.5$ \\
Cuflon\textsuperscript{\textregistered} 10 mil (51\,cm)      & 2.7 &$21\pm6$ & $<15$ & $300\pm80$ &  $<9.2$ & $<6.5$   \\
4-ch front end PCB          & $17$& $230\pm30$  & $70\pm40$ & $1300\pm400$ &  $<27$ & $< 67$  \\
\noalign{\smallskip}\hline
\end{tabular}
\end{center}
\end{table*}

\subsection{Coaxial Cables}
Table~\ref{tab:cableact} compares the length-specific activities of the
coaxial cables deployed in the lock in Phase~I and II.  In Phase~II, the high
voltage cables (RG179) are factors of about 15 to 25 better in $^{226}$Ra and
$^{228}$Th, and a factor of 50 better in $^{40}$K than in Phase~I (TR11.18kV);
compared to the standard production SAMI cables, they are better by factors of
about 20 to 30 in $^{226}$Ra and $^{228}$Th and $^{40}$K.  The $^{108m}$Ag and
$^{110m}$Ag isotopes are not present in the Phase~II cables due to the bare
copper choice for the conductor.

\subsection{Electronic front-end}
Table~\ref{tab:feact} reports the measured activities of the FE
devices. Compared to Phase~I, the new FE circuits and contacts exhibit a
significantly lower contamination, which is reduced for $^{226}$Ra and
$^{228}$Th by a factor of 1.5 and 30 per channel, respectively. For a BI of
\dctsper\ before LAr veto and PSD cuts, Monte Carlo (MC) simulations predict
maximum allowed activities of 2\,mBq and 0.5\,mBq, respectively, if the
electronics boards are mounted 30\,cm above the top detectors. The total
activity of the 10 boards amounts to 2.3(3)\,mBq and 0.7(4)\,mBq for
$^{226}$Ra and $^{228}$Th, respectively, being thus close to the radiopurity
limit.

For the front-end cables the corresponding limits are 0.5\,$\upmu$Bq/cm for
$^{226}$Ra and 0.04\,$\upmu$Bq/cm for $^{228}$Th.  The activity of the
Pyralux\textsuperscript{\textregistered} cables is well below the limit for
$^{226}$Ra and almost meets the limit for $^{228}$Th.  The
Cuflon\textsuperscript{\textregistered} cables just meet the limit for
$^{226}$Ra and miss it for $^{228}$Th by about a factor of 10. With 60\,\% of
the deployed front-end cables being of the
Cuflon\textsuperscript{\textregistered} type this excess reduces to a factor
of $\sim$4; this is still tolerable since the LAr veto alone suppresses the
$^{228}$Th induced background by about a factor of 100 (see
Section~\ref{sec:performance}).

\subsection{Detector holders}
Table~\ref{tab:detholder} compares the radiopurity of Phase~I and Phase~II
detector holders.  The substitution of copper by silicon mounting material
results in significantly reduced $^{228}$Th and $^{226}$Ra activities. The
corresponding contribution to the BI is less than 3$\cdot 10^{-5}$\ctsper
\ according to simulations before LAr veto and PSD cuts.
\begin{table*}[htb]
\begin{center}
\caption{\label{tab:detholder}
          Comparison of the radiopurity of the Phase~I and II detector holders
          (for masses of construction materials see
          Table~\ref{tab:massperkg}).  Specific activities from neutron
          activation analysis (NAA) \cite{PhDTGold} and ICPMS assume secular
          equilibrium. The background contribution BI at Q$_{\beta\beta}$ is
          estimated after anti-coincidences but before LAr veto and PSD.
}
\begin{tabular}{lcccccrrrcrr}
\hline
Material & method & \multicolumn{3}{c}{specific activity}& &\multicolumn{3}{c}{total activity}& &\multicolumn{2}{c}{background index} \\              
         &        & $^{228}$Th   & $^{226}$Ra & $^{40}$K  & & $^{228}$Th   & $^{226}$Ra & $^{40}$K   & &  $^{208}$Tl   & $^{214}$Bi \\            
  \cline{3-5} \cline{7-9} \cline{11-12} \\[-2ex]
	 &        &\multicolumn{3}{c}{($\upmu$Bq/kg)}      & & \multicolumn{3}{c}{($\upmu$Bq)} & &  \multicolumn{2}{c}{(10$^{-6}$cts/(keV$\cdot$kg$\cdot$yr)}\\   
\hline
Phase I: \\
Cu          &\gspec       &  $<20$ & $<20$           &  -    & &  $<1.6$  & $<1.3$     & - & & $<500$ & $<200$ \\
Si          &NAA         &  $<10^{-3}$ & $<10^{-4}$  &  -    & &     -     & -         & - & & - & - \\  
PTFE        &\gspec       &  25(9) & 30(14)          & 600   & &   0.18(6)  & 0.21(10) & 4.2 & & 63(23) & 45(21) \\
\hline
Phase II: \\
Cu          &\gspec       &  $<20$ & $<20$           & -  & &  $<0.26$  & $<0.26$      & - & & $<10$ & $<10$ \\    
Si          &NAA         & $<10^{-3}$ & $<10^{-4}$   & -  & &     -     & -            & - & & - & - \\
PTFE        &\gspec      & $50(20)$  & $67(25)$      & -   & &  $0.05(2)$  & $0.07(3)$ & 0.6 & & $0.9(4)$ & $0.7(3)$ \\ 
CuSn6       &ICPMS       & $<300$  & $<300$          & 600 & & $<0.15$   & $<0.15$     & - & & $<3$ & $<1$ \\
\hline
\end{tabular}
\end{center}
\end{table*}
		    
\subsection{LAr veto system}
\begin{table*}[htb!]
\begin{center}
\caption{\label{tab:pmt_activity_BI}
        Activity of the components of the PMT LAr veto system and their
        estimated background contribution BI at Q$_{\beta\beta}$ after
        anti-coincidences but before the LAr veto and PSD.
}
\begin{tabular}{lrrrcrr}
\hline
Component                                &  \multicolumn{1}{c}{$^{228}$Th}   & \multicolumn{1}{c}{$^{226}$Ra} & $^{40}$K & & BI($^{208}$Tl) & BI($^{214}$Bi) \\
\cline{2-4} \cline{6-7} \\ [-2ex]
	  		                 & \multicolumn{3}{c}{($\upmu$Bq)} & &  \multicolumn{2}{c}{(10$^{-6}$cts/(keV$\cdot$kg$\cdot$yr))}  \\
\hline
PMTs                                     & $<$1940/pc  & $<$1700/pc  & $<$9100/pc & & $<$245 & $<$33 \\
voltage dividers                         & $<$500/pc   & $<$1140/pc  & $<$11500/pc & & $<$63  & $<$22 \\
SAMI RG178 cables                        & $<$14.4/m   & $<$11.2/m   & 81(16)/m   & &               \\ 
\hspace{0.5cm} - along Cu shrouds        &             &             &            & & $<$16 & $<$2 \\
\hspace{0.5cm} - along Fiber shroud      &             &             &            & & $<$227  & $<$37 \\
copper shrouds                           & 37/kg       & 148/kg      &  -~~~~~~   & & 8.6(1)& 5.7(1) \\
Tetratex\textsuperscript{\textregistered} coated& 70/m$^2$   & 150/m$^2$    & 9800/m$^2$       & & 18.2(2) & 6.4(1)\\
\hline
\end{tabular}
\end{center}
\end{table*}
\begin{table*}[htb!] 
\tabcolsep=0.11cm
\begin{center}
\caption{\label{tab:activity}
 Activity of the components of the fiber-SiPM LAr veto system and their estimated background contribution BI at 
 Q$_{\beta\beta}$ after anti-coincidences but before the LAr veto and PSD.}
\begin{tabular}{lccccrrr}
\hline
Component & ~~method~~	& $^{228}$Th        	& $^{226}$Ra	& $^{40}$K       & mass		& \multicolumn{1}{c}{BI($^{208}$Tl)}     & BI($^{214}$Bi) \\
 \cline{3-5}  \cline{7-8} \\ [-2ex]		
	&		& \multicolumn{3}{c}{(mBq/kg)}  & \multicolumn{1}{c}{(g)} &  \multicolumn{2}{c}{(10$^{-6}$cts/(keV$\cdot$kg$\cdot$yr))}  \\
\hline
fiber BCF-91A 	& ICPMS		& 0.058			& 0.042		& 0.46	& 765		& 238	 	& 175~(1)
 		 	\\

plastic opt.coupl.& \gspec 	& 0.15(8)		& $< 0.19$ 	&3.0	& 32		& 7(3)       & $<$0.64	\\

SiPM		& ICPMS		& $<$1	     	        & $<$3		&-        & 1.3		& $<$1.8    & $<$0.41 \\

Cuflon\textsuperscript{\textregistered}		& \gspec	& 0.8(5)  & 1.3(4) &18		& 15		& 16(10)     & 2.1(6)  \\

pins 		& \gspec	& $<$5.8	        & 15(3) 	&220	& 9.4		& $<74$	     & 15(3)  \\

screws		& \gspec	& $<$5.8		& 15(4) 	&$<$120	& 12.7		& $<$100    & 20(5)  \\

glue EP601	& \gspec	& $<$1.5		& $<$0.55	&$<$13	& 4.0		& $<$8.1    & $<$0.23 \\
\hline
\end{tabular}
\end{center}
\end{table*}
The radiopurity of the components of the LAr veto system is critical since
they add a large amount of new material close to the detector array. The
activities of the components as well as the estimated background contributions
of the PMT and fiber systems are given in Tables~\ref{tab:pmt_activity_BI} and
\ref{tab:activity}, respectively.

The specification was that both systems contribute about equally to the
BI. The sixteen 3" PMTs and voltage dividers exhibit the largest specific
activities although the radiopurity of the PMTs had been further improved in
collaboration with the manufacturer, and the potted voltage dividers have been
fabricated from selected materials, 0.5\,mm
Cuflon\textsuperscript{\textregistered} as printed circuit board and film
capacitors rather than ceramic ones. Nevertheless, the PMTs with their voltage
dividers had to be mounted at a minimum distance of more than 1\,m from the
closest Ge detector in order to stay within the background budget. On the
other hand, the very low specific activity of the fibers allows to deploy
them, almost 1\,kg of material, rather close to the detector array.
Altogether, both subdetectors contribute about
1.2$\cdot$10$^{-3}$cts/(keV$\cdot$kg$\cdot$yr) to the BI at \qbb\ before LAr
veto and PSD; that is more than the Phase~II goal.  As shown in
Section~\ref{sec:performance} the LAr veto alone suppresses, however, the
backgrounds at \qbb\ from \Th\ and \Ra\ sources by factors of about 100 and 6,
respectively, resulting in BI values that are fully acceptable for Phase~II.
Preliminary MC simulations with photon tracking predict indeed an even larger
suppression yielding a total contribution of the LAr veto hardware to the BI
at \qbb\ of about 10$^{-5}$cts/(keV$\cdot$kg$\cdot$yr) after LAr veto and PSD
cuts.

\section{Performance}
\label{sec:performance}
Phase~II physics data taking started on December 25, 2015.  A first data
release PIIa has been presented after about 6 months of running at the
Neutrino Conference 2016, and the results have been published in
Nature~\cite{g2-nature}. The following discussion considers the Phase~II data
taking period up to April 15, 2017.  About 6.7\,\% of the total of 477
calendar days were devoted to calibration or maintenance, the rest to physics
data taking. A series of nearby earthquakes with magnitudes up to 5.5 slightly
affected the ongoing measurements.  The exposure taken on tape with the
enriched detectors amounts to 43.4\,\kgyr\ which corresponds to a duty factor
of more than 93\,\%. Data validation reduces this exposure by 9\,\% to
34.4\,\kgyr. As of April 15 2017, all deployed detectors (see Appendix
Fig.~\ref{fig:stringcontent}) are operational except the enriched detector
GD91C which has a damaged JFET.

The performance of the detector system is characterized in particular by the
stability of its energy calibration as well as by the resolutions of the
various detector components. It is assessed by regular calibrations and by
monitoring continuously the flow of physics data and test pulser events.  The
following sections present representative examples for the various components
and an evaluation of the full system performance including the achieved level
of background suppression.

\subsection{Ge detectors}
The operation of bare Ge crystals in LAr is a non-standard technique, and the
stability of leakage currents was a major issue in the preparations for
Phase~I~\cite{g1-instr}. It was found that passivated detectors - those with a
passivation layer in the insulating groove between p$^+$ contact and the
conductive lithium n$^+$ layer - showed constantly increasing leakage current
from repeated calibrations, presumably due to the build up of charges in the
groove. Non-passivated detectors did not show this effect, and hence all
semi-coaxial detectors deployed in Phase~I had the passivation layer
removed. To speed up the refurbishment and delivery of the Phase~II detectors,
the removal of the standard passivation was given up for quite a number of
detectors (see yellow colored detectors in Appendix
Fig.~\ref{fig:stringcontent}).

\begin{figure}[htb]
\begin{center}
\includegraphics[width=0.98\columnwidth]{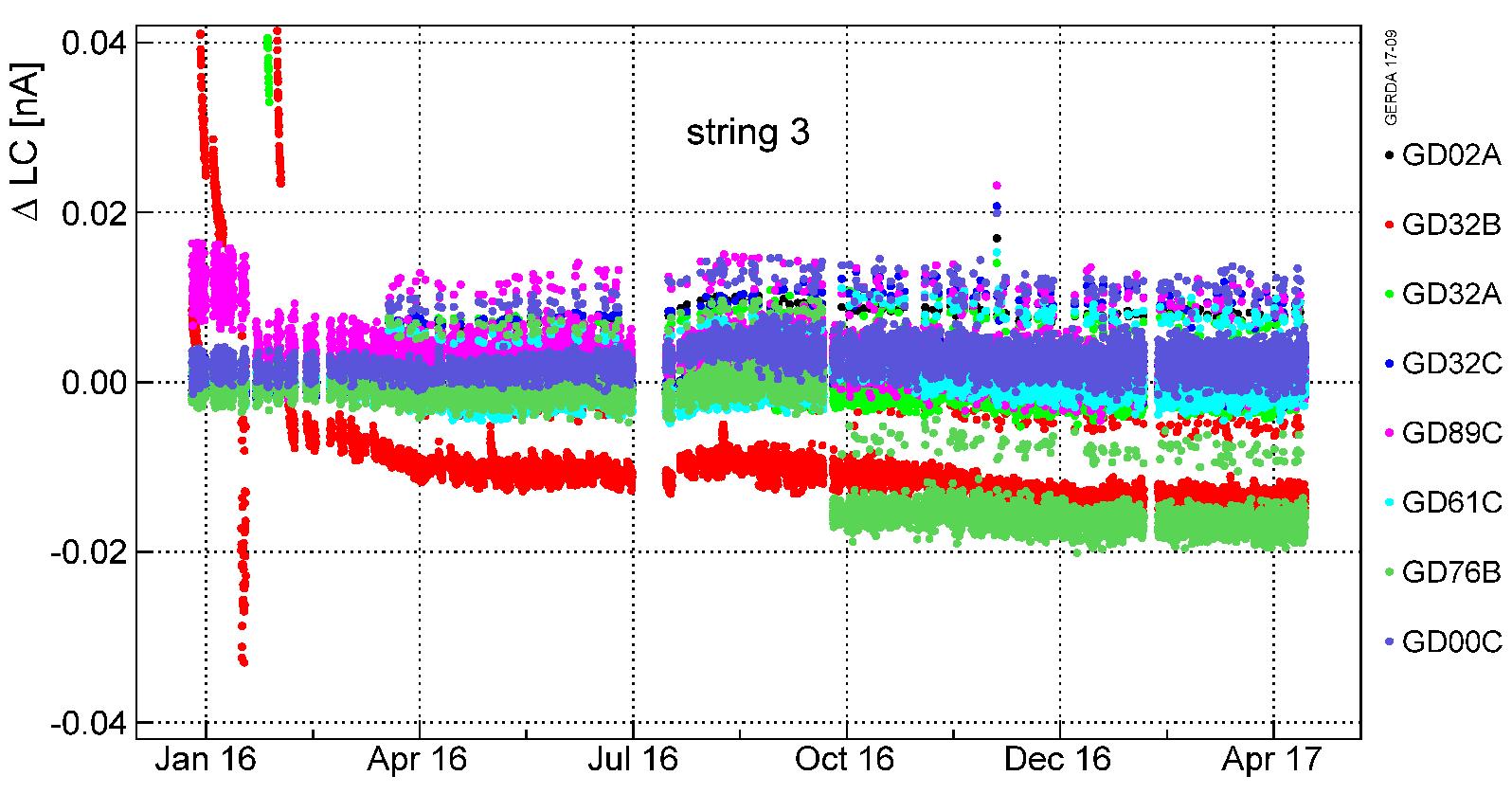}
\caption{\label{fig:bslvst}
       Leakage currents of indicated detectors of string 3 in the period from
       December 2015 to April 2017.  For the other detector strings the
       corresponding data are shown in Fig.~\ref{fig:bslvst1} of the
       Appendix.
}
\end{center}
\end{figure}
Fig.~\ref{fig:bslvst} and Fig.~\ref{fig:bslvst1} of the Appendix show the
evolution of the leakage current for all Phase~II Ge detectors in the time
span between December 2015 and April 2017. In the first months, a temporary
increase of leakage currents during calibration has been observed for several
detectors, passivated and non-passivated ones. This is visible by current
spikes reaching up to 300\,pA immediately after irradiation; they return to
the previous current values within half a day. Except for one detector, the
leakage currents of all detectors, both passivated and non-passivated ones,
are stable since May 2016 or even reduced compared to the beginning. Hence the
number of stable detectors included in the analysis has increased in
time. Thus, our experience is strikingly contrary to a former report about
`the limited long-term stability of naked detectors in LN as result of
increasing leakage current'~\cite{KlaKri06}.

\begin{figure*}[htb!]
\begin{center}
\includegraphics[width=1.9\columnwidth]{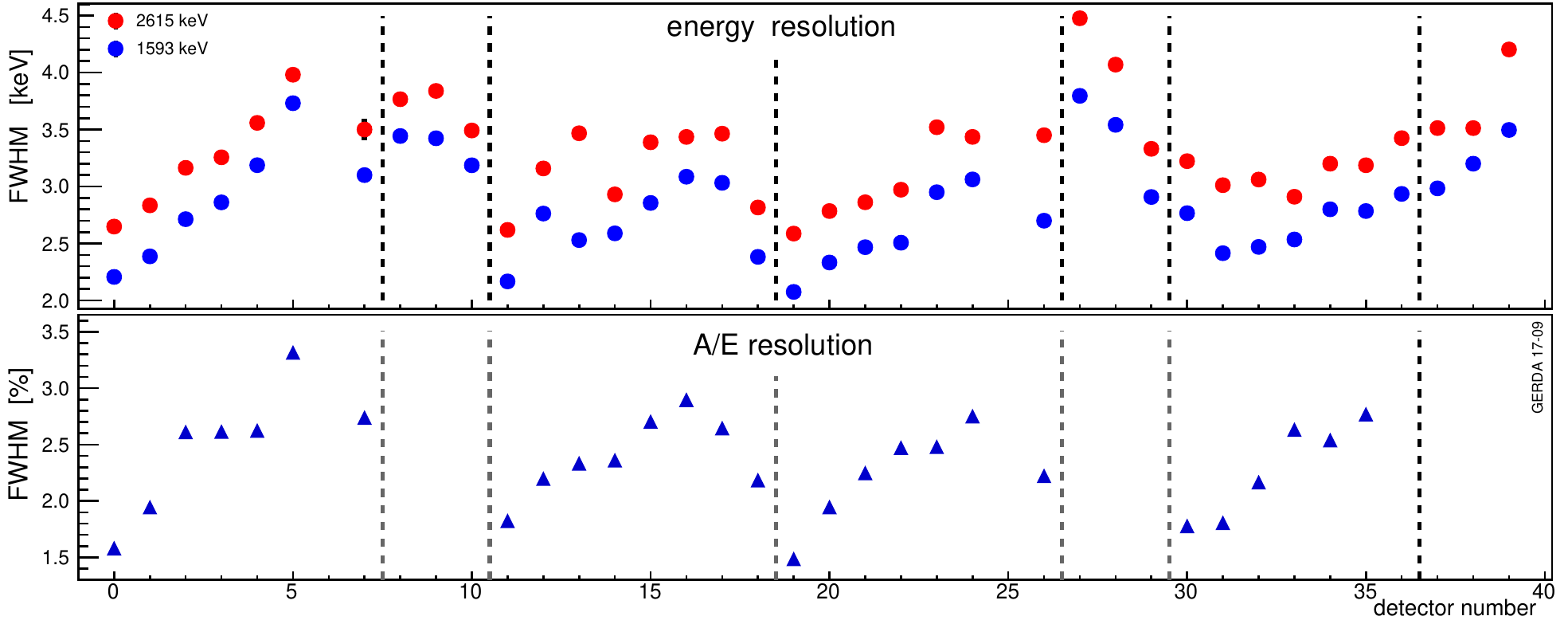}
\caption{\label{fig:resolutions}
       FWHM energy resolutions (top) of the 2615\,keV line and its DEP at
       1593\,keV, and A/E resolutions (bottom) of individual Ge detectors
       deduced from the DEP~\cite{PhDvici}.  The dashed lines separate the 7
       detector strings; within each string the detector number increases from
       top to bottom detector.
}
\end{center}
\end{figure*}

\subsubsection{Energy scale stability and resolution}
Energy calibrations with a \Th\ source are performed regularly, typically once
per week; a calibration spectrum is shown in subsection~\ref{sec:sslar}
(Fig.~\ref{fig:LArThSpectrum}).  The energy reconstruction is performed in the
off-line analysis of the digitized charge pulses using either a standard
semi-Gaussian shaping or, for improved energy resolution, a finite-length
zero-area cusp filter~\cite{ZAC}. The position of the various peaks are used
to establish the energy scale (gain) as well as the resolution as a function
of energy. The \Tl\ $\gamma$ line at 2614.5\,keV provides a convenient measure
to monitor the resolution close to the \qbb\ energy. The upper part of
Fig.~\ref{fig:resolutions} shows for all detectors the measured FWHM
resolutions for this line as well as for its double escape peak (DEP) at
1592.5\,keV.  The coaxial detectors are in strings 2, 5, and 7. Most of them
have a resolution of less than 4.0\,keV FWHM. In the period from December 2015
to April 2017 the interpolated average energy resolution at \qbb\ is
3.90(7)\,keV FWHM, i.e.  better than in Phase~I (4.8(2)\,keV). The resolutions
of the BEGe detectors show unexpectedly strong variations between 2.5 and
5.6\,keV. At \qbb\ the interpolated mean is 2.93(6)\,keV which is also better
than in Phase~I (3.2(2)\,keV).  The top detector in each string exhibits
indeed the expected resolution well below 3\,keV but there is a consistent
trend of decreasing resolution when going down in the string.  This effect has
not been observed in Phase~I where single wires were used for the signal
readout instead of the FFCs in Phase~II. Thus larger stray capacitances with
increasing cable length and capacitive coupling within the readout cable
bundle of a string might explain the observation. A good understanding of the
resolution pattern is, however, still lacking and further work on this issue
is in progress.
 
\begin{figure}[htb!]
\begin{center}
\includegraphics[width=0.98\columnwidth]{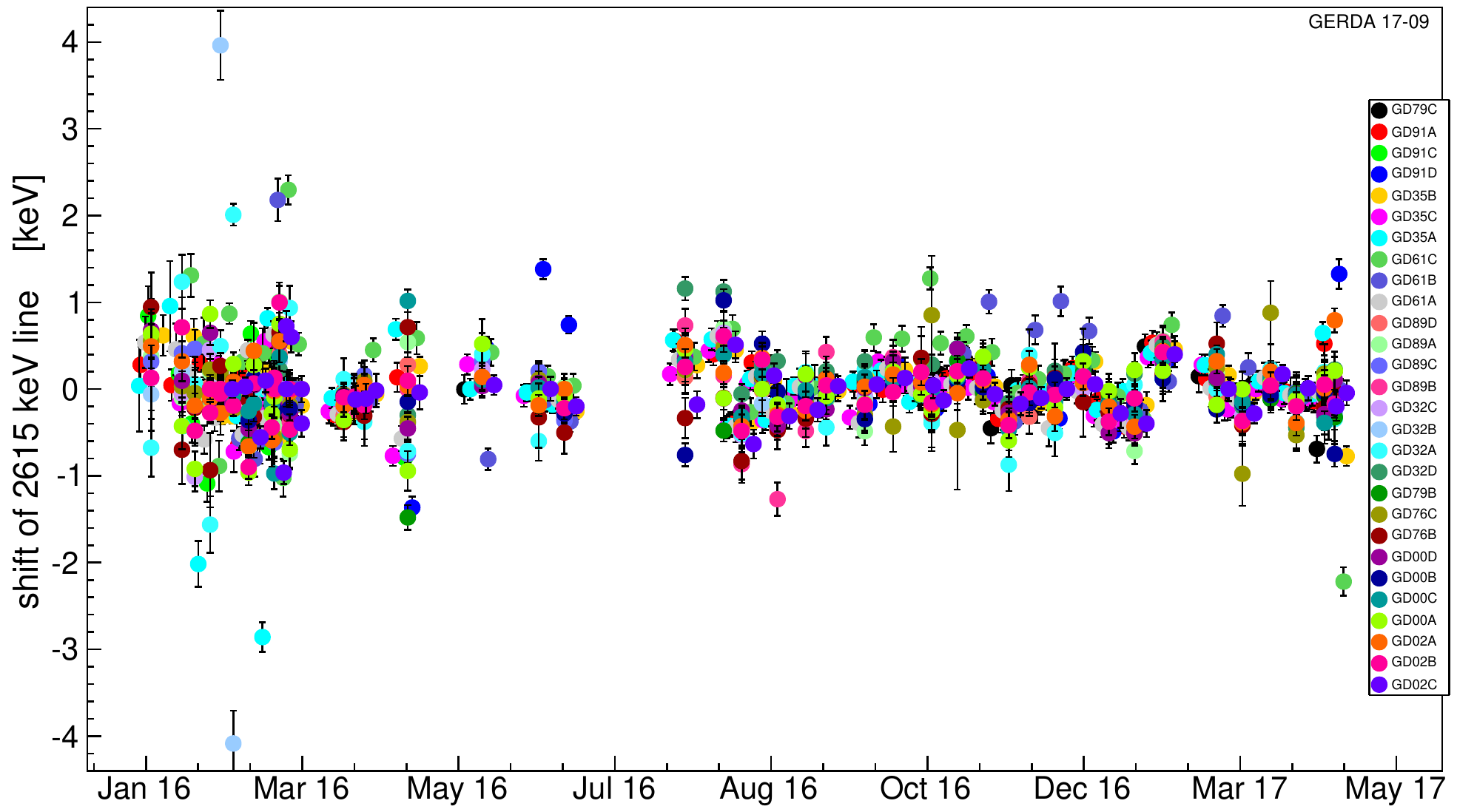}
\includegraphics[width=0.98\columnwidth]{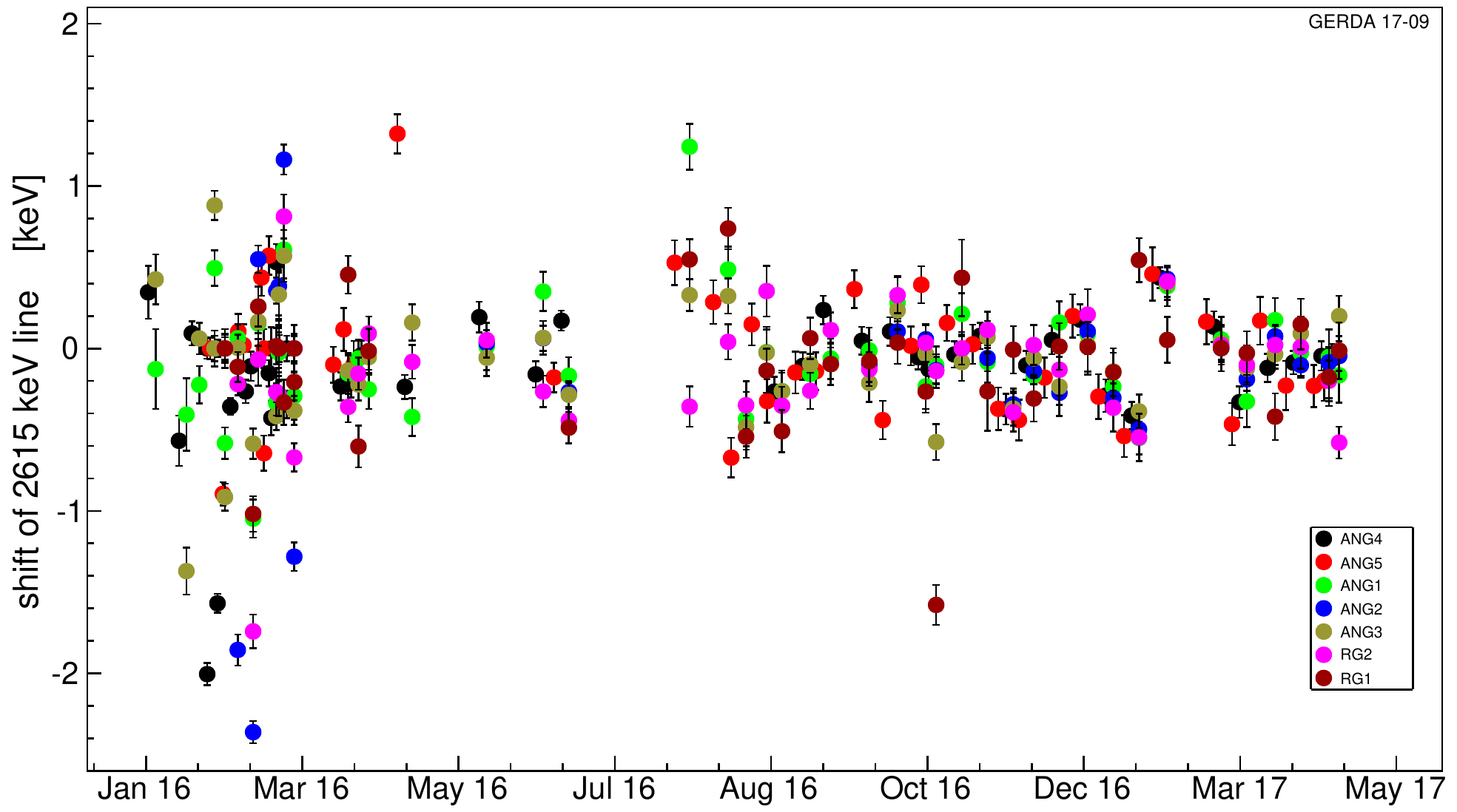}
\end{center}
\caption{\label{fig:shift-after-calibs}
        Shift of the 2615\,keV peak of \Tl\ between consecutive calibrations
        for indicated BEGe  (top) and enriched coaxial detectors (bottom). 
}
\end{figure}

The stability of the energy scale is not only verified by calibration
measurements but continuously monitored by injecting every 20\,s a test charge
into the electronic front ends. If the gain variation is larger than 0.1\%, an
extra unscheduled calibration with the \Th\ source is
taken. Fig.~\ref{fig:shift-after-calibs} shows for the period from December
2015 to April 2017 the shifts of the 2615\,keV calibration peak between
successive calibrations.  Most data points are within the range of
$\pm1\sigma\approx\pm$1.5\,keV; this is sufficient to allow the merging of the
data from all periods (periods with shifts exceeding 2\,keV are discarded from
analysis).

\subsubsection{Pulse shape discrimination efficiencies}
The new Ge detectors for Phase~II have been chosen to be of the BEGe type
because the time profile of its current pulse allows for a powerful but simple
discrimination between $0\nu\beta\beta$-like events, i.e. localized
single-site events (SSEs), from background events which have often multiple
energy depositions (multi-site events, MSEs) or occur at the detector
surface~\cite{g1-psd}. It is based on a single parameter, the ratio A/E of the
maximum of the current pulse A over the total energy E.  SSEs are identified
by an (normalized) A/E value of about 1, MSEs and events on the n$^+$ surface
by a value lower than 1, and surface events at the p$^+$ contact like $\alpha$
particles by A/E$>$1. Like in Phase~I the A/E cut is calibrated with the data
from the weekly calibrations with the \Th\ source. Events in the DEP are used
as a proxy for $0\nu\beta\beta$ events. The inset in \Th\ calibration spectrum
(Fig.~\ref{fig:LArThSpectrum}) illustrates the effect of the A/E$\sim$1 cut
(yellow colored spectrum): the obvious survival of the DEP at 1593\,keV and
the distinct suppression of the single gamma line of $^{212}$Bi at 1621\,keV.
\begin{figure*}[htb]
\begin{center}
\includegraphics[width=1.8\columnwidth]{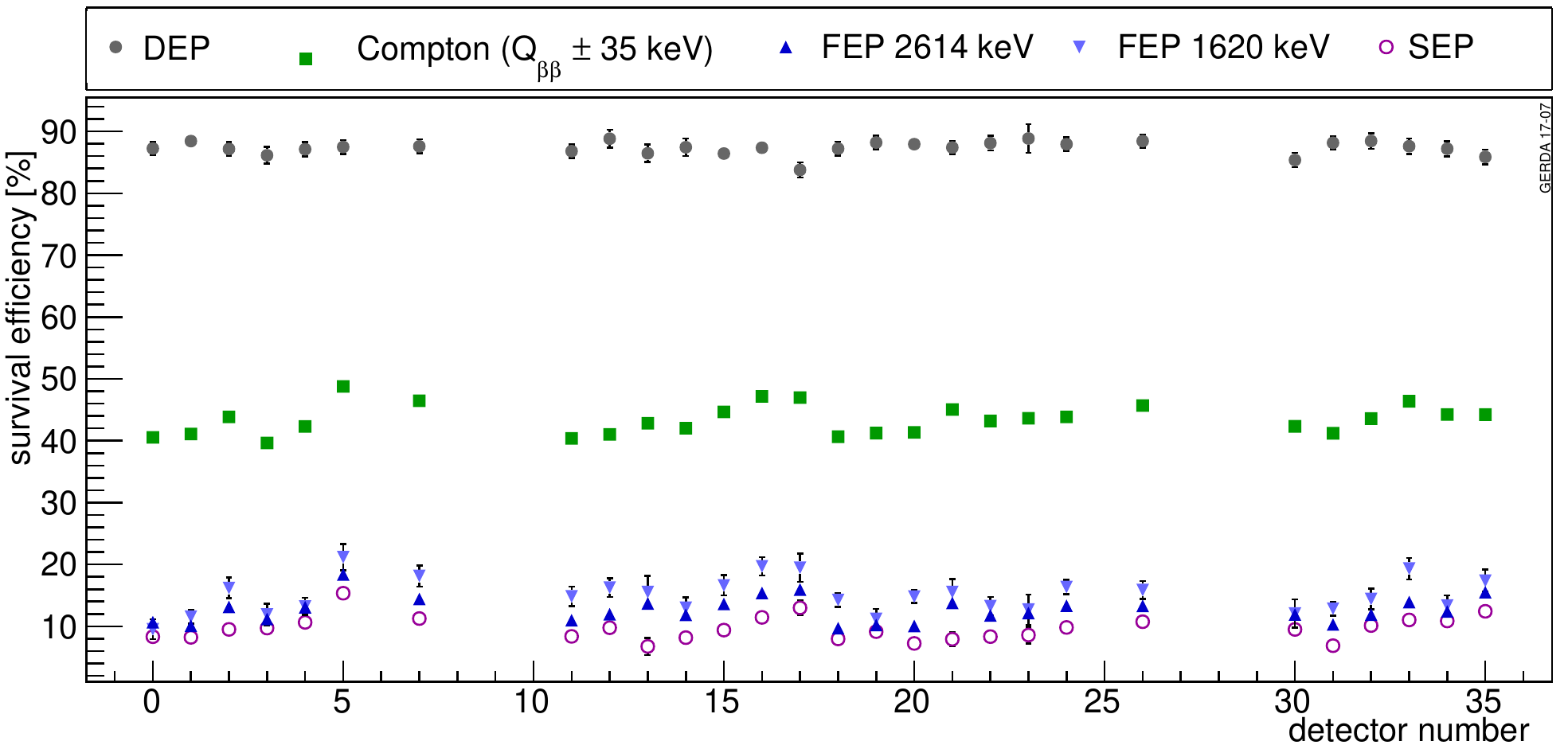}
\caption{\label{fig:psdeffi}
      Survival efficiencies for the double escape (DEP) and single escape peak
      (SEP) of the 2615\,keV $\gamma$ line of $^{208}$Tl, for the full energy
      $\gamma$ peaks (FEPs) at 1620\,keV and 2615\,keV, and the Compton events
      around Q$_{\beta\beta}$ \cite{PhDvici}.  Uncertainties are partly
      smaller than the symbols.
}
\end{center}
\end{figure*}

While a detailed account of the PSD analysis of Phase~II data is in
preparation we discuss here just some differences compared to the analysis of
the Phase~I data.  To select signal-like events a two-sided A/E cut is
applied: MSE and n$^{+}$ surface events are removed by a cut on the low A/E
side, while events on the p$^+$ contact are rejected by a high A/E cut.  The
lower part of Fig.~\ref{fig:resolutions} shows the strong variation of the A/E
resolution of individual BEGe detectors following a similar trend as the
energy resolution.  The A/E resolutions differ for individual detectors by up
to a factor of 2, and they are often significantly worse than in Phase~I where
they varied between 1.5\,\% and 1.9\,\%.  Moreover, to reach the Phase~II BI
goal of \dctsper, a stronger low A/E cut for higher suppression of MSE
compared to Phase~I is applied.  Hence, other than in Phase~I, individual cuts
have been applied to each detector accounting for the increased noise.

The low A/E cut is chosen according to 90\,\% acceptance of the DEP events.
Fig.~\ref{fig:psdeffi} shows for each BEGe detector and its respective A/E
cuts the survival fractions for various energy and peak regions in the
$^{228}$Th spectrum. The survival fractions in the peak and Compton regions
increase from top to bottom in the respective detector strings, reflecting the
deterioration of the A/E resolution from top to bottom.  For further
suppression of surface events, a high A/E cut with twice the separation from
A/E\,=\,1 as the lower cut is applied.

Table~\ref{tab:effiall} provides a summary from the combined data of all BEGe
detectors and calibration runs.  The survival fraction of Compton events from
\Tl\ is 45.3\,\% at \qbb. Full energy $\gamma$ peaks (FEPs) are suppressed to
less than 15\,\%. The strongest suppression (11\,\%) is achieved for single escape
peaks (SEPs) which exhibit a relatively large MSE component. The high A/E cut
reduces the survival fractions of events in the DEP and \qbb\ region by less
than 5\,\%.

\begin{table}[htb!]
\begin{center}
\caption{\label{tab:effiall}
         Rejected event fractions by the low and high A/E cuts, and the
         survival fractions (s.f.) from all BEGe detectors and calibration
         runs~\cite{PhDvici}.}
\begin{tabular}{lccc}
\hline
region    & low cut & high cut & s.f. \\
          &   (\%)    &   (\%)     & (\%)   \\
\hline
DEP(1593) & 10.0(2) & 2.68(6)  & 87.3(2)\\
FEP(1621) & 83.5(3) & 1.62(8)  & 14.9(3) \\
FEP(2615) & 83.58(3)& 1.82(1)  & 14.60(2) \\
SEP(2104) & 87.8(2) & 1.55(4)  & 10.6(2) \\
(2039$\pm$35)\,keV & 52.5(1) & 2.20(2) & 45.3(1) \\
\hline
\end{tabular}
\end{center}
\end{table}

\subsection{LAr veto}\label{sec:sslar}
\begin{figure}[htb!]
\begin{center}
\includegraphics[width=0.9\columnwidth]{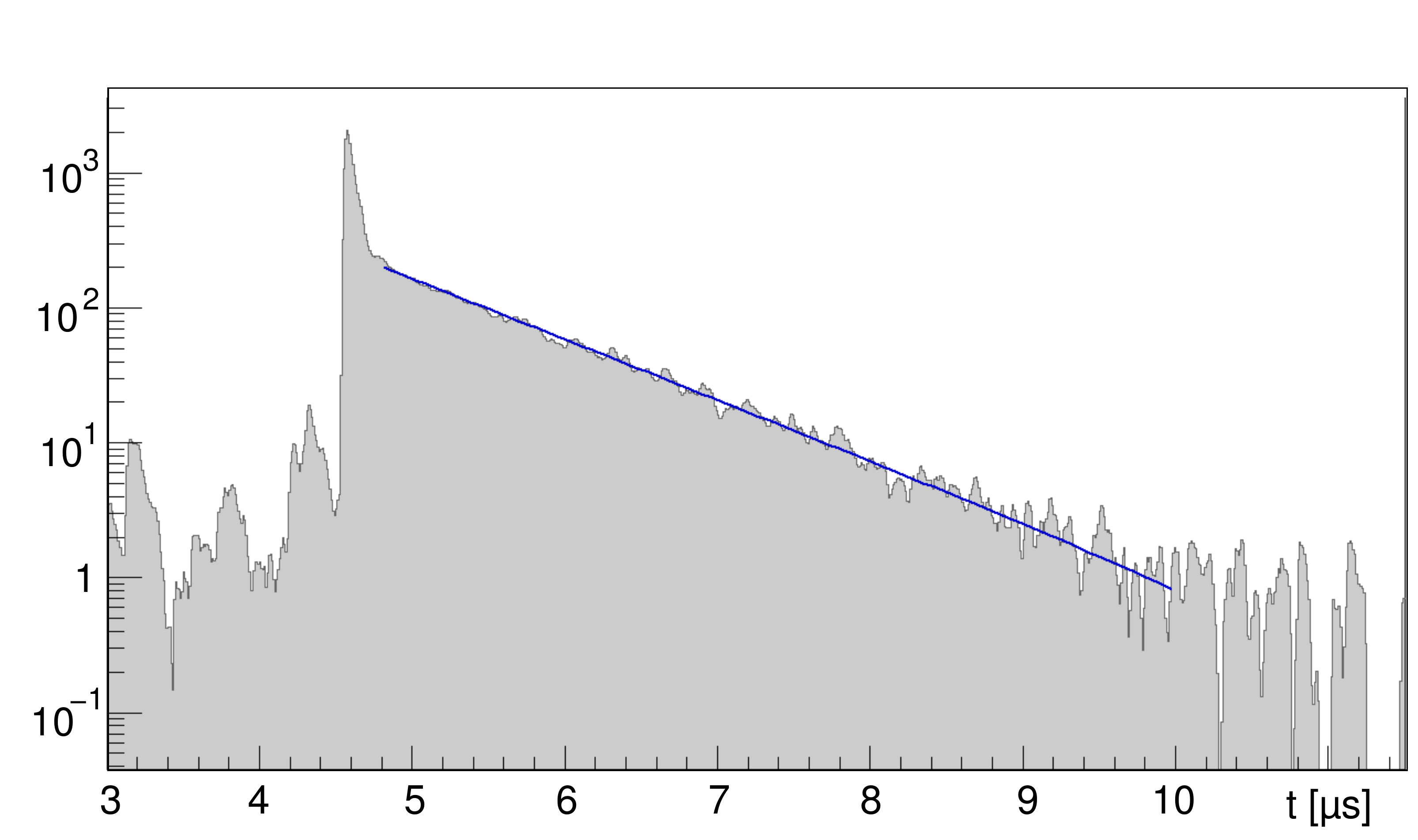}
\caption{\label{fig:tripletLFT}
      Average PMT waveform recorded on 30.09.2015 in the LAr of the
      \gerda\ cryostat. The solid line indicates the triplet lifetime fit with
      a decay time of 0.97\,$\upmu$s~\cite{PhDanne}.
}
\end{center}
\end{figure}

In ultra-pure LAr the scintillation light yield is about 41$\cdot10^3$ photons
per 1\,MeV electron-equivalent energy deposition for the fast and slow
(triplet) component of the Ar excimer~\cite{ArSciPhotoY}.  The purity of LAr
has a strong influence on the light yield affecting predominantly the light
emitted from the triplet state whose lifetime can thus be taken as a good
indication for the LAr purity. For ultra-pure LAr, it is
1590(100)\,ns~\cite{triplifetime}.  The triplet lifetime of the LAr contained
in the \gerda\ cryostat has been deduced by averaging the waveforms (see
Fig.~\ref{fig:GePMTSiPMT}) acquired with the PMTs of the LAr veto system and
fitting the slow exponential tail (Fig.~\ref{fig:tripletLFT}). The fit yields
a value of 968(11)\,ns, indicating a still acceptable LAr purity after almost
6 years of operation without a significant refill and any purification.

The LAr veto system is running stably with all PMT and SiPM channels fully
functional since March 2016.  Before this date, 2 PMTs were not operational
due to contact problems outside the lock system.  During commissioning the LAr
veto system has been immersed into the cryostat (and warmed up) more than 20
times. No ageing effects have been observed since its first deployment in
April 2015.

\begin{figure}[htb]
\begin{center}
\includegraphics[width=0.8\columnwidth]{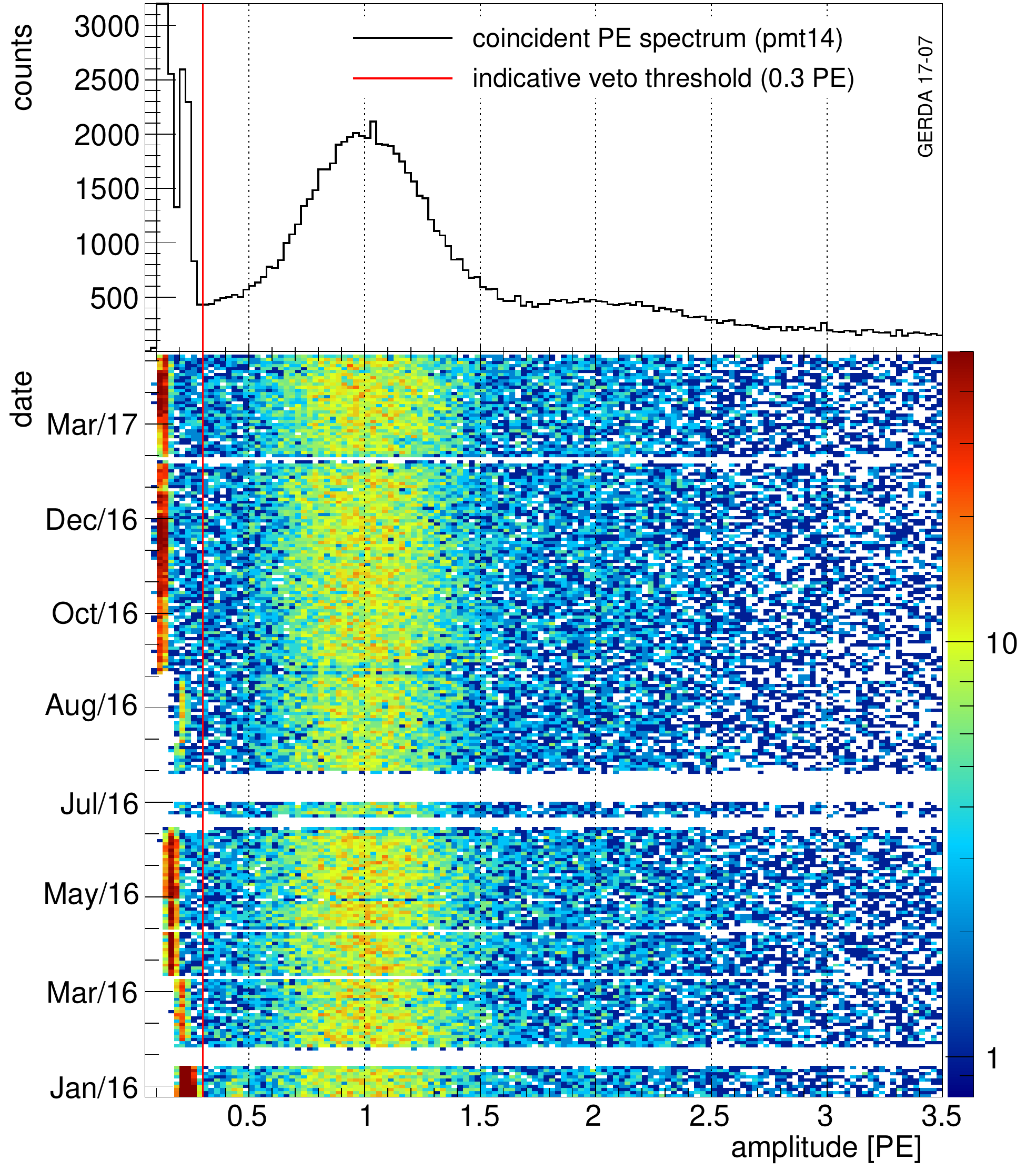}
\caption{\label{fig:PHSTABpmt}
    Pulse height spectrum of a PMT (top) deduced from the scatter plot
    (bottom) which has been accumulated between January 2016 and April 2017.
}
\end{center}
\end{figure}
\begin{figure}[htb]
\begin{center}
\includegraphics[width=0.8\columnwidth]{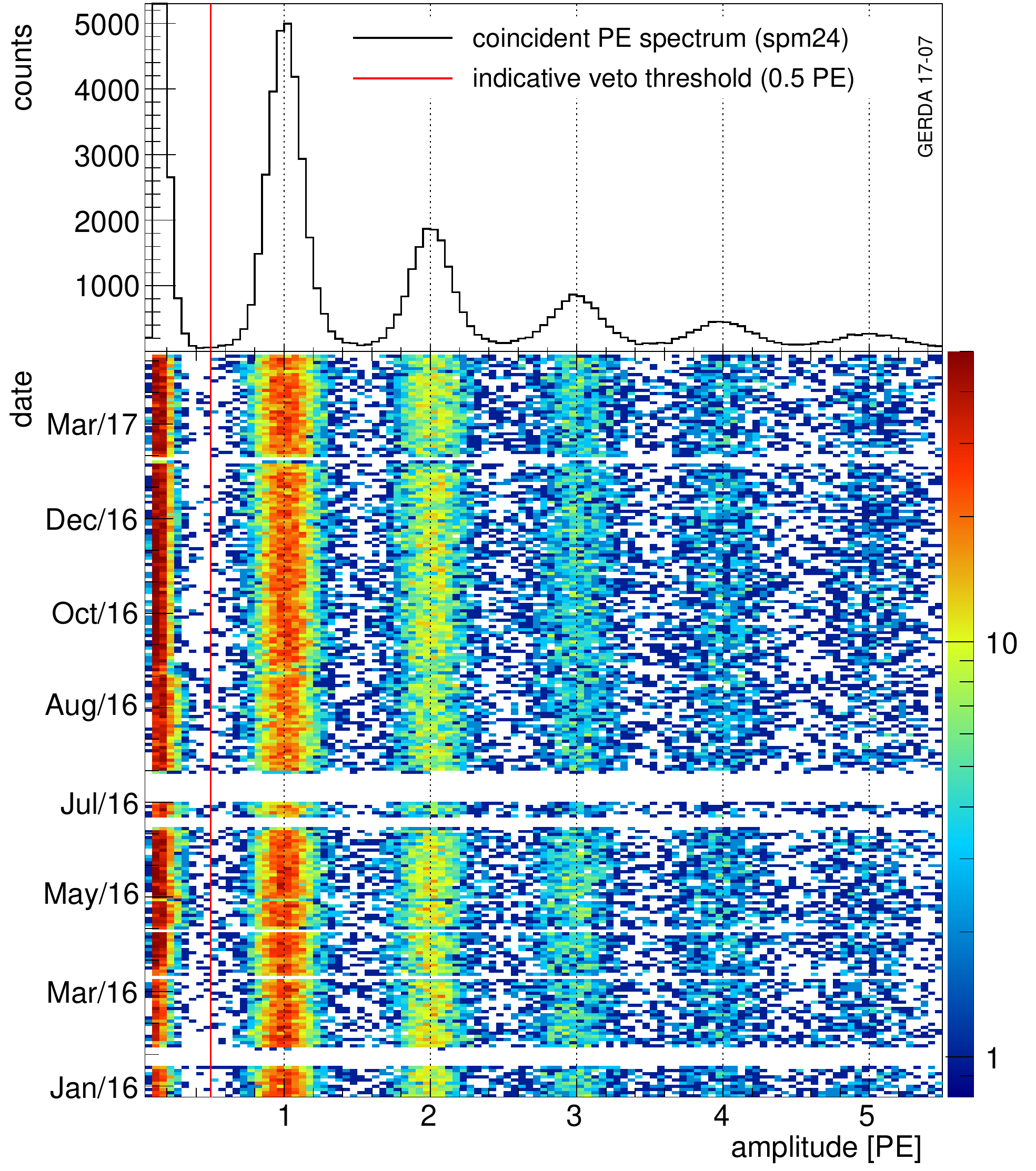}
\caption{\label{fig:PHSTABsipm} 
      Same as Fig.~\ref{fig:PHSTABpmt} but for a SiPM channel representing 6
      SiPMs read out in parallel.
}
\end{center}
\end{figure}

The dark rate in the individual PMTs and SiPM channels is 300-500\,Hz and
$<$200\,Hz, respectively.  It is mostly due to the $\beta$ decay
(Q$_\beta$=565\,keV) of $^{39}$Ar (8$\cdot10^{-16}$ abundance relative to the
stable $^{40}$Ar) with an activity of $\sim$1\,Bq/kg of
$^{40}$Ar~\cite{ar39act}. The resulting dead time is about 2\,\%.  Note that
the radon shroud~\cite{g1-instr} limits the field of view of the SiPM-fiber
assembly in the cryostat to radii of less than 37.5\,cm; our preliminary
estimate for the attenution length is $\sim$15\,cm.

The stability and noise level of the LAr veto system is continuously monitored
with physics data.  Figs.~\ref{fig:PHSTABpmt} and \ref{fig:PHSTABsipm} show on
top a typical pulse height spectrum for a PMT and SiPM channel, respectively;
and at bottom the scatter plots from which these pulse height spectra have
been deduced.  These data have been accumulated in coincidence with Ge
triggers during a time span of about 16 months.  The noise levels and gains
are obviously stable, and the clear separation between noise and single
photoelectron (PE) peak leads to a straightforward setting of the veto
thresholds, indicated by the red lines, that can be easily readjusted if the
gain jumps.  These settings are 0.2\,-\,0.35\,PE for PMTs, and 0.4\,-\,0.6 PE
for SiPMs; the threshold is higher for only one PMT channel (0.6\,PE) and one
SiPM channel (0.9\,PE) where the individual SiPMs exhibit slightly different
gains.

\begin{figure*}[htb!]
\begin{center}
\includegraphics[width=0.98\columnwidth]{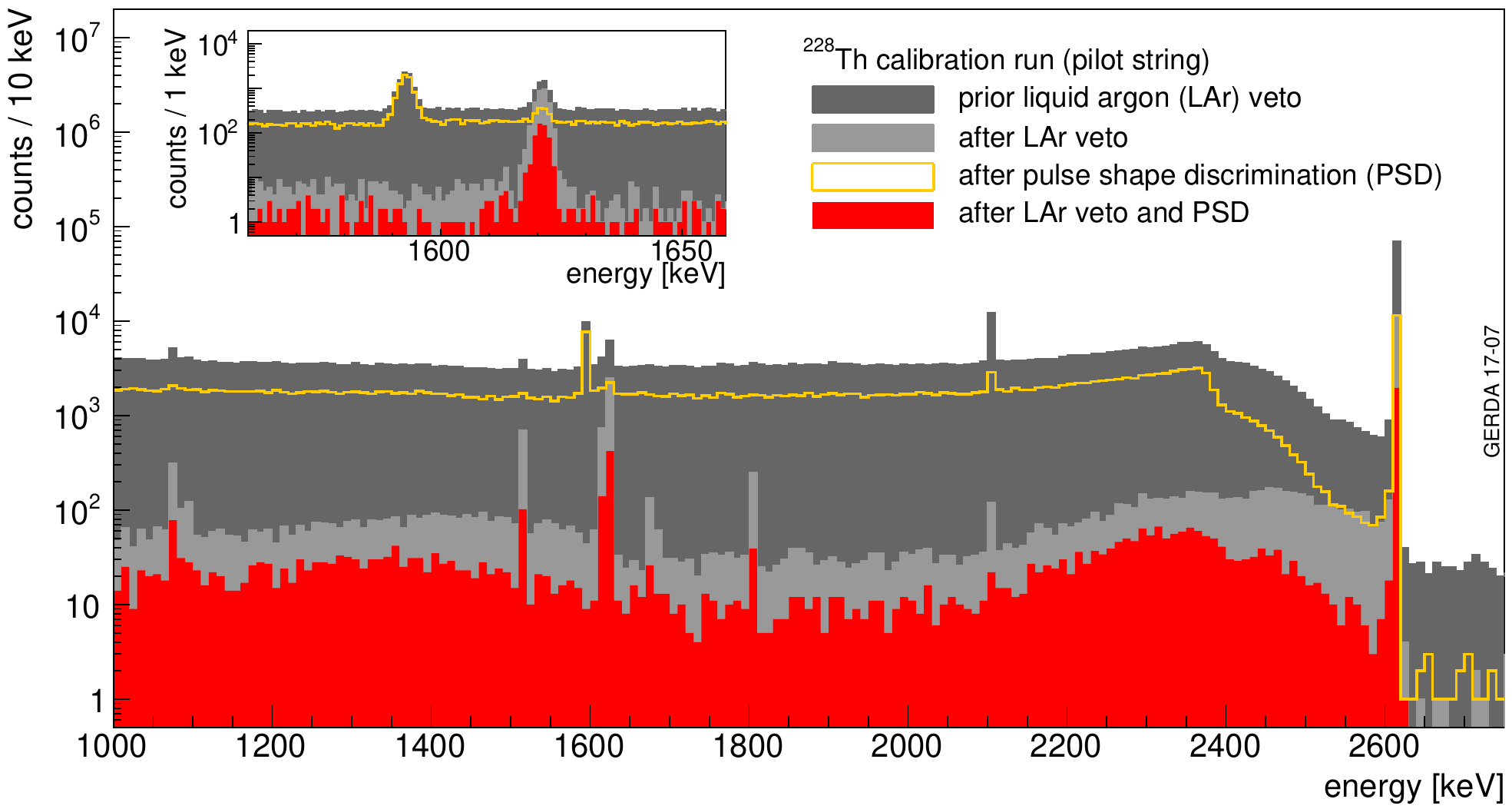}
\includegraphics[width=0.98\columnwidth]{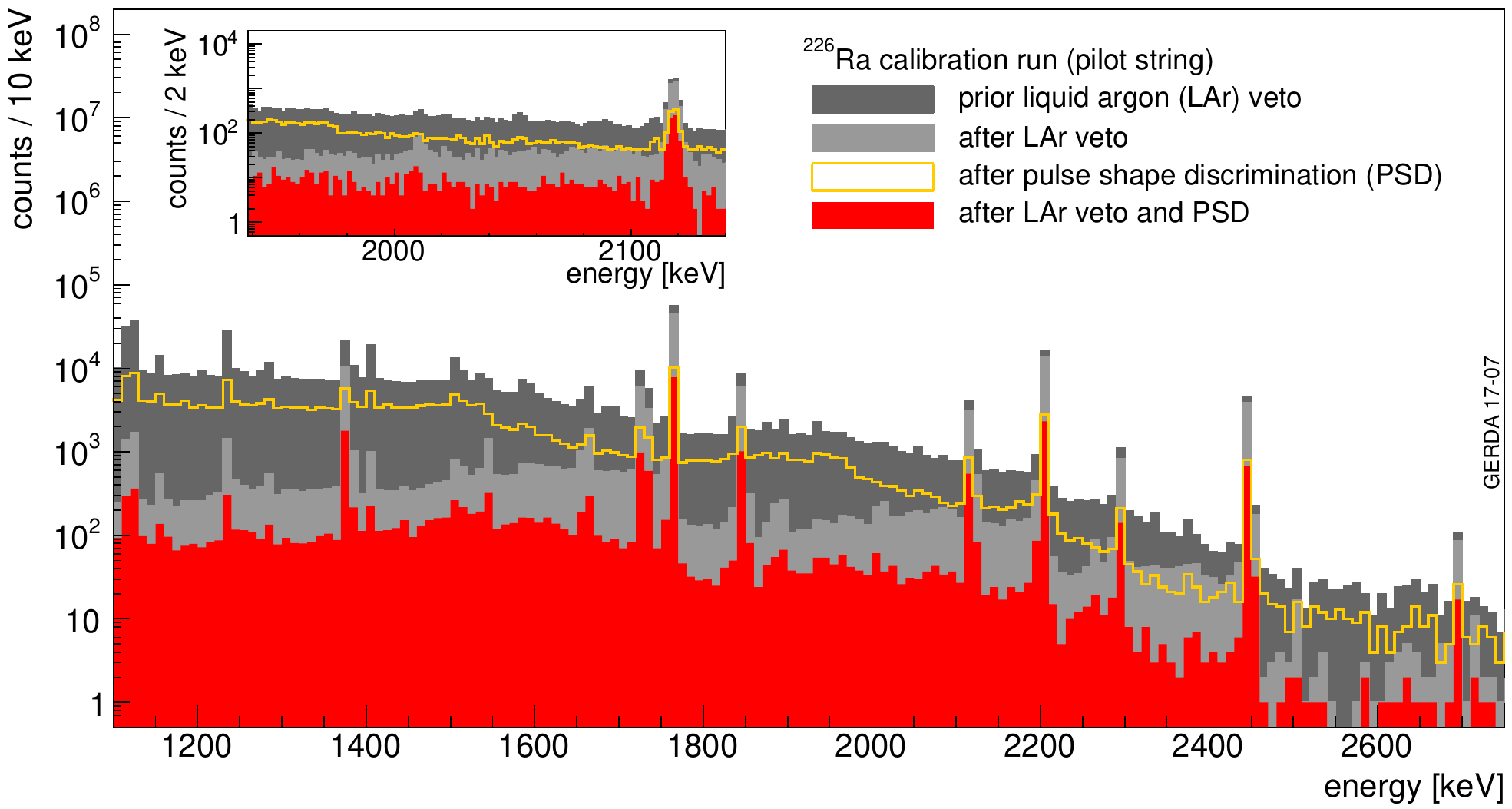}
\caption{\label{fig:LArThSpectrum}
      Calibration spectra for \Th\ (left) and \Ra\ (right) with detector
      anti-coincidence and muon veto (dark grey), with suppression by the LAr
      veto system and pulse shape discrimination (PSD), and by the combination
      of LAr veto system and PSD. The inset on the l.h.s. shows the double
      escape peak of \Tl\ at 1593\,keV and the 1621\,keV single gamma line of
      $^{212}$Bi, on the r.h.s the 2118\,keV single gamma line of $^{214}$Bi.
}
\end{center}
\end{figure*}

The performance of the LAr veto system has been tested with both a \Th\ and
\Ra\ source.  Fig.~\ref{fig:LArThSpectrum} shows the measured spectra as
obtained after standard quality cuts, and after indicated additional cuts.
Both spectra exhibit the same specific features that have been observed in the
\LArGe\ test stand; for a detailed discussion see~\cite{LArGe}. One highlight
is displayed in the inset of the \Th \ spectrum (Fig.~\ref{fig:LArThSpectrum},
l.h.s.). The DEP of the \Tl\ 2615\,keV $\gamma$ line at 1593\,keV is barely
suppressed by PSD ($\approx$90\,\% acceptance) while the LAr veto cut
eliminates it completely due to the two escaping 511\,keV annihilation
quanta. On the other hand, the $^{212}$Bi single $\gamma$ line at 1621\,keV is
largely accepted by the LAr veto because the full energy is deposited in the
Ge detector.

The \Tl\ 2615\,keV $\gamma$ line as well as the various \Bi\ $\gamma$ lines in
the \Ra\ spectrum (Fig.~\ref{fig:LArThSpectrum}, r.h.s.) provide significant
contributions to the background around \qbb~\cite{g1-back}, and hence the
suppression of the continuum around \qbb\ is of special interest.  The
respective suppression factors are collected in Table~\ref{tab:sup_factors}.
\begin{table}[htb!]
\begin{center}
\caption{\label{tab:sup_factors}
       Suppression factors obtained in the ROI with a $^{228}$Th and
       $^{226}$Ra source from Ge anti-coincidences (anti), the LAr veto (LAr),
       pulse shape discrimination (PSD), and after all cuts (all); the
       acceptance (acc.) values refer to all cuts.
}
\begin{tabular}{lccccc}
\hline
Source &         anti  &  LAr       &  PSD  &        all        &  acc. \\
\hline
$^{228}$Th  &   1.26(1)   & 98(4)  &  2.19(1)   &   345(25) &  0.868   \\
$^{226}$Ra &    1.26(1)   & 5.7(2) &  2.98(6)   &   29(3)   &  0.899    \\
\hline
\end{tabular}
\end{center}
\end{table}
While the suppression by PSD is comparable, a much stronger suppression by the
LAr veto for \Th\ than for \Ra\ is noticed.  This can be understood since for
\Ra\ the main contribution to the Compton continuum is due to the 2204\,keV
single $\gamma$ line so that less than 200\,keV are available for deposition
in the LAr; in case of \Th, however, the more energetic 2615\,keV $\gamma$ ray
is accompanied in 86\,\% of the cases by a 583\,keV $\gamma$ ray so that more
energy is available for deposition in the LAr.  In general, our measured
suppression factors are smaller than the ones obtained in
\LArGe~\cite{LArGe}. This might be due to several reasons, e.g.  different
geometry with different shadowing conditions, different radioactive sources
with different leakage of $\beta$ particles, and, definitely, lower purity of
the LAr.

\begin{figure}[h!]
\begin{center}
\includegraphics[width=0.7\columnwidth]{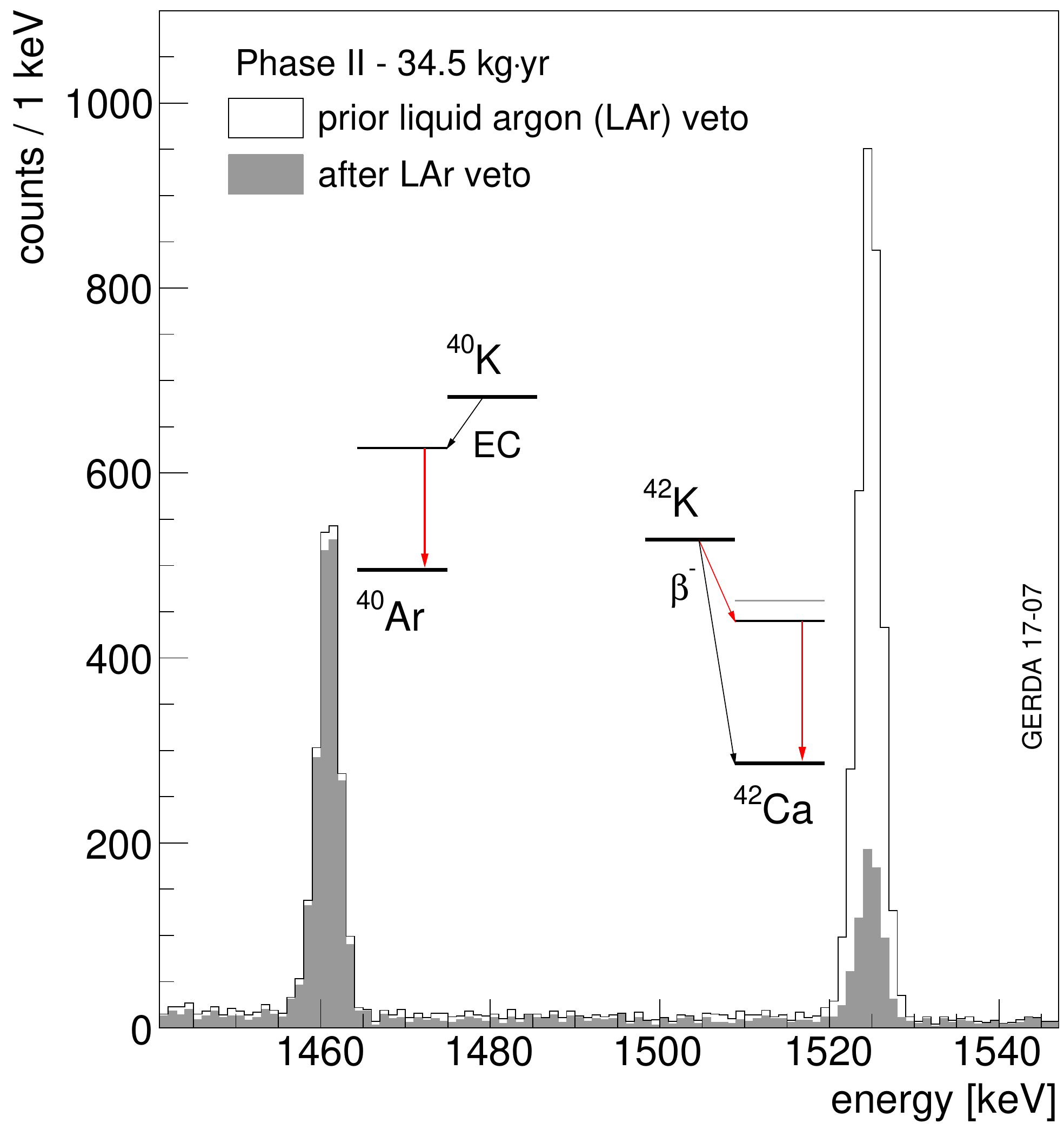}
\caption{\label{fig:k40k42larveto}
               Monitoring the LAr veto performance with the $^{40}$K and $^{42}$K
               $\gamma$ lines (see text). 
}
\end{center}
\end{figure}
The performance of the LAr veto system can be monitored continuously with the
two strongest $\gamma$ lines of the physics data
(Fig.~\ref{fig:k40k42larveto}, see also Fig.~\ref{fig:g2-bgndspectra}). The
$\gamma$ line at 1525\,keV is due to the decay of $^{42}$K, a $\beta$-$\gamma$
cascade, in which the $\beta$ particle can deposit up to 2\,MeV in the
LAr. The LAr veto system suppresses this line by typically a factor of 5.  On
the other hand, the $^{40}$K line at 1461\,keV, a single $\gamma$ line, is not
suppressed since it follows electron capture of $^{40}$K without any energy
deposition in the LAr. Hence, no suppression is expected apart from random
coincidences so that this line can be used to determine the LAr veto
acceptance independently of the pulser.

\subsection{Muon veto system}	       
\begin{figure}[h]
\begin{center}
\includegraphics[width=0.99\columnwidth]{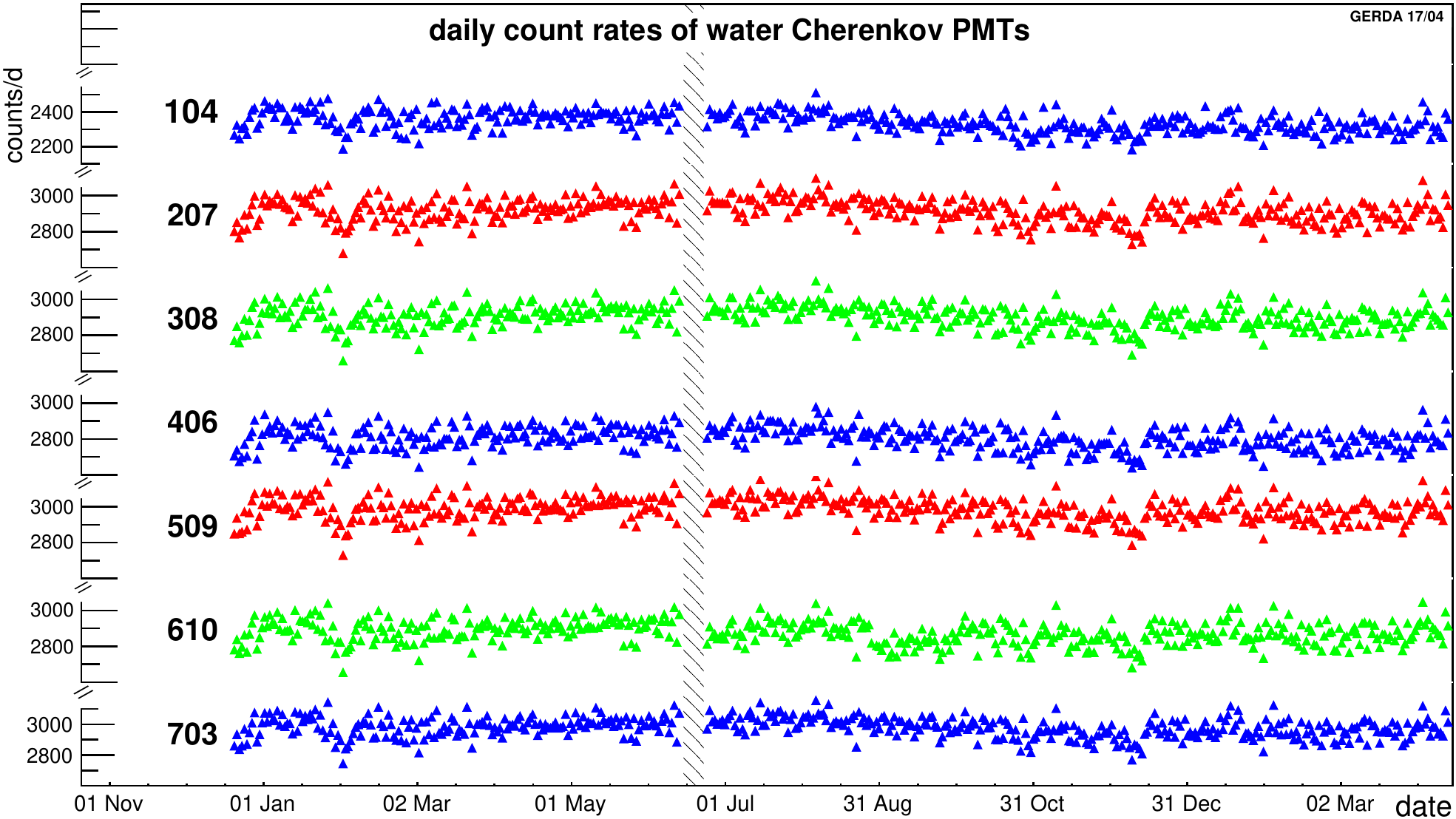}
\caption{\label{fig:muon_stable}
         Daily rates of water Cherenkov PMTs, one from each ring
         (c.f. Ref.~\cite{mv_knies}).
}
\end{center}
\end{figure}

Fig.~\ref{fig:muon_stable} shows for 16 months in 2016/17 the daily rate of one PMT 
of each of the seven PMT rings in the water tank~\cite{g1-instr},
i.e. of varying height in the water.  Only the standard trigger requirements
are requested: either 5 Cherenkov PMTs within 60\,ns with a signal above
threshold of 0.5\,PE, or a triple coincidence within the plastic veto.  A
mean stability of 4\,\% can be observed. No readjustment of the HV was
necessary during the 477 days of operation of Phase~II. 
The mean daily rate of 3164(6) muons was measured which translates into a
rate of 3.54$\cdot 10^{-4}$/(m$^2\cdot$s).  This is a 1.8\,\% deviation from
the mean of Phase~I~\cite{mv_modu}, still consistent and giving hints on
systematic uncertainties.
		  
A further proof of the reliability of the muon veto performance comes through
the analysis of the muon seasonal variation. Despite the shorter measurement
period essentially the same parameters were derived as
previously~\cite{mv_modu}.

\subsection{Background levels}
\subsubsection{Spectra}
\begin{figure*}[htb!]
\begin{center}
\includegraphics[width=2.0\columnwidth]{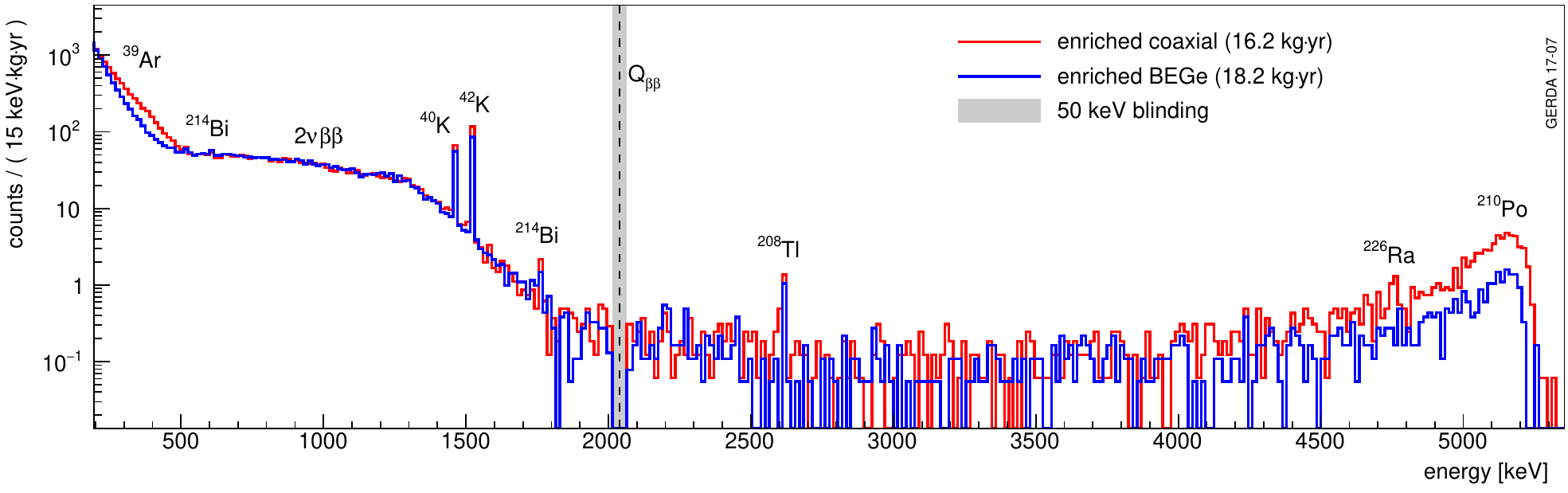}
\caption{\label{fig:g2-bgndspectra}
        Normalized \gerda\ Phase~II background spectra from BEGe and enriched
        coaxial detectors before pulse shape discrimination and LAr veto. The
        grey bar indicates the blinded energy region around \qbb. 
}
\end{center}
\end{figure*}

Fig.~\ref{fig:g2-bgndspectra} shows the exposure-normalized Phase~II
background spectra obtained with the BEGe and enriched semi-coaxial detectors
after quality cuts but before LAr veto and PSD.
Both spectra agree well, exhibiting the same prominent features as observed in
Phase~I: below 500\,keV the tail of the $^{39}$Ar $\beta$-spectrum, between
600\,keV and 1600\,keV the broad structure from the $2\nu\beta\beta$ decays,
individual $\gamma$-lines between 400 and 2650\,keV, and $\alpha$-structures
above 3500\,keV predominantly due to $^{210}$Po and $^{226}$Ra decays.

The observed differences for the tail of the $^{39}$Ar $\beta$-spectrum
originate from the different dead layers in BEGe and coaxial detectors, in
particular from the large and thin p$^+$ contact in the bore hole of the
latter detector type.

The pulse shapes of the observed $\alpha$-events reveal them to be
predominantly located at the surface of the p$^+$ contact. While the count
rate is relatively low and similar in case of the BEGe detectors, it is larger
for the coaxial ones where one detector (ANG4) contributes more than 50\% to
the total rate.  The origin of the contamination is not understood, but there
is evidence that it occurred during the fabrication process.

\subsubsection{Intensity of $\gamma$ lines}
The intensities of the $\gamma$ lines carry information about the nature and
location of various contamination sources~\cite{g1-back}. While the detailed
model study of the background is in progress, we note here that the count
rates of the $^{40}$K and $^{42}$K lines are increased strongly in Phase~II,
by factors of 4 and 2 compared to Phase~I, respectively.  Since the lines are
located well below Q$_{\beta\beta}$ they will not affect the background index.
The increased $^{40}$K intensity might be due to the different kind and
increased number of cables which run in Phase~II along the detector array as
well as to the LAr veto system. The increased $^{42}$K intensity is understood
qualitatively by the change from metallic to plastic mini-shrouds. The latter
no longer shield the electric fields of the Ge detectors so that more $^{42}$K
ions can be attracted closer to the detectors.  For all other $\gamma$ lines
the count rates are similar to those observed in Phase~I; the intensities of
the $^{214}$Bi $\gamma$-lines is even lower than in Phase~I. This is a
remarkable success considering the additional amount of material that has been
deployed in Phase~II close to the detector array.

\subsubsection{Background index at \qbb }
Table~\ref{tab:BI} collects data in the analysis window\footnote{The analysis
window extends from 1930\,keV to 2190\,keV excluding the intervals
2104(5)\,keV and 2119(5)\,keV of known $\gamma$ lines from \Tl\
(SEP) and \Bi.  In addition, the 50\,keV wide blinded interval centered at
\qbb\ is excluded for the Phase~IIb coaxial data which are not yet
unblinded.  } around \qbb\ which are used to extract the BI in Phase~II.
In addition, to the Phase~IIa data (December 2015 - June 2016) of the first
data release~\cite{g2-nature}, data of subsequent physics runs until April
2017 (Phase~IIb) are shown which exhibit a more than 50\,\% larger exposure.
\begin{table}[htb!]
\centering
\caption{\label{tab:BI}
        Counts in the analysis window observed in Phase~IIa and IIb with
        indicated exposures ${\cal E}$ after anti-coincidences within the
        detector array (AC), after anti-coincidences and LAr veto (LAr) or
        pulse shape discrimination (PSD), and after all cuts (all).  The
        quoted background indices BI in units (biu) of \dctsper\ have been
        determined after all cuts; the uncertainties are statistical and
        deduced for a Poissonian signal. 
}
\begin{tabular}{lrrrrrl}
\hline
det/phase      &  \multicolumn{1}{c}{${\cal E}$} & AC & LAr& PSD & all & \multicolumn{1}{c}{BI} \\
              &  \multicolumn{1}{c}{(kg$\cdot$yr)}& cts & cts & cts & cts & (biu)  \\
\hline	      
Coax/IIa             & 5.0    & 19 & 12 &  8 & 4 & 3.5$^{+2.1}_{-1.5}$ \\
\hphantom{Coax}/IIb*  &11.2    & 38 & 14 & 17 & 5 & \\
\cline{2-7} \\[-2ex]
total*                &16.2    & 57 & 26 & 25 & 9 & 2.7$^{+1.0}_{-0.8}$ \\
\hline

BEGe/IIa          & 5.8   & 21 &  6 &  5 & 1 & 0.7$^{+1.1}_{-0.5}$ \\
\hphantom{BEGe}/IIb&12.4  & 30 & 11 &  6 & 3 & \\
\cline{2-7} \\[-2ex]
total             &18.2   & 51 & 17 & 11 & 4 & 1.0$^{+0.6}_{-0.4}$ \\
\hline
\end{tabular}
* Values preliminary since PSD cuts are not yet finalized.
\end{table}
These data with higher statistics confirm the BI reported from the first data
release with increased significance yielding in total
(2.7$^{+1.1}_{-0.8}$)$\cdot$\dctsper\ for the coaxial and
(1.0$^{+0.6}_{-0.4}$)$\cdot$\dctsper\ for the BEGe detectors. This
corroborates that the goal of the Phase~II upgrade, an improvement of the BI
by a factor of 10 compared to Phase~I, has been achieved.

Table~\ref{tab:BI} shows also the counts in the analysis window after various
cuts. After anti-coincidences within the detector array these counts
correspond to a BI below (1-2)$\cdot$10$^{-2}$\,cts/(keV$\cdot$kg$\cdot$yr)
for both BEGe and coaxial detectors.  Within the limited statistics, the LAr
veto and PSD seem to suppress background events with comparable strength. As
first evidence for various contributing suppression mechanisms it is noted
that of the 108 events surviving the anti-coincidence cut, 23 resp. 30 events
are cut exclusively by the LAr veto and PSD, while 42 events are cut by both
LAr veto and PSD.

\section{Conclusion}
\gerda\ continues to exploit novel technologies in order to study neutrinoless
double beta decay of \gesix\ with unprecedented low background rate. Operating
in Phase~I for the first time about 18\,kg of bare enriched Ge detectors in
LAr, it achieved a background index of \pIbi, reaching a sensitivity of
2.4$\cdot 10^{25}$ yr after an exposure of 21.6\,kg$\cdot$yr, and setting the
most stringent \onbb\ half-life limit for \gesix.

In order to reach the goal of Phase~II, a further reduction of the background
index by another order of magnitude to \pIIbi, two major pioneering measures
have been taken: (i) the instrumentation of the LAr surrounding the detectors
with PMTs and fibers read out by SiPMs; this allows the identification and
vetoing of background events by the scintillation light from their energy
deposition in the LAr, and (ii) the deployment of an additional 20\,kg of
enriched BEGe detectors that exhibit superior pulse shape discrimination and
energy resolution compared to the coaxial detector type prevailing in Phase~I.
These actions necessitated an upgrade of several infrastructure components
respecting thereby the strict requirements on the radiopurity of the newly
introduced materials. Affected components include lock, cabling, detector
mounts and replacement of the copper mini-shrouds by transparent ones.

The upgrade to Phase~II including commissioning was concluded at the end of
2015 and the experiment is taking physics data with a duty factor of more than
90\,\% since.  Both the Ge detectors as well as the LAr veto system exhibit
stable operation within specifications.  The leakage currents of 39 out of the
40 deployed Ge detectors are stable since more than one year of continuous
operation in LAr, and often even reduced compared to the beginning.  At
\qbb\ the interpolated energy resolution of coaxial and BEGe detectors is on
average 3.90(7)\,keV and 2.93(6)\,keV FWHM, respectively. Regular calibrations
establish the gain drifts of the Ge detector readout chains to be smaller than
1.5\,keV around \qbb, i.e. to stay within $\pm1\sigma$ of the energy
resolution; this allows to combine data sets accumulated during more than one
year without significant deterioration of energy resolution.  The combined
data from these calibrations runs are used to optimize event selection
criteria and efficiencies of pulse shape discrimination.  The performance of
the LAr veto system has been established by dedicated calibrations runs with
\thzza\ and \Ra\ sources yielding suppression factors at \qbb\ of about 100
and 6, respectively. Quasi-continuous monitoring during the physics runs is
achieved by analysis of the relatively strong \kvn\ and $^{42}$K $\gamma$
lines at 1461\,keV and 1525\,keV.

Based on an exposure of 10.8\,kg$\cdot$yr results of a first period of the
Phase~II physics run have been published recently~\cite{g2-nature}. They
demonstrate that the goal of Phase~II, a background index of about \pIIbi, has
been achieved. This finding is statistically corroborated here with data from
a more than doubled exposure of 23.6\,kg$\cdot$yr.  \gerda\ will thus remain
background-free up to its design exposure of 100\,kg$\cdot$yr reaching thereby
a sensitivity beyond $10^{26}$ yr.

\begin{acknowledgements}
The \gerda\ experiment is supported by the German 
Federal Ministry for Education and Research (BMBF), the German Research 
Foundation (DFG) via the Excellence Cluster Universe, the Italian Istituto 
Nazionale di Fisica Nucleare (INFN), the Max Planck Society (MPG), the Polish 
National Science Centre (NCN), the Foundation for Polish Science (TEAM/2016-2/17),
the Russian Foundation for Basic Research 
(RFBR) and the Swiss National Science Foundation (SNF). These research 
institutions acknowledge internal financial support.

This project has received funding/support from the European Union's Horizon
2020 research and innovation programme under the Marie Sklodowska-Curie grant
agreements No 690575 and No 674896.

The \gerda\ Collaboration thanks the directors and the staff of the 
\lngs\ for their support of the \gerda\ experiment.
\end{acknowledgements}

\vspace*{50mm}

\begin{table*}[htb!]
\begin{center}
\caption{
\label{table:detproperties}
    Main parameters of all detectors used in \textsc{Gerda} Phase~II.  The GD,
    ANG and RG detectors are made of germanium enriched in $^{76}$Ge from
    85.5\,\% to 88.3\,\%.  The three GTF detectors are made from natural
    germanium. The operational voltages recommended by the manufacturer are
    quoted.  A `y' marks in column 4 the detectors with a passivation layer
    (PL) in the groove.  The position number in a given string increases from
    top to bottom.  The active masses of the newly produced BEGe detectors
    include a correction that considers a full charge collection depth growth
    occurred during storage at room temperature in the three years before
    deployment in \textsc{Gerda}. Finally, the full energy peak detector
    efficiencies $\epsilon_{fep}$ for the $0\nu\beta\beta$ decay in
    $^{76}${Ge} are quoted.
}
\begin{tabular}{|l|l|ccccccc|}
\hline
Nr.     & Detector      & V$_{rec}$	  & With& String \&     &      $f_{Ge76}$              & {$M_{diode}$} & {$M_{av}$}\,$^{{+ucorr+corr}}_{{-ucorr-corr}}$ & $\epsilon_{fep}\pm{ucorr}\pm{corr}$\\
	&    		& [kV]		  & PL	& Position      &                              & [g] 		&    [g]                                         &             \\
\hline
13	&	GD32A	&	3.0	  &	&III-2 	&      0.877	$\pm$	0.013	&	458	&	404	$^{+	10	+	4	}_{-	10	-	2	}$	&	0.888	$\pm$	0.001	$\pm$	0.002	\\
12	&	GD32B	&	4.0	  &	&III-1	&      0.877	$\pm$	0.013	&	716	&	632	$^{+	10	+	4	}_{-	10	-	2	}$	&	0.900	$\pm$	0.001	$\pm$	0.002	\\
14	&	GD32C	&	4.0	  &	&III-3	&      0.877	$\pm$	0.013	&	743	&	665	$^{+	10	+	4	}_{-	10	-	2	}$	&	0.901	$\pm$	0.001	$\pm$	0.002	\\
34	&	GD32D	&	4.0	  &	&VI-4	&      0.877	$\pm$	0.013	&	720	&	657	$^{+	10	+	5	}_{-	10	-	2	}$	&	0.900	$\pm$	0.001	$\pm$	0.002	\\
24	&	GD35A	&	4.0	  &	&IV-5	&      0.877	$\pm$	0.013	&	768	&	693	$^{+	13	+	3	}_{-	13	-	2	}$	&	0.904	$\pm$	0.001	$\pm$	0.002	\\
1	&	GD35B	&	4.0	  &	&I-1	&      0.877	$\pm$	0.013	&	810	&	740	$^{+	11	+	5	}_{-	11	-	2	}$	&	0.902	$\pm$	0.001	$\pm$	0.002	\\
19	&	GD35C	&	3.5	  & y	&IV-0	&      0.877	$\pm$	0.013	&	634	&	572	$^{+	9	+	4	}_{-	9	-	3	}$	&	0.893	$\pm$	0.001	$\pm$	0.002	\\
4	&	GD61A	&	4.5	  & y	&I-4	&      0.877	$\pm$	0.013	&	731	&	652	$^{+	12	+	4	}_{-	11	-	3	}$	&	0.902	$\pm$	0.001	$\pm$	0.002	\\
26	&	GD61B	&	4.0	  & y	&IV-7	&      0.877	$\pm$	0.013	&	751	&	666	$^{+	12	+	5	}_{-	12	-	2	}$	&	0.899	$\pm$	0.001	$\pm$	0.002	\\
16	&	GD61C	&	4.0	  &	&III-5	&      0.877	$\pm$	0.013	&	634	&	562	$^{+	10	+	5	}_{-	10	-	3	}$	&	0.892	$\pm$	0.001	$\pm$	0.002	\\
17	&	GD76B	&	3.5	  & y	&III-6	&      0.877	$\pm$	0.013	&	384	&	326	$^{+	7	+	3	}_{-	7	-	2	}$	&	0.883	$\pm$	0.001	$\pm$	0.002	\\
20	&	GD76C	&	3.5	  & y	&IV-1	&      0.877	$\pm$	0.013	&	824	&	723	$^{+	12	+	5	}_{-	12	-	2	}$	&	0.902	$\pm$	0.001	$\pm$	0.002	\\
32	&	GD79B	&	3.5	  &	&VI-2	&      0.877	$\pm$	0.013	&	736	&	648	$^{+	13	+	5	}_{-	13	-	2	}$	&	0.897	$\pm$	0.001	$\pm$	0.002	\\
23	&	GD79C	&	3.5	  &	&IV-4	&      0.877	$\pm$	0.013	&	812	&	713	$^{+	11	+	5	}_{-	11	-	2	}$	&	0.900	$\pm$	0.001	$\pm$	0.002	\\
35	&	GD89A	&	4.0	  &	&VI-5	&      0.877	$\pm$	0.013	&	524	&	462	$^{+	10	+	3	}_{-	9	-	2	}$	&	0.893	$\pm$	0.001	$\pm$	0.002	\\
5	&	GD89B	&	3.5	  & y	&I-5	&      0.877	$\pm$	0.013	&	620	&	533	$^{+	12	+	4	}_{-	12	-	2	}$	&	0.890	$\pm$	0.001	$\pm$	0.002	\\
15	&	GD89C	&	4.0	  & y	&III-4	&      0.877	$\pm$	0.013	&	595	&	520	$^{+	12	+	5	}_{-	11	-	2	}$	&	0.889	$\pm$	0.001	$\pm$	0.002	\\
21	&	GD89D	&	4.0	  &	&IV-2	&      0.877	$\pm$	0.013	&	526	&	454	$^{+	9	+	5	}_{-	9	-	2	}$	&	0.884	$\pm$	0.001	$\pm$	0.002	\\
0	&	GD91A	&	3.5	  &	&I-0	&      0.877	$\pm$	0.013	&	627	&	557	$^{+	10	+	3	}_{-	11	-	2	}$	&	0.898	$\pm$	0.001	$\pm$	0.002	\\
25	&	GD91B	&	3.5	  &	&IV-6	&      0.877	$\pm$	0.013	&	650	&	578	$^{+	10	+	5	}_{-	10	-	2	}$	&	0.897	$\pm$	0.001	$\pm$	0.002	\\
7	&	GD91C	&	4.0	  & y	&I-7	&      0.877	$\pm$	0.013	&	627	&	556	$^{+	11	+	4	}_{-	11	-	2	}$	&	0.896	$\pm$	0.001	$\pm$	0.002	\\
33	&	GD91D	&	4.5	  &	&VI-3	&      0.877	$\pm$	0.013	&	693	&	615	$^{+	12	+	5	}_{-	12	-	2	}$	&	0.899	$\pm$	0.001	$\pm$	0.002	\\
30	&	GD00A	&	2.5	  & y	&VI-0	&      0.877	$\pm$	0.013	&	496	&	439	$^{+	8	+	3	}_{-	9	-	2	}$	&	0.888	$\pm$	0.001	$\pm$	0.002	\\
3	&	GD00B	&	3.5	  &	&I-3	&      0.877	$\pm$	0.013	&	697	&	613	$^{+	12	+	5	}_{-	12	-	2	}$	&	0.897	$\pm$	0.001	$\pm$	0.002	\\
18	&	GD00C	&	3.5	  & y	&III-7	&      0.877	$\pm$	0.013	&	815	&	727	$^{+	14	+	5	}_{-	13	-	2	}$	&	0.903	$\pm$	0.001	$\pm$	0.002	\\
22	&	GD00D	&	3.5	  & y	&IV-3	&      0.877	$\pm$	0.013	&	813	&	723	$^{+	13	+	5	}_{-	13	-	2	}$	&	0.902	$\pm$	0.001	$\pm$	0.002	\\
11	&	GD02A	&	2.5	  & y	&III-0	&      0.877	$\pm$	0.013	&	545	&	488	$^{+	8	+	3	}_{-	8	-	2	}$	&	0.893	$\pm$	0.001	$\pm$	0.002	\\
2	&	GD02B	&	3.0	  &	&I-2	&      0.877	$\pm$	0.013	&	625	&	553	$^{+	10	+	4	}_{-	10	-	2	}$	&	0.895	$\pm$	0.001	$\pm$	0.002	\\
31	&	GD02C	&	3.5	  &	&VI-1	&      0.877	$\pm$	0.013	&	788	&	700	$^{+	13	+	5	}_{-	13	-	2	}$	&	0.901	$\pm$	0.001	$\pm$	0.002	\\
6	&	GD02D$^a$&	4.0	  & y	&I-6	&      0.877	$\pm$	0.013	&	662	&	552	$^{+	11	+	0	}_{-	11	-	2	}$	&	not defined, see remark			\\
\hline																												
36	&	ANG1	&	4.0	  &	&VI-6	&      0.859	$\pm$	0.029	&	958	&	795	$^{+	43	+	26	}_{-	43	-	26	}$	&	0.889	$\pm$	0.018			\\
27	&	ANG2	&	4.0	  & y	&V-0	&      0.866	$\pm$	0.025	&	2833	&	2468	$^{+	121	+	80	}_{-	121	-	80	}$	&	0.918	$\pm$	0.018			\\
10	&	ANG3	&	3.5	  & y	&II-2	&      0.883	$\pm$	0.026	&	2391	&	2070	$^{+	118	+	60	}_{-	118	-	67	}$	&	0.916	$\pm$	0.018			\\
29	&	ANG4	&	3.0	  & y	&V-2	&      0.863	$\pm$	0.013	&	2372	&	2136	$^{+	116	+	69	}_{-	116	-	69	}$	&	0.916	$\pm$	0.018			\\
8	&	ANG5	&	2.5	  &	&II-0	&      0.856	$\pm$	0.013	&	2746	&	2281	$^{+	109	+	74	}_{-	109	-	74	}$	&	0.918	$\pm$	0.018			\\
9	&	RG1	&	5.0	  &	&II-1	&      0.855	$\pm$	0.015	&	2110	&	1908	$^{+	109	+	62	}_{-	109	-	62	}$	&	0.915	$\pm$	0.018			\\
28	&	RG2	&	4.0	  &	&V-1	&      0.855	$\pm$	0.015	&	2166	&	1800	$^{+	99	+	58	}_{-	99	-	58	}$	&	0.912	$\pm$	0.018			\\
\hline
38	&	GTF32	&	3.5	  & y	&VII-1	&      0.078	$\pm$	0.001	&	2321	&	2251	$^{+	116			}_{-	116			}$	&	0.92	$\pm$	0.018			\\
39	&	GTF45\_2&	3.5	  &	&VII-2	&      0.078	$\pm$	0.001	&	2312	&	1965										&	0.92	$\pm$	0.018			\\
37	&	GTF112	&	3.5	  & y	&VII-0	&      0.078	$\pm$	0.001	&	2965	&	2522										&	0.92	$\pm$	0.018			\\
\hline
\end{tabular}
\end{center}
{\footnotesize
Remarks:\\
$^a$ Detector GD02D does not deplete due to an unsuitable impurity concentration. The material is 
rather a pn junction than of p-type.}
\end{table*} 

{\bf Appendix}
\vfill

\hphantom{a}
\vskip1truecm

\begin{figure*}[h!]
\begin{center}
\includegraphics[width=120mm]{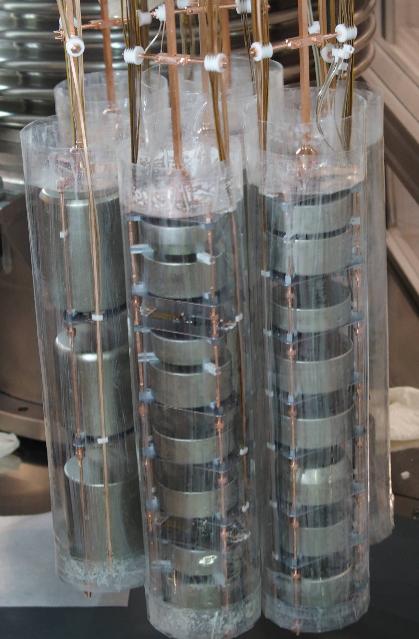}
\caption{\label{fig:arrayfoto}
        Photo of \gerda\ Phase~II detector array, showing from left to right
        string 2 with three semi-coaxial detectors, and strings 3 and 4 with
        eight BEGe detectors, respectively. Each of the seven strings is
        enclosed by a transparent mini-shroud.
}
\end{center}
\end{figure*}

\newpage

\begin{figure*}[ht!]
\begin{center}
\includegraphics[width=80mm]{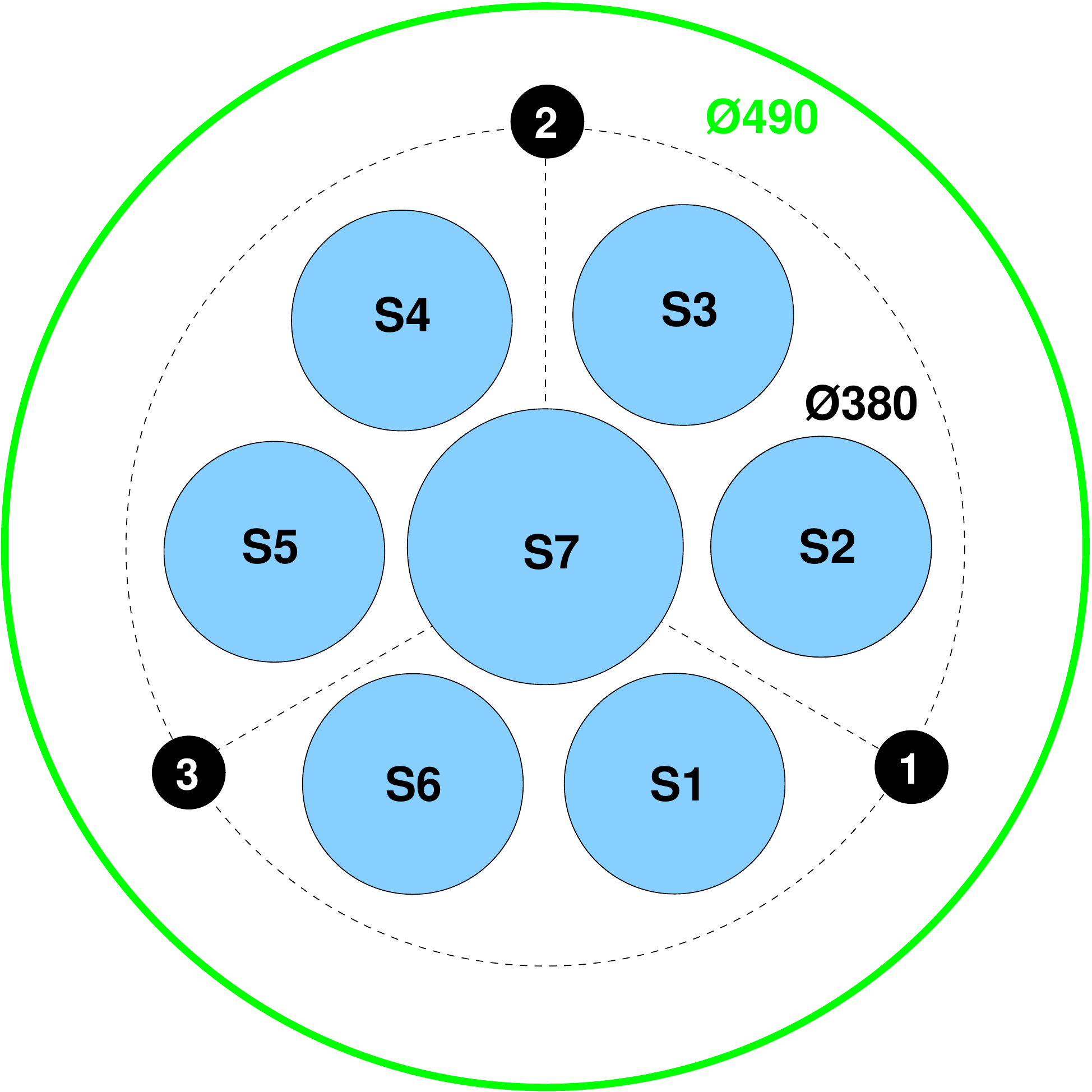}
\caption{\label{fig:arfromtop}
      View of the array from top. The positions of the three calibration
      sources (1, 2, 3) and the boundaries of the LAr veto system
      (\O\,490\,mm) are indicated. The surrounding radon shroud with a
      diameter of 750\,mm~\cite{g1-instr} is not shown.
}
\end{center}
\end{figure*}

\begin{figure*}[h!]
\begin{center}
\includegraphics[scale=0.2]{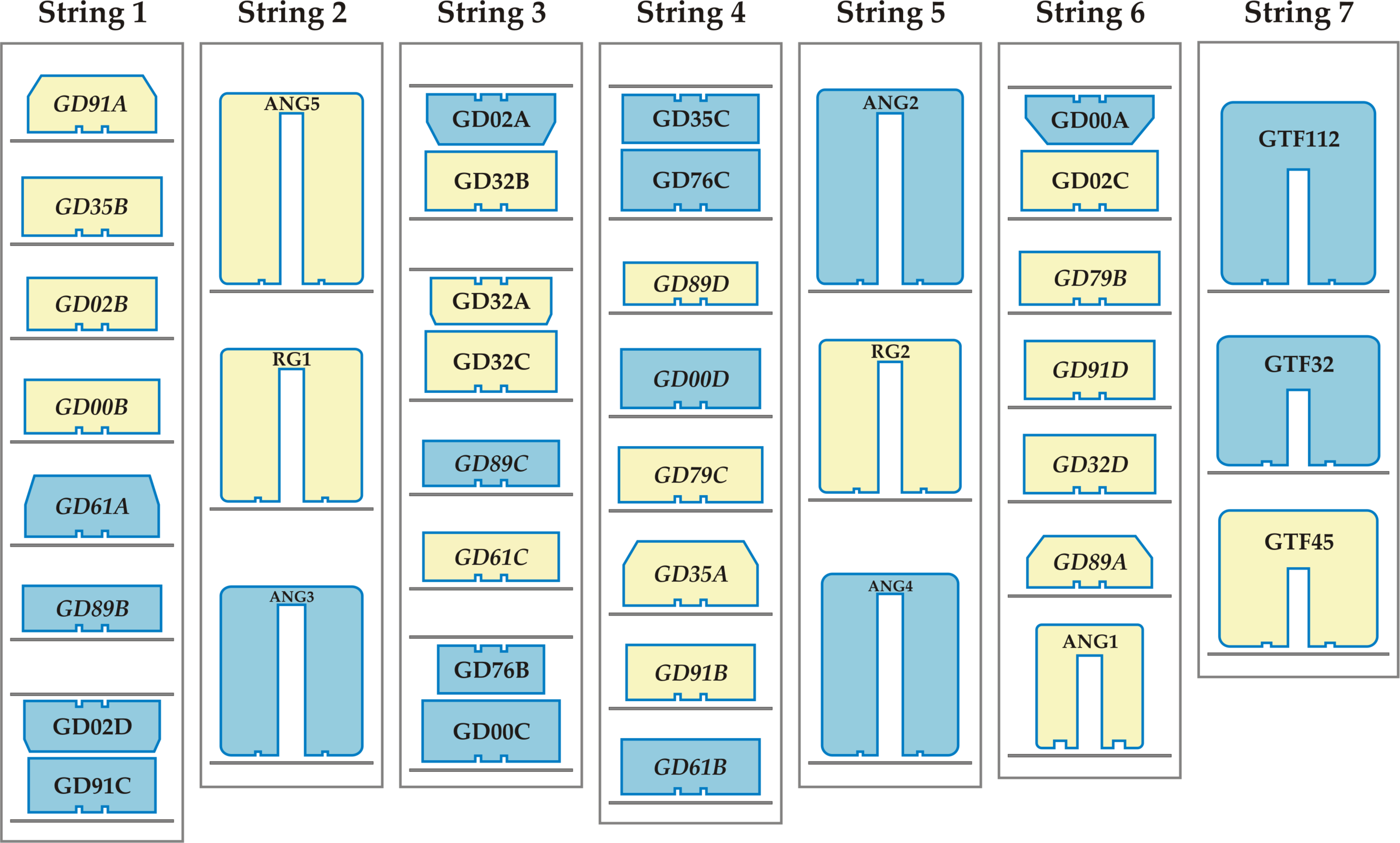}
\caption{\label{fig:stringcontent}
      Distribution of the enriched  BEGe (GDxxx) and semi-coaxial (RGx and
      ANGx) detectors in the various strings of the \gerda\ Phase~II detector
      array; the natural low-background semi-coaxial detectors carry the
      labels GTFxx. Blue colored detectors carry the manufacturer's
      passivation on the insulating groove between the p$^+$ and n$^+$
      contact; the yellow colored ones have this layer removed. Horizontal
      grey lines indicate the positions of the silicon plates (see
      Fig.~\ref{fig:bege-mount}). \newline
      Detectors are numbered from top to bottom string by string, starting
      with nr.\,0 in string 1 (GD91A) and ending with nr.\,39 in string 7 (GTF45).
}
\end{center}
\end{figure*}

\newpage

\begin{figure*}[htb!]
\begin{center}
\includegraphics[width=170mm]{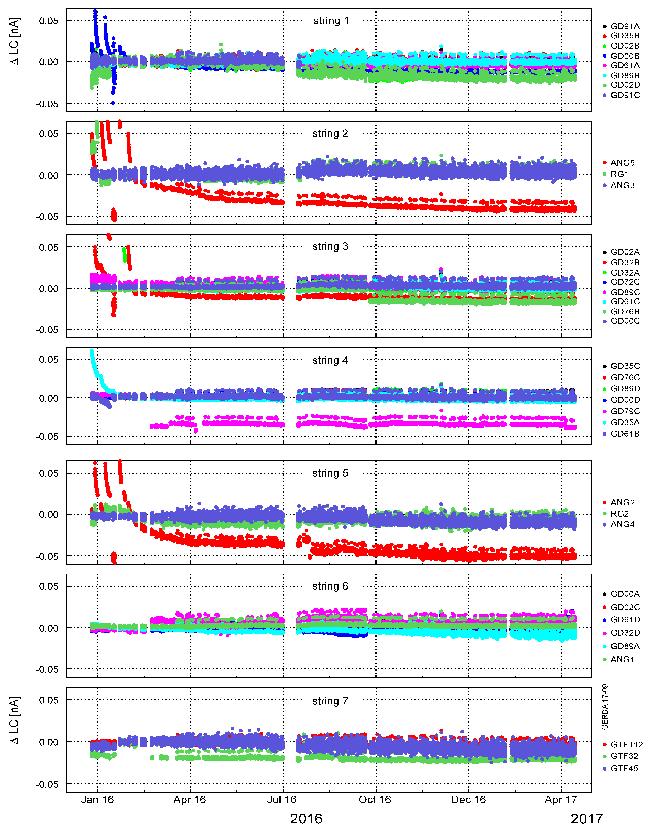}
\caption{\label{fig:bslvst1}
         Leakage currents of indicated detectors in the period from
         December 2015 to April 2017. }
\end{center}
\end{figure*}

\end{document}

%% file: gerda-abbreviations.tex
\usepackage{upgreek}     
\newcommand{\gspec}      {$\gamma$\,\,spec.}
\newcommand{\ctsper}      {cts/(keV$\cdot$kg$\cdot$yr)}

\newcommand{\pIbi}        {{$10^{-2}$~cts/(keV$\cdot$kg$\cdot$yr)}}
\newcommand{\dctsper}     {{$10^{-3}$~cts/(keV$\cdot$kg$\cdot$yr)}}
\newcommand{\pIIbi}       {{$10^{-3}$~cts/(keV$\cdot$kg$\cdot$yr)}}

\newcommand{\vctsper}     {{$10^{-4}$~cts/(keV$\cdot$kg$\cdot$yr)}}

\newcommand{\kgyr}        {{kg$\cdot$yr}}

\newcommand{\mum}         {{$\upmu$m}}
\newcommand{\mus}         {{$\upmu$s}}

\newcommand{\qbb}         {{$Q_{\beta\beta}$}}

\newcommand{\thalfzero}   {${T^{0\nu}_{1/2}}$}

\newcommand{\onbb}        {{$0\nu\beta\beta$}}

\newcommand{\nnbb}        {{$2\nu\beta\beta$}}

\newcommand{\gerda}       {\textsc{Gerda}}

\newcommand{\GERDA}       {\mbox{\textsc{Gerda}}}  
\newcommand{\lngs}        {{\mbox{\textsc{Lngs}}}}
\newcommand{\LNGS}        {{\mbox{\textsc{Lngs}}}}

\newcommand{\LArGe}       {\textsc{LArGe}}

\newcommand{\borex}       {\mbox{\textsc{Borexino}}}
\newcommand{\borexino}       {\mbox{\textsc{Borexino}}}

\newcommand{\gelatio}     {\textsc{Gelatio}}

\newcommand{\gesix}       {{$^{76}$Ge}}

\newcommand{\natcoax}     {{$^{\rm nat}$Coax}}

\newcommand{\thzza}       {{$^{228}$Th}}

\newcommand{\kvn}         {{$^{40}$K}}

\newcommand{\Ra}          {$^{226}$Ra}

\newcommand{\Bi}          {$^{214}$Bi}

\newcommand{\Th}          {$^{228}$Th}
\newcommand{\Tl}          {$^{208}$Tl}

